\documentclass[structabstract]{aa} 
\usepackage{graphicx}
\usepackage{txfonts}
\usepackage{natbib}
\usepackage{longtable}
\usepackage{lscape}
\usepackage{multirow}
\usepackage{color}
\usepackage{lscape}

\bibpunct{(}{)}{;}{a}{}{,}

\newcommand{\kms}{km\,s$^{-1}$}

\newcommand{\degree}{$^{\circ}$}

\newcommand{\nodata}{...}

\begin{document}
\title{Molecular clouds and star formation toward the Galactic plane within 216.25\degree $\le$ $l$ $\le$ 218.75\degree and $-$0.75\degree$\le$ $b$ $\le$ 1.25\degree}
\author{Y.~Gong\inst{1,2}, R.~Q.~Mao\inst{1}, M.~Fang\inst{1}, S.~B.~Zhang\inst{1}, Y.~Su\inst{1}, J.~Yang\inst{1}, Z.~B.~Jiang\inst{1}, Y.~Xu\inst{1}, M.~Wang\inst{1}, Y.~Wang\inst{1}, D.~R.~Lu\inst{1}, J.~X.~Sun\inst{1}}
\institute{Purple Mountain Observatory \& Key Laboratory of Radio Astronomy, Chinese Academy of Sciences, 2 West Beijing Road, 210008 Nanjing, PR China 
\and 
University of Chinese Academy of Sciences, No. 19A Yuquan Road, 100049 Beijing, PR China
}

   \date{}

\abstract{Molecular clouds trace the spiral arms of the Milky Way and all its star forming regions. Large-scale mapping of molecular clouds will provide an approach to understand the processes that govern star formation and molecular cloud evolution.}
{As a part of the Milky Way Imaging Scroll Painting (MWISP) survey, the aim is to study the physical properties of molecular clouds and their associated star formation toward the Galactic plane within 216.25\degree $\le$ $l$ $\le$ 218.75\degree and $-$0.75\degree$\le$ $b$ $\le$ 1.25\degree, which covers the molecular cloud complex S287.}
{Using the 3$\times$3 Superconducting Spectroscopic Array Receiver (SSAR) at the PMO-13.7m telescope, we performed a simultaneous $^{12}$CO (1--0), $^{13}$CO (1--0), C$^{18}$O (1--0) mapping toward molecular clouds in a region encompassing 3.75 square degrees. We also make use of archival data to study star formation within the molecular clouds.}
	{We reveal three molecular clouds, the 15~\kms\,cloud, the 27~\kms\,cloud, and the 50~\kms\,cloud, in the surveyed region. The 50~\kms\,cloud is resolved with an angular resolution of $\sim$1\arcmin\,for the first time. Investigating their morphology and velocity structures, we find that the 27~\kms\,cloud is likely affected by feedback from the stellar association Mon OB3 and the 50~\kms\,cloud is characterized by three large expanding molecular shells. The surveyed region is mapped in C$^{18}$O (1--0) for the first time. We discover seven C$^{18}$O clumps that are likely to form massive stars, and 15 dust clumps based on the Bolocam Galactic Plane Survey (BGPS) archive data. Using infrared color-color diagrams, we find 56 Class I and 107 Class II young stellar object (YSO) candidates toward a slightly larger region of 5.0 square degrees. Based on the distribution of YSO candidates, an overdensity is found around the H{\scriptsize II} region S287 and the intersection of two shells; this is probably indicative of triggering. The star formation efficiency (SFE) and rate (SFR) of the 27~\kms\,cloud are discussed. Comparing the observed values of the filament S287-main with fragmentation models, we suggest that turbulence controls the large-scale fragmentation in the filament, while gravitational fragmentation plays an important role in the formation of YSOs on small scales. We find that star-forming gas tends to have a higher excitation temperature, a higher $^{13}$CO (1--0) opacity, and a higher column density than non-star-forming gas, which is consistent with the point that star formation occurs in denser gas and star-forming gas is heated by YSOs. Using the 1.1 mm dust emission to trace dense gas, we obtain a dense gas fraction of 2.7\%--10.4\% for the 27~\kms\,cloud.}
	{}

	 \maketitle

	\section{Introduction}\label{t.intro}
	Molecular clouds are also known as stellar nurseries because they are the densest parts of the interstellar medium that may gravitationally collapse, and thus create protostars \citep[e.g.,][]{1987ARA&A..25...23S,2012ARA&A..50..531K}. Molecular clouds comprise a significant fraction of the total mass of interstellar matter in the Milky Way \citep[e.g.,][]{2005pcim.book.....T}. CO is regarded as a good tracer for molecular clouds. CO observations of the face-on spiral galaxy M51 have shown that molecular clouds correlate well with the spiral arms \citep[e.g.,][]{2013ApJ...779...42S}. Resolving molecular clouds in the Milky way can provide information about the spiral arms, while similar physical spatial resolution cannot be easily achieved toward most of external galaxies. A very recent large-scale CO survey has shown the fact that the Scutum-Centaurus Arm might be extended to the outer second quadrant \citep{2015ApJ...798L..27S}. Furthermore, a large-scale survey is essential to improve our understanding of the properties of molecular clouds and large-scale processes that govern star formation, molecular cloud evolution, feedback, etc. \citep[e.g.,][]{1991MNRAS.252..210B,2001ApJ...547..792D,2013A&A...559A..34L,2014AJ....147...46Z,2014ApJ...791..109Z,2014A&A...568A..73R,2014ApJ...788..122S,2015arXiv150807898S}. Therefore, it is very important to study the large-scale properties of molecular clouds and their associated star-forming activities.
	 
\begin{figure*}[!htbp]
\centering
\includegraphics[width = 0.9\textwidth]{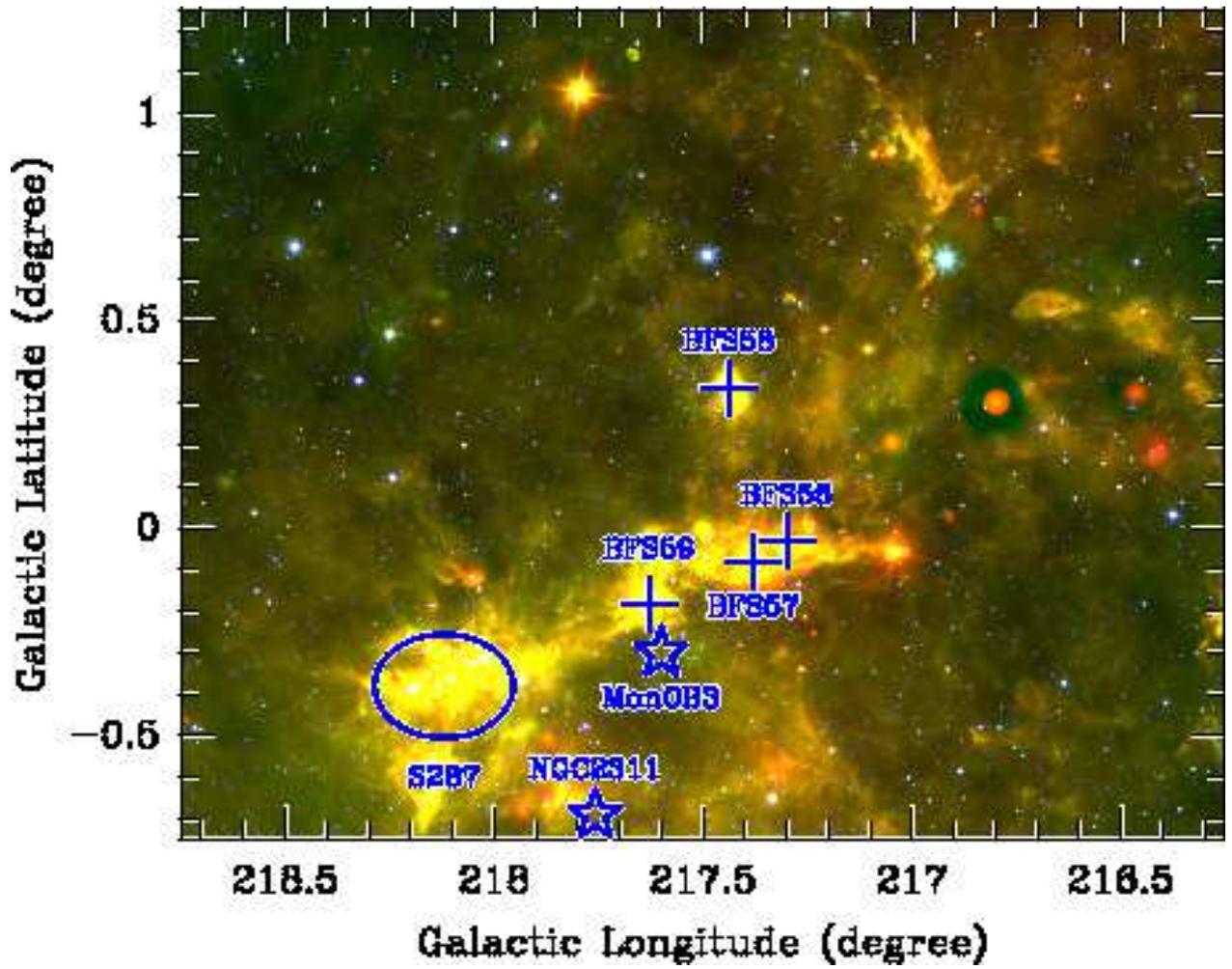}
\caption{{Overview of the surveyed region in a three-color composite image of the 22 $\mu$m (red), the 12 $\mu$m (green) and 4.6 $\mu$m emission (blue) from WISE. The optical H{\scriptsize II} region S287, four BFS sources, the stellar association Mon OB3, and the open cluster NGC 2311 are marked with symbols, near which source names are given. (Descriptions of these objects are presented in Sect.~\ref{t.intro}.)} \label{Fig:guide}}
\end{figure*}

We here present a new survey toward molecular clouds within 216.25\degree\,$\le$ $l$ $\le$ 218.75\degree\, and $-$0.75\degree\,$\le$ $b$ $\le$ 1.25\degree\,in the $J$=1--0 rotational transitions of three main CO isotopologs ($^{12}$CO, $^{13}$CO and C$^{18}$O). This region is part of a molecular cloud complex to the north of the Maddalena cloud \citep{1985ApJ...294..231M}. Figure~\ref{Fig:guide} shows an overview of our surveyed region. This region hosts the bright optical H{\scriptsize II} region S287, which was first discovered by \citet{1959ApJS....4..257S}. It is ionized by the O9.5V star LS VI $-$04 19 \citep{1984NInfo..56...59A}. Another four molecular clouds associated with optical H{\scriptsize II} regions BFS 56, BFS 57, BFS 58, and BFS 59, were found by \citet{1982ApJS...49..183B}. Of these, BFS 57, also known as the bipolar nebula NS 14 \citep[with a size of 80\arcsec;][]{1984A&A...131..200N}, is illuminated by a trapezium system of four stars with spectral types B0.5, B1, B9, and A5 at a distance of 2.3 kpc \citep{1998ApJ...509..749H}. Moreover, the stellar association Mon OB3 lies south of the region \citep[][]{1995AstL...21...10M,2009MNRAS.400..518M}, and the open cluster NGC 2311 is located at ($l$~=~217.7579\degree , $b$~=~$-$0.6916\degree) \citep{2005A&A...438.1163K,2013MNRAS.436.1465B}. There have been some previous CO surveys toward this region. Five CO wings, indicative of molecular outflows, have also been found in the region \citep{1989ESOC...33...95F,2002ApJS..141..157Y}. The area was first mapped by a large, unbiased $^{12}$CO (1--0) Galactic survey with a beam width of $\sim$8\arcmin\,\citep{2001ApJ...547..792D}. The region was further investigated by the Nagoya-4m $^{13}$CO (1--0) survey with a beam width of 2\arcmin$\!\!$.7\ \citep{2004PASJ...56..313K}, and three peaks with $^{13}$CO (1--0) integrated intensities $>$ 1~K~\kms\,were identified. An area of 1.5 square degrees associated with the H{\scriptsize II} region S287 has been mapped in $^{12}$CO (1--0) and $^{13}$CO (1--0) with the FCRAO-14m telescope \citep{1994JKAS...27..147L}. The cloud associated with S287 was highly disturbed, and it was inferred that triggered star formation might occur in the cloud. Here, we observe not only a larger area of 3.75 square degrees with a higher sensitivity and a wider velocity coverage than the previous FCRAO survey, but we also simultaneously provide three main CO isotopologs ($^{12}$CO, $^{13}$CO, and C$^{18}$O; details are presented in Sect.~\ref{Sec:obs}). The three $J$=1--0 rotational transitions can be used to study molecular gas from low- ($\sim$10$^{2}$~cm$^{-3}$) to high-density regions ($\sim$10$^{4}$~cm$^{-3}$) because of their different abundances with respect to H$_{2}$. Furthermore, the region is mapped in C$^{18}$O (1--0) for the first time, and the distribution of young stellar objects (YSOs) in this region has not been studied before.

\section{Observations and data reduction}\label{Sec:obs}
\subsection{PMO-13.7m CO observations}
This work is based on the Milky Way Imaging Scroll Painting (MWISP\footnote{http://www.radioast.nsdc.cn/mwisp.php}) project which is dedicated to the large-scale survey of molecular gas along the northern Galactic Plane. The simultaneous observations of $^{12}$CO (1--0), $^{13}$CO (1--0), and C$^{18}$O (1--0) toward molecular clouds within 216.25\degree $\le$ $l$ $\le$ 218.75\degree and $-$0.75\degree$\le$ $b$ $\le$ 1.25\degree\,were carried out with the PMO-13.7m telescope\footnote{The PMO-13.7 m telescope is operated by the Purple Mountain Observatory (PMO).}, located at Delingha (3200~m altitude) in China, from April 20 to May 9 2012. Supplementary observations were conducted on September 28 2012. The observations took about 51~hours in total to cover 15~cells (3.75 square degrees), each of which is $30$\arcmin $\times$ $30$\arcmin. The $3 \times 3$-beam  Superconducting Spectroscopic Array Receiver (SSAR) system was used as front end. Each fast Fourier transform spectrometer (FFTS) with a bandwidth of 1~GHz provides 16384 channels and a spectral resolution of 61 kHz, equivalent to a velocity coverage of $\sim$2600~\kms\,and a velocity resolution of $\sim$0.17 $\rm{km~s}^{-1}$ at 110~GHz. The detailed properties of this system are described in \citet{pmo2}. A specific local oscillator (LO) frequency was carefully selected so that the upper sideband is centered at the $^{12}$CO (1--0) line and the lower sideband is able to cover the $^{13}$CO (1--0) and C$^{18}$O (1--0) lines. Observations are conducted in position-switch on-the-fly (OTF) mode, scanning the region at a rate of 50\arcsec\,per second with a dump time of 0.3 seconds. Each cell was scanned at least in two orthogonal directions, along the Galactic longitude and the Galactic latitude, to reduce scanning effects. 

All cells were reduced with the GILDAS\footnote{http://www.iram.fr/IRAMFR/GILDAS} software including CLASS and GreG. We used the routine XY\_MAP in CLASS to regrid raw data and then convert them into fits files. The pixel size of these fits files is 30\arcsec$\times$30\arcsec. Using these standard fits files from the survey, we combined related data to make up the final region.

The antenna temperature ($T_{\rm A}$) was calibrated by the standard chopper-wheel method \citep{1976ApJS...30..247U}. All the intensities throughout the paper are converted to a scale of main beam temperatures with the relation $T_{\rm mb}=T_{\rm A}/B_{\rm eff}$, where the beam efficiency $B_{\rm eff}$ is 46\% at 115~GHz and 49\% at 110~GHz according to the status report\footnote{http://www.radioast.nsdc.cn/zhuangtaibaogao.php} of the PMO-13.7m telescope. The calibration accuracy is estimated to be within 10\%. Typical system temperatures were 191-387~K at the upper sideband, and 142-237~K at the lower sideband. The typical sensitivity is about 0.5~K ($T_{\rm mb}$) for $^{12}$CO (1--0), and 0.3~K ($T_{\rm mb}$) for $^{13}$CO (1--0) and C$^{18}$O (1--0) at a channel width of $\sim 0.17~\rm{km~s^{-1}}$. The beam widths are about 55\arcsec\,and 52\arcsec\,at 110~GHz and 115~GHz, respectively. The pointing of the telescope has an rms accuracy of about 5\arcsec. Throughout this paper, velocities are all given with respect to the local standard of rest (LSR).

\subsection{Archival data}
 We obtained infrared data from the Wide-field Infrared Survey Explorer (WISE) which mapped the full sky at 3.4, 4.6, 12, and 22 $\mu$m (W1, W2, W3, W4) \citep{2010AJ....140.1868W}. Using the WISE infrared data in combination with the $J$, $H$, $K_{\rm s}$ band data of the Two Micron All Sky Survey (2MASS) \citep{2006AJ....131.1163S}, we searched for the disk-bearing YSO candidates in the studied region (details are described in Sect.~\ref{SF}). We made use of the Bolocam Galactic Plane Survey (BGPS) of 1.1 mm dust continuum emission \citep{2010ApJS..188..123R,2011ApJS..192....4A,2013ApJS..208...14G} to study dust clumps and the dense gas mass fraction of molecular clouds (details are described in Sects.~\ref{t:clumps} and \ref{t:frac}). In the following text, the dust emission image is convolved with a Gaussian kernel of 44\arcsec\,to achieve a better signal-to-noise ratio, but the final angular resolution becomes 55\arcsec, coarser than the original beam size of 33\arcsec.
\section{Results}\label{Sec:res}
\subsection{Properties of molecular clouds}\label{t:cloud}
\begin{figure*}[!htbp]
\centering
\includegraphics[width = 0.9\textwidth]{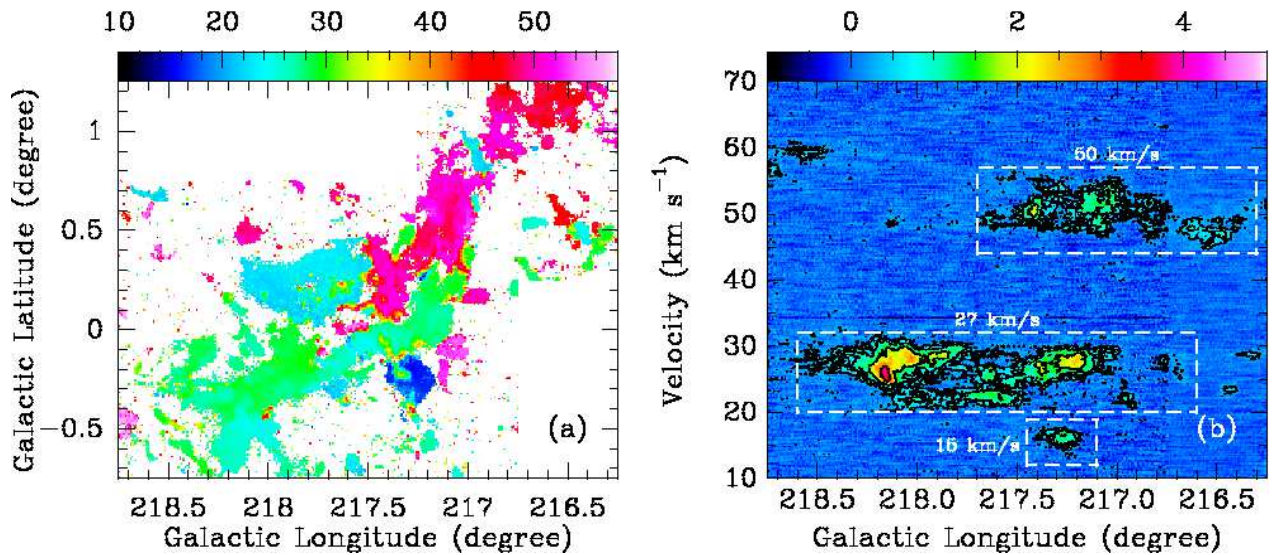}
\caption{{(a): $^{12}$CO (1--0) intensity-weighted velocity map. The color bar represents the velocity in units of \kms. (b): The position-velocity diagram of the $^{12}$CO emission averaged over the observed range in latitude and plotted against its longitude. The color bar represents intensities in units of K. The lowest contour is 0.4~K, and each contour is twice the previous one. The clouds are also marked in this panel.} \label{Fig:pvdecomp}}
\end{figure*}

The $^{12}$CO (1--0) intensity-weighted velocity map (see Fig.~\ref{Fig:pvdecomp}a) shows multiple clouds overlapping along the light of sight toward the surveyed region. In the longitude-velocity diagram (see Fig.~\ref{Fig:pvdecomp}b), we identify three molecular clouds, designated as the 15 \kms\,cloud, the 27 \kms\,cloud, and the 50 \kms\,cloud. It is worth noting that the 50 \kms\,cloud has not been reported by the previous FCRAO study \citep{1994JKAS...27..147L}, but was detected in previous H{\scriptsize I} and CO studies with poor angular resolution\citep[$>$8\arcmin,][]{1996ApJ...464..247W,2001ApJ...547..792D}. Thus, the 50~\kms\,cloud is resolved with an angular resolution of $\sim$1\arcmin\,for the first time. Following the method of \citet{1987ApJ...319..730S}, the Galactic coordinates of the three clouds were calculated with $l=\Sigma T_{\rm i} l_{\rm i}/\Sigma T_{\rm i}$ and $b=\Sigma T_{\rm i} b_{\rm i}/\Sigma b_{\rm i}$, where $l_{\rm i}$ and $b_{\rm i}$ are the Galactic coordinates of pixel i, and $T_{\rm i}$ is the intensity of pixel i. The velocities were taken to be the Gaussian fit velocities of the average $^{13}$CO (1--0) spectra. Based on the Galactic coordinates and the measured velocities of the clouds, we calculated their kinematics distances with the Galactic rotation curve \citep[the spatial-kinematic model A5 of ][]{2014ApJ...783..130R}. We also estimated their Galactocentric radii ($D_{\rm GC}$) and heights ($H$) from the Galactic mid-plane. These results are listed in Table~\ref{Tab:clouds}. Our distance estimate of the 27 \kms\,cloud is roughly consistent with values in previous studies \citep[e.g.,][]{1984NInfo..56...59A,1994JKAS...27..147L,1998ApJ...509..749H}. Their small heights from the Galactic mid-plane suggest that they are within in the molecular disk of the Milky Way \citep[the FWHM thickness of the molecular disk is larger than 100 pc at a Galactocentric radius larger than 9 kpc,][]{2015ARA&A..53..583H}. Based on the fitted spiral arms of the Milky Way \citep[Table 2 of][]{2014ApJ...783..130R} and the derived positions of the three clouds (see Table~\ref{Tab:clouds}), we find that the 15~\kms\,cloud is located at the interarm between the Local arm (predicted to be $D_{\rm GC}$=8.53$\pm$0.17~kpc at the Galactocentric azimuth, where the error value is half the width of the arm; this is the same for the following cases) and the Perseus arm (predicted to be $D_{\rm GC}$=10.17$\pm$0.19~kpc at the Galactocentric azimuth); the 27~\kms cloud is located at the Perseus arm (predicted to be $D_{\rm GC}$=10.07$\pm$0.19~kpc at the Galactocentric azimuth); the 50~\kms cloud belongs to the Outer arm (predicted to be $D_{\rm GC}$=13.24$\pm$0.32~kpc at the Galactocentric azimuth).

We used the intensity-weighted radius to characterize the size of the molecular clouds. The intensity-weighted radius is defined as $<r>=\Sigma W_{\rm i}r_{\rm i}/\Sigma W_{\rm i}$, introduced by \citet{2000ApJ...537..221U}, where $ W_{\rm i}$ is the integrated intensity of pixel i, and $r_{\rm i}$ is the measured distance relative to the centroid. Here, we took the $^{13}$CO (1--0) intensity-weighted radius for the following analysis. The FWHM line widths of clouds were taken to be those of the $^{13}$CO (1--0) average spectra. The line widths are supersonic, suggesting that turbulent motions are dominant in the molecular clouds. To derive the physical properties of the clouds, we employed two methods to obtain the H$_{2}$ column density. For the first method, we used the $X$-factor ($X_{\rm CO}$) to convert the $^{12}$CO (1--0) integrated intensity into the H$_{2}$ column density. Although $X_{\rm CO}$ can vary from 0.9$\times10^{20}$ cm$^{-2}$~(K~\kms)$^{-1}$ to 4.8$\times10^{20}$ cm$^{-2}$~(K~\kms)$^{-1}$ \citep[e.g.,][]{2010ApJ...723.1019H}, we chose a $X_{\rm CO}$ of 2.8$\times10^{20}$ cm$^{-2}$~(K~\kms)$^{-1}$ to be consistent with \citet{1998ApJ...498..541K} for the following discussion (see Sect.~\ref{t:sfe}). We then derived the total cloud masses with formula~(\ref{f.mass}) by integrating over the areas (Col. 5 in Table~\ref{Tab:cloudproperty}) of the $^{12}$CO (1--0) emission with signal-to-noise ratios higher than 3. In the second method, we assumed that all molecular gas are in local thermodynamic equilibrium (LTE). We derived the excitation temperature from the optically thick $^{12}$CO (1--0) line. With this value and the $^{13}$CO (1--0) line parameters, we then also determined the opacity and the $^{13}$CO column density. \citet{2005ApJ...634.1126M} found a [$^{12}$C/$^{13}$C] isotopic ratio gradient in our Galaxy,
\begin{equation}\label{f.ratio}
[^{12}{\rm C}/^{13}{\rm C}]=6.21D_{\rm GC}+18.71\;.
\end{equation} 
We applied this relationship to calculate the [$^{12}$C/$^{13}$C] isotopic ratios of the three clouds that were used to convert the $^{13}$CO column density into the $^{12}$CO column density. We then derived the H$_{2}$ column density with a constant [H$_{2}$/$^{12}$CO] abundance ratio of 1.1$\times 10^{4}$ \citep{1982ApJ...262..590F}. Following the formula~(\ref{f.mass}) and assuming that $^{13}$CO (1--0) is optically thin (based on the result of Appendix~\ref{s.statistics}), we estimated the LTE masses of the clouds by integrating over the areas (Col. 8 in Table~\ref{Tab:cloudproperty}) of the $^{13}$CO (1--0) emission with signal-to-noise ratios higher than 3. These properties are listed in Table~\ref{Tab:cloudproperty}. Taking the distance difference into account, the LTE mass of the 27~\kms\,cloud derived from $^{13}$CO agrees with the previous FCRAO study \citep{1994JKAS...27..147L}. The surface density was calculated by dividing the mass by the area. The difference of the surface densities derived from $^{12}$CO and $^{13}$CO is within a factor of 2. Furthermore, the derived H$_{2}$ surface densities of the 27~\kms\,cloud are similar to those (49~$M_{\odot}$~pc$^{-2}$--105~$M_{\odot}$~pc$^{-2}$) of nearby clouds \citep{2010ApJ...723.1019H} and slightly higher than the median value (42~$M_{\odot}$~pc$^{-2}$) of molecular clouds in the Galactic Ring Survey \citep{2009ApJ...699.1092H}. The 15~\kms\,cloud, which lies in the interarm, has an excitation temperature, line width, surface density, and cloud mass similar to some clouds in the spiral arms \citep{2009ApJS..182..131R,2013MNRAS.431.1587E}, indicating that the spiral-arm structure may have little effect in altering properties of molecular clouds.

\begin{table*}[!hbt]
\caption{Molecular clouds identified in this region.}\label{Tab:clouds}             
\scriptsize
\centering                                      
\begin{tabular}{cccccccccc}          
\hline\hline                        
Cloud      & $l$        &  $b$      & R.A.(J2000)   & Dec.(J2000) & $\upsilon_{\rm lsr}$\tablefootmark{a}  &  $D$\tablefootmark{b} & $D_{\rm GC}$ & $H$ & Location \\
           & (\degree)&(\degree)& (hh:mm:ss.s) & (dd:mm:ss)  & (\kms)         & (kpc)     & (kpc)      & (pc)    &         \\
\hline
  15 \kms\,cloud & 217.278  & $-$0.237 & 06:58:30.5 & $-$03:56:17 & 15.92 & 1.35 & 9.47  & $-$5.58 & interarm\\
  27 \kms\,cloud & 217.752  & $-$0.148 & 06:59:17.2 & $-$03:59:16 & 27.17 & 2.42 & 10.38 & $-$6.25 & Perseus Arm     \\
  50 \kms\,cloud & 217.145  & 0.523    & 07:00:59.8 & $-$03:30:35 & 50.60 & 5.40 & 13.05 & 49.29   & Outer Arm       \\
\hline
 \end{tabular}
 \tablefoot{\\
 \tablefoottext{a}{The velocity is taken to be the Gaussian fit velocity of the $^{13}$CO average spectrum.}\\
 \tablefoottext{b}{The distance is estimated with the A5 model of \citet{2014ApJ...783..130R}.}\\
 }
 			        \normalsize
\end{table*}

\begin{table*}[!hbt]
\caption{Derived cloud properties.}\label{Tab:cloudproperty}             
\scriptsize
\centering                                      
\begin{tabular}{ccccccccccc}          
\hline\hline                        
                      &             &         &                       & \multicolumn{3}{c}{$^{12}$CO}  &  & \multicolumn{3}{c}{$^{13}$CO}   \\
\cline{5-7}\cline{9-11}
Cloud              & $T_{\rm ex}$  & $<r>$   & $\Delta \upsilon_{\rm cloud}$  & Area      & $\Sigma$   & $M$ &  & Area     & $\Sigma$   & $M$   \\
                   & (K)         &  (pc)   & (\kms)  & pc$^{2}$ & ($M_{\odot}$~pc$^{-2}$) & ($M_{\odot}$) &  & pc$^{2}$ & ($M_{\odot}$~pc$^{-2}$)  & ($M_{\odot}$)\\
\hline
 15 \kms\,cloud & 10.4 & 1.9	 &1.81 & 29   & 48 & 1.4$\times10^{3}$ & & 12  &  27  &   3.4$\times10^{2}$      \\
 27 \kms\,cloud & 10.1 & 19.1 &3.90 & 1566 & 53 & 8.3$\times10^{4}$ & & 414 &  64  &   2.6$\times10^{4}$	\\
 50 \kms\,cloud & 7.4  & 45.5 &3.13 & 4093 & 29 & 1.4$\times10^{5}$ & & 732 &  18  &   1.3$\times10^{4}$	\\
\hline
 \end{tabular}
 \tablefoot{\\ Column 1 gives the name of the selected clouds. Column 2 gives the median of excitation temperature (the detailed statistical results are described in Appendix~\ref{s.statistics}). Column 3 gives the intensity-weighted radius of the clouds. Column 4 gives the line width of the $^{13}$CO average spectrum. The areas, surface densities, cloud masses derived from $^{12}$CO and $^{13}$CO are listed in Column 5--10.
 }
 			        \normalsize
\end{table*}

\subsection{Spatial morphology and velocity structure}\label{t:struc}
\begin{figure*}[!htbp]
\centering
\includegraphics[width = 0.9\textwidth]{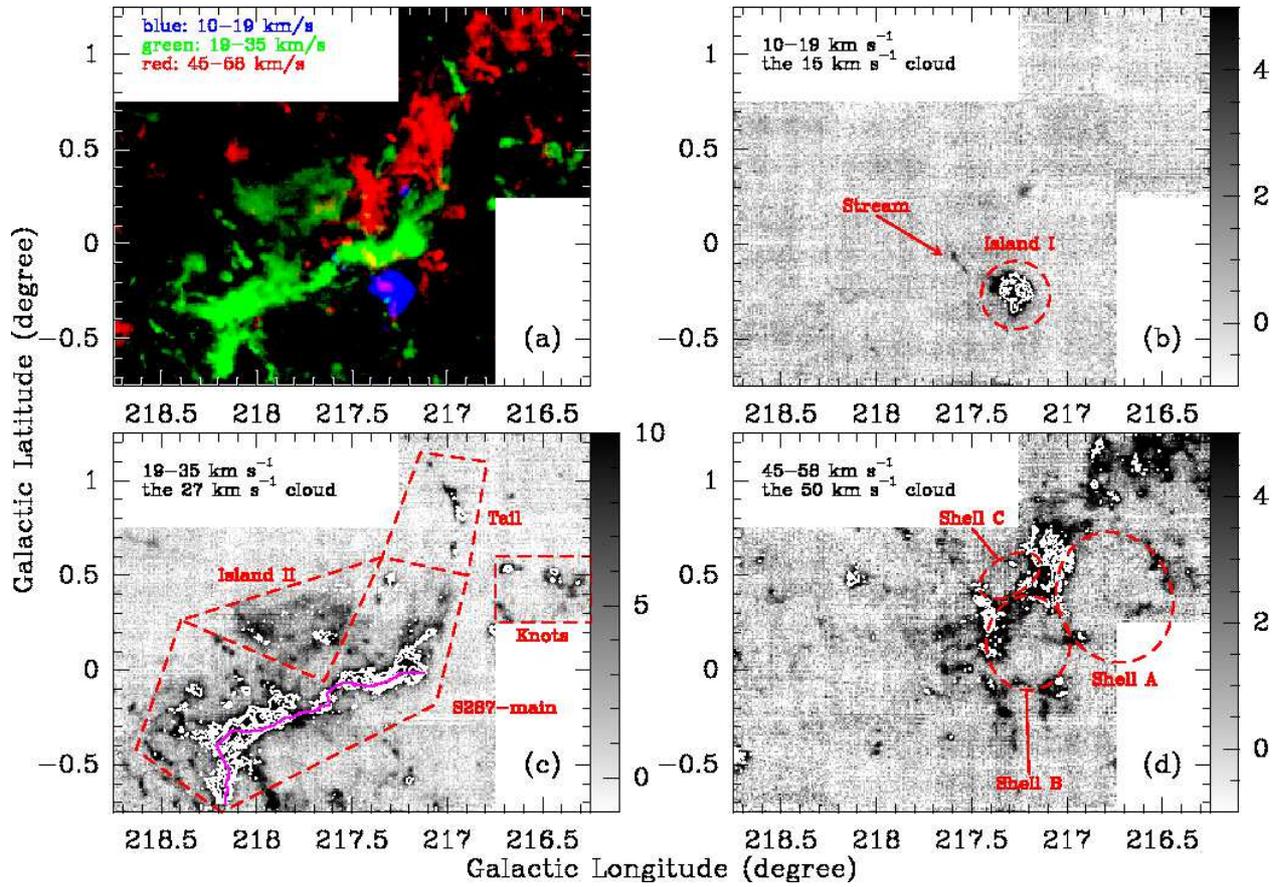}
\caption{{(a): Three-color integrated intensity $^{12}$CO (1--0) image of the surveyed region (red: integrated between 45 and 58~\kms; green: integrated between 19 and 35~\kms; blue: integrated between 10 and 19~\kms). (b): $^{12}$CO (1--0) integrated intensity map (background) overlaid with a $^{13}$CO (1--0) integrated intensity map (contours). The two integrated velocity ranges extend from 10 to 19~\kms. The grayscale colors correspond to a linear stretch of $^{12}$CO (1--0) integrated intensities in units of K~\kms. The contours represent $^{13}$CO (1--0) integrated intensities that start at 1.5~K~\kms\,and increase by 2.5~\kms. (c) Same as Fig.~\ref{Fig:allclouds}b, but integrated velocity ranges extend from 19 to 35~\kms. The contours start at 2.0~K~\kms\,, and each contour is twice the previous one. The purple solid line marks the highest H$_{2}$ column density line of the filament S287-main along its long axis. (d) Same as Fig.~\ref{Fig:allclouds}b, but integrated velocity ranges reach from 45 to 58~\kms. The contours start at 1.8~K~\kms\,and increase by 1.8~K~\kms. The subregions of the three clouds are marked and labeled in the respective panels.} \label{Fig:allclouds}}
\end{figure*}

Figure~\ref{Fig:allclouds}a displays the three clouds that overlap along the light of sight. Figures~\ref{Fig:allclouds}b--\ref{Fig:allclouds}d show the decomposed spatial distribution of the 15~\kms\,cloud, the 27~\kms\,cloud, and the 50~\kms cloud. Overall, the morphologies of the $^{13}$CO (1--0) integrated intensity resemble those of the $^{12}$CO (1--0) integrated intensity, but the $^{12}$CO emission is more extended, tracing more diffuse molecular gas. The three clouds are characterized by several subregions whose names are labeled in Figs.~\ref{Fig:allclouds}b--\ref{Fig:allclouds}d for further analyses of their spatial distributions and velocity structures below. Furthermore, we identify four molecular shells in the mapped regions. In this work, only large-scale shell-like structures are taken into account. Molecular shells are identified through visual investigation of their morphology and velocity structures. The shells are only identified if they show shell-like morphology in their integrated-intensity maps and show blue- and red-shifted velocity components in p-v diagrams.

\begin{figure*}[!htbp]
\centering
\includegraphics[width = 0.9\textwidth]{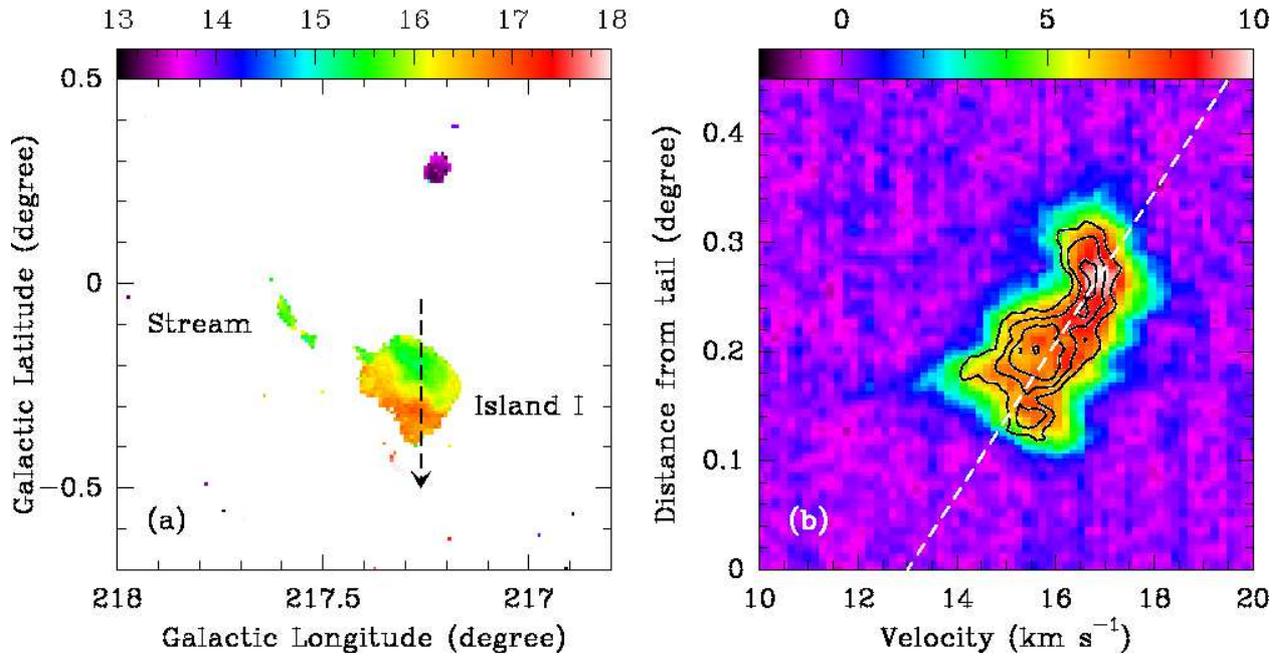}
\caption{{(a): $^{12}$CO (1--0) intensity-weighted velocity map toward the 15~\kms\,cloud. The color bar represents velocities in units of \kms. (b): position-velocity diagram along the cut indicated in Fig.~\ref{Fig:pvstream}a ($^{13}$CO black contours overlaid on the $^{12}$CO image). The color bar represents $^{12}$CO intensities in units of K. $^{13}$CO contours start at 0.9~K and increase by 0.6~K. The white dashed line represents the velocity gradient seen in island I.} \label{Fig:pvstream}}
\end{figure*}

\emph{The 15 \kms\,cloud}.-- This cloud has two subregions, island I and stream (see Fig.~\ref{Fig:allclouds}b). Island I is located south of the 27~\kms\,cloud. The p-v diagram across island I (see Fig.~\ref{Fig:pvstream}) shows a velocity gradient of about 0.6~\kms~pc$^{-1}$ from north to south, which is presumably due to rotation or shear motion. On the other hand, Fig.~\ref{Fig:pvstream}b shows that the velocity dispersion becomes smaller from north to south, indicating that its southern part might be more perturbed. Stream has nearly the same velocity as the north of Island I, indicating that they are physically related and may be connected by more diffuse gas (e.g., H{\scriptsize I} gas). 

\begin{figure*}[!htbp]
\centering
\includegraphics[width = 0.9\textwidth]{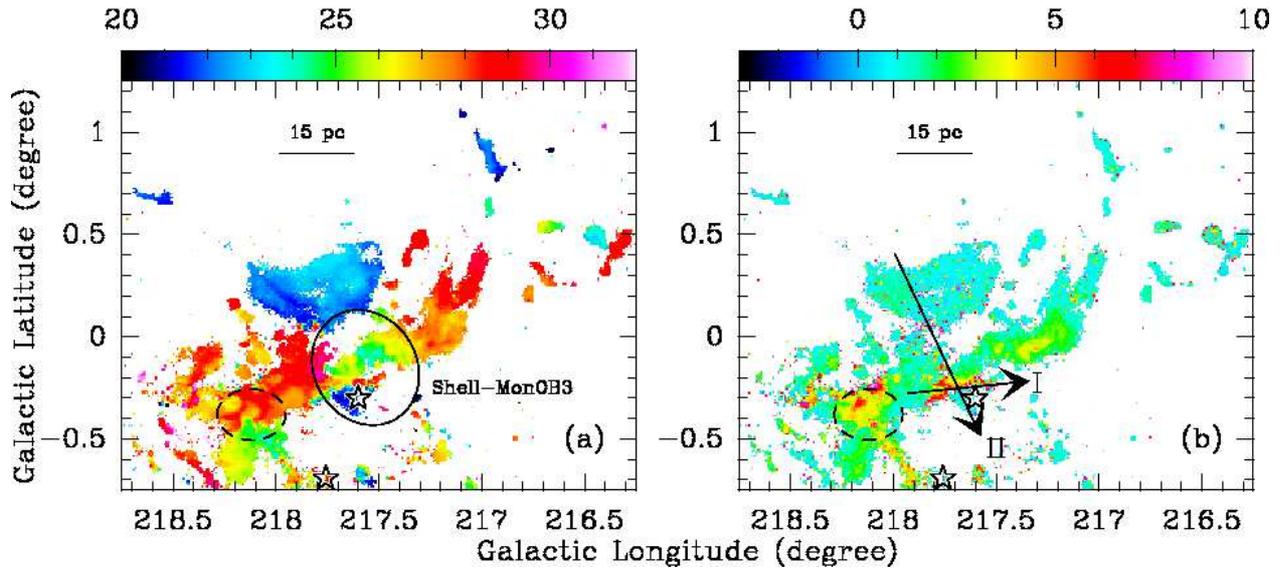}
\caption{{(a): $^{12}$CO (1--0) intensity-weighted velocity map toward the 27~\kms\,cloud. The color bar represents velocities in units of \kms. The H{\scriptsize II} region S287, the stellar association Mon OB3, and the open cluster NGC 2311 are marked with a black dashed ellipse and open pentagrams. The solid black ellipse marks the cavity-like feature. (b): Same as Fig.~\ref{Fig:cloud2-mom}a but for the intensity-weighted $^{12}$CO (1--0) line width map of the 27~\kms\,cloud. The arrows represent the p-v cuts of the interacting interface I and II, indicated by I and II in the panel. Their corresponding position-velocity maps are shown in Fig.~\ref{Fig:cloud2-pv}} \label{Fig:cloud2-mom}}
\end{figure*}

\emph{The 27 \kms\,cloud}.-- The 27~\kms\,cloud has been studied by \citet{1994JKAS...27..147L}. Nevertheless, we present an analysis of the cloud with a new perspective. We separated this cloud into four subregions, S287-main, island II, tail, and knots (see Fig.~\ref{Fig:allclouds}c). Its $^{12}$CO (1--0) intensity-weighted velocity map (see Fig.~\ref{Fig:cloud2-mom}a) presents a very complex velocity structure, which agrees with the conclusion that the cloud is highly disturbed \citep{1994JKAS...27..147L}.

\begin{figure*}[!htbp]
\centering
\includegraphics[width = \textwidth]{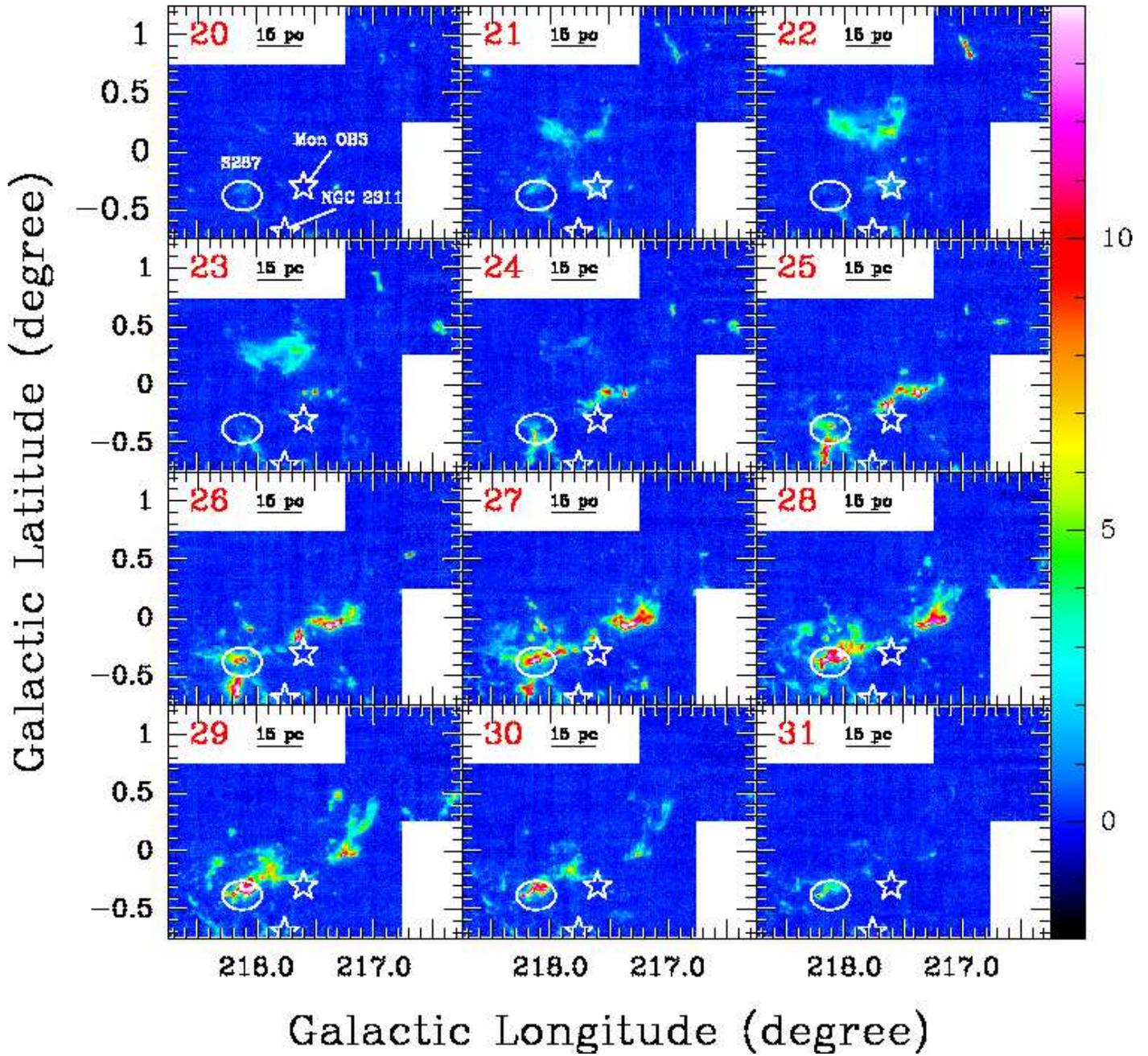}
\caption{{$^{12}$CO velocity channel maps of the 27 \kms\,cloud. Each panel consists of the intensity integrated over a 1.0 \kms wide range, and the velocity, in \kms, is labeled in the upper left corner of each panel. The H{\scriptsize II} region S287, the stellar association Mon OB3, and the open cluster NGC 2311 are marked with a white ellipse and pentagrams, and their names are also given in the upper left panel. The color bar represents integrated intensities in units of K~\kms.} \label{Fig:s287ch}}
\end{figure*}

\begin{figure*}[!htbp]
\centering
\includegraphics[width = 0.9 \textwidth]{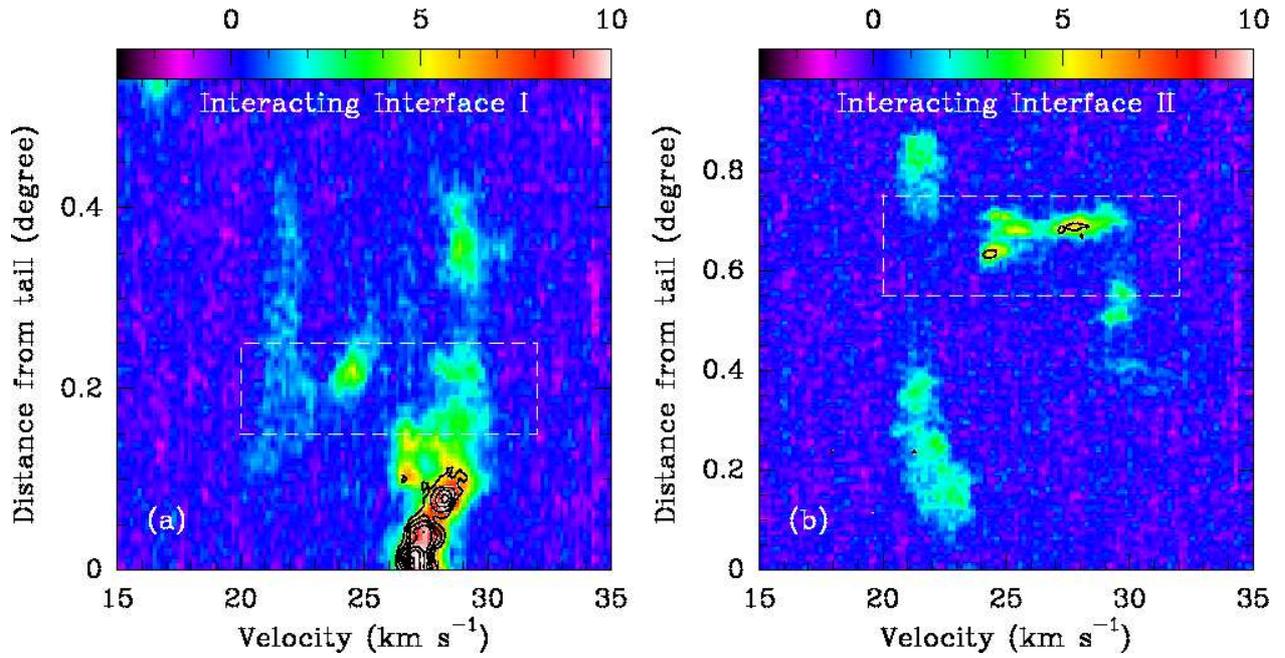}
\caption{{Position-velocity diagram along the cuts across the interacting interface I (a) and II (b) indicated in Fig.~\ref{Fig:cloud2-mom}b ($^{13}$CO black contours overlaid on the $^{12}$CO image). The color bars represent $^{12}$CO intensities in units of K. $^{13}$CO contours start at 0.9~K and increase by 0.6~K. The white dashed boxes display the interacting regions. } \label{Fig:cloud2-pv}}
\end{figure*}

S287-main shows a filamentary structure and is the densest of the four subregions in the 27~\kms\,cloud. Based on the H$_{2}$ column density map derived from $^{13}$CO (1--0), we extracted the highest H$_{2}$ column density line of the filament indicated by the purple solid line in Fig.~\ref{Fig:allclouds}c. S287-main measures 71 pc in length and 1.8 pc in width, resulting in an aspect ratio of about 39:1. These properties are similar to those of the filaments which are believed to be associated with Milky Way spiral arms \citep{2015MNRAS.450.4043W}. The velocity map of S287-main shows a blueshifted velocity component in the center of the filamentary structure where the redshifted part shows a shell-like feature (shell-MonOB3, indicated by the solid black ellipse in Fig.~\ref{Fig:cloud2-mom}a). The stellar association Mon OB3 lies inside shell-MonOB3 and has a radial velocity of $\sim$27~\kms ~\citep{1995AstL...21...10M,2009MNRAS.400..518M}. Mon OB3 is therefore confirmed to be associated with S287-main. Shell-MonOB3 is likely due to the formation of Mon OB3 out of S287-main and the disruption by the feedback of Mon OB3. From Fig.~\ref{Fig:s287ch}, we find that the interacting interface I (indicated in the intensity-weighted $^{12}$CO (1--0) line width map, see Fig.~\ref{Fig:cloud2-mom}b) is swept up to the blueshifted panels (23--25~\kms), while S287-main has clearly been split into two parts in the red-shifted panels (28--30~\kms). Figure~\ref{Fig:cloud2-mom}b shows large line widths around the S287 H{\scriptsize II} region and Mon OB3, which is likely due to the perturbation by the feedback of the S287 H{\scriptsize II} region and Mon OB3. Position-velocity diagrams across the interacting interface I and II (indicated by the arrow in Fig.~\ref{Fig:cloud2-mom}b) demonstrate that the line widths of the perturbed molecular gas are around 4~\kms\,(indicated by the white dashed boxes in both panels of Fig.~\ref{Fig:cloud2-pv}), which is much larger than those ($\sim$~1--2~\kms) of ambient gas. Two distinct velocity components at $\sim$22~\kms\,and $\sim$29~\kms, as shown in Fig.~\ref{Fig:cloud2-pv}, could just be the projection of swept-up clumpy gas on either side (front and back) of Mon OB3. This would be expected in the proposed feedback scenario mentioned above. We assume that the two velocity components in the position-velocity diagrams arise from the front and backside of expanding shells, and the expansion velocity is thus half the velocity separation. With this method, the expansion velocity of shell-MonOB3 is found to be 3.5~\kms. The uncertainty in the estimate of the expansion velocity is approximate $\pm$1~\kms. We also note that the expansion velocity is a lower limit since they are only based on the velocity along the light of sight. Position and size of shell-MonOB3 is manually fitted with the ellipse shown in Fig.~\ref{Fig:cloud2-mom}a. The properties of shell-MonOB3 are given in Table~\ref{Tab:shell}. Figure~\ref{Fig:allclouds}c shows that most of the molecular gas in island II does not show much $^{13}$CO (1--0) emission, suggestive of a low-density region. Island II might be fragmenting since its $^{12}$CO (1--0) emission in the center is weaker than that farther outside. In the 21--22~\kms\,panels of Fig.~\ref{Fig:s287ch}, island II is connected to S287-main by weak $^{12}$CO (1--0) emission, indicating that Island II might be affected by the feedback of Mon OB3. Tail and knots are not covered by the previous FCRAO study because of their smaller sky coverage \citep{1994JKAS...27..147L}. Figure~\ref{Fig:cloud2-mom}a shows that tail has a velocity of $\sim$20~\kms\,and is not aligned with the velocity gradient of the northern part of S287-main, indicating that tail and S287-main have different origins. Knots also displays a velocity gradient from north to south (see Fig.~\ref{Fig:cloud2-mom}a).

\begin{figure*}[!htbp]
\centering
\includegraphics[width = 0.9 \textwidth]{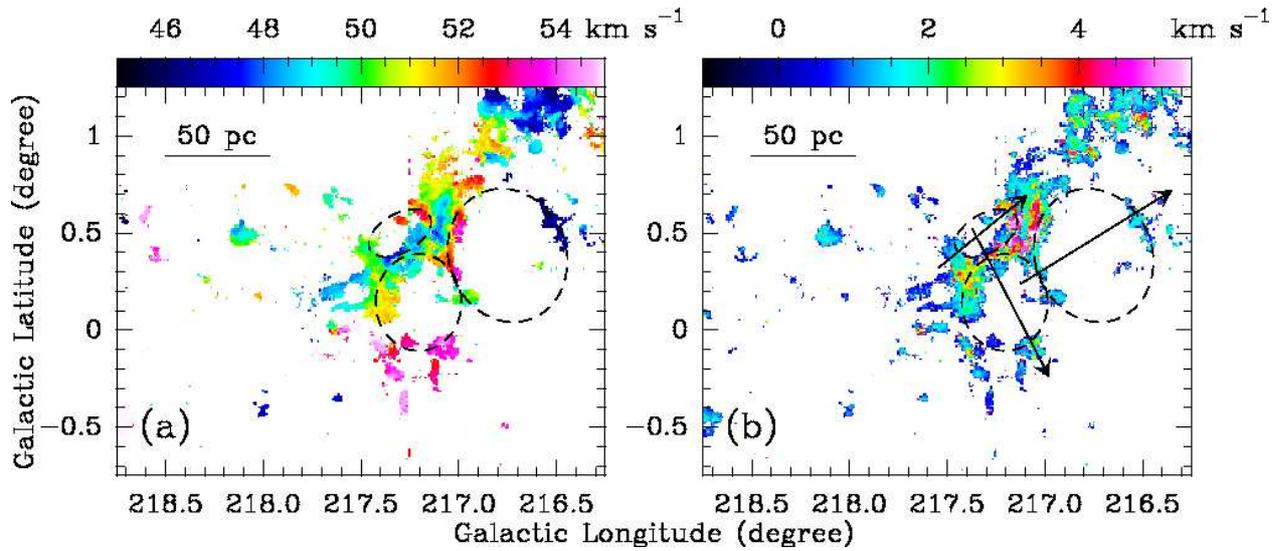}
\caption{{(a): Intensity-weighted $^{12}$CO velocity map of the 50~\kms\,cloud. Shells A, B, and C are marked with black dashed ellipses. The color bar shows the intensity-weighted velocity scale in units of \kms.  (b): Same as Fig.~\ref{Fig:shell-mom}a but for the intensity-weighted $^{12}$CO (1--0) line width map of the 50~\kms\,cloud. The arrows show the pv-diagram cuts presented in Fig.~\ref{Fig:shell-pv}.} \label{Fig:shell-mom}}
\end{figure*}

\begin{figure*}[!htbp]
\centering
\includegraphics[width = 0.9 \textwidth]{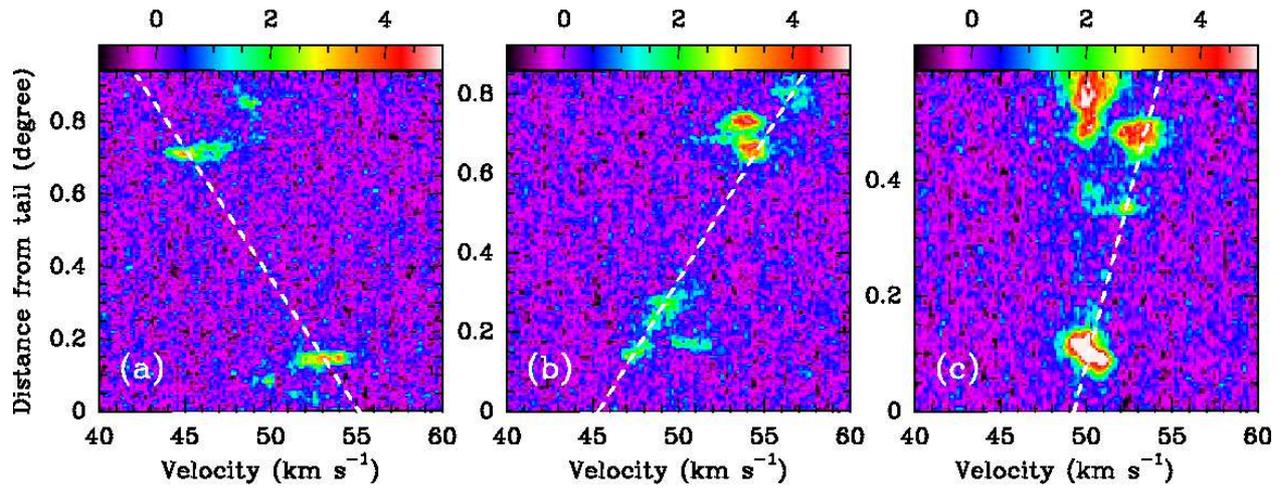}
\caption{{$^{12}$CO (1--0) position-velocity diagram along the cuts across shell A (Fig.~\ref{Fig:shell-pv}a), B (Fig.~\ref{Fig:shell-pv}b), and C (Fig.~\ref{Fig:shell-pv}c) indicated in Fig.~\ref{Fig:shell-mom}b. The white dashed lines stand for the velocity gradients across the three molecular shells. The color bars represent intensities in units of K.} \label{Fig:shell-pv}}
\end{figure*}

\emph{The 50 \kms\,cloud}.-- The 50~\kms\,cloud in Fig.~\ref{Fig:allclouds}d shows many cavity-like features, indicating that the cloud is highly perturbed by past activities. Around these cavities, we reveal three large molecular shells, labeled as shell A, shell B, and shell C.  Figure~\ref{Fig:shell-mom}a shows that there are no red- or blue-shifted components toward the center of the three shells, which suggests that they are ring-like rather than spherical. This is readily explained if their natal molecular clouds are approximately sheet-like \citep[e.g.,][]{2009apsf.book.....H,2010ApJ...709..791B}. From its $^{12}$CO (1--0) intensity-weighted velocity map (Fig.~\ref{Fig:shell-mom}a) and p-v diagrams (Fig.~\ref{Fig:shell-pv}), we can see that there are two velocity components (blue- and red-shifted) toward the three molecular shells, which is indicative of expansion. Using the same method for shell-MonOB3, we roughly estimate the expansion velocities of shell A, shell B, and shell C to be 4.0~\kms, 2.5~\kms, and 1.5~\kms, respectively. The ellipses used to fit the positions and sizes of the three shells are shown in Fig.~\ref{Fig:allclouds}d. The properties of the three molecular shells are given in Table~\ref{Tab:shell}. The three shells have sizes larger than those driven by low-mass stars in the Perseus clouds \citep{2011ApJ...742..105A} and those driven by H{\scriptsize II} regions \citep{2010ApJ...709..791B}, but they are similar to some supernova remnants (SNRs) \citep{2014BASI...42...47G}. This indicates that they are driven by energetic sources. On the other hand, the difference of expansion velocities and sizes between the three molecular shells may result from their different dynamic ages or different energy inputs of driving sources. Furthermore, we find that $^{13}$CO (1--0) emission mainly lies in the intersections between the three molecular shells (see Fig.~\ref{Fig:allclouds}d), where molecular gas is more likely to be compressed. The intensity-weighted $^{12}$CO (1--0) line width map (Fig.~\ref{Fig:shell-mom}b) shows large velocity dispersions at the edges of the shells, supporting the assumption that the edges are highly perturbed by the expansion of the three shells.

\begin{table*}[!hbt]
\caption{Properties of molecular shells.}\label{Tab:shell}             
\centering                                      
\begin{tabular}{cccccccc}          
\hline\hline                        
Name      & $l$        &  $b$       & angular size              & physical size    & PA        & $\upsilon_{\rm exp}$ \\
          & (\degree)  &(\degree)   & (~\arcmin~$\times$~\arcmin)  & (pc~$\times$~pc)  & (\degree) &  (\kms)        \\
\hline
\multicolumn{7}{c}{The 27 \kms\,cloud}\\
\hline
Shell-MonOB3 & 217.565    & $-$0.150      & 30$\times$35              & 21$\times$25 & 30        & 3.5     \\
\hline
\multicolumn{7}{c}{The 50 \kms\,cloud}\\
\hline
 Shell A     & 216.750    & 0.385         & 42$\times$36              & 66$\times$57 & 20	 & 4.0	   \\
 Shell B     & 217.300    & 0.500	  & 30$\times$26	      & 47$\times$41 & 0	 & 2.5	   \\
 Shell C     & 217.204    & 0.143	  & 21$\times$12	      & 33$\times$19 & $-$60	 & 1.5	   \\
\hline
 \end{tabular}
 			        \normalsize
\end{table*}

\subsection{Clumps}\label{t:clumps}
\begin{figure*}[!htbp]
\centering
\includegraphics[width = 0.9\textwidth]{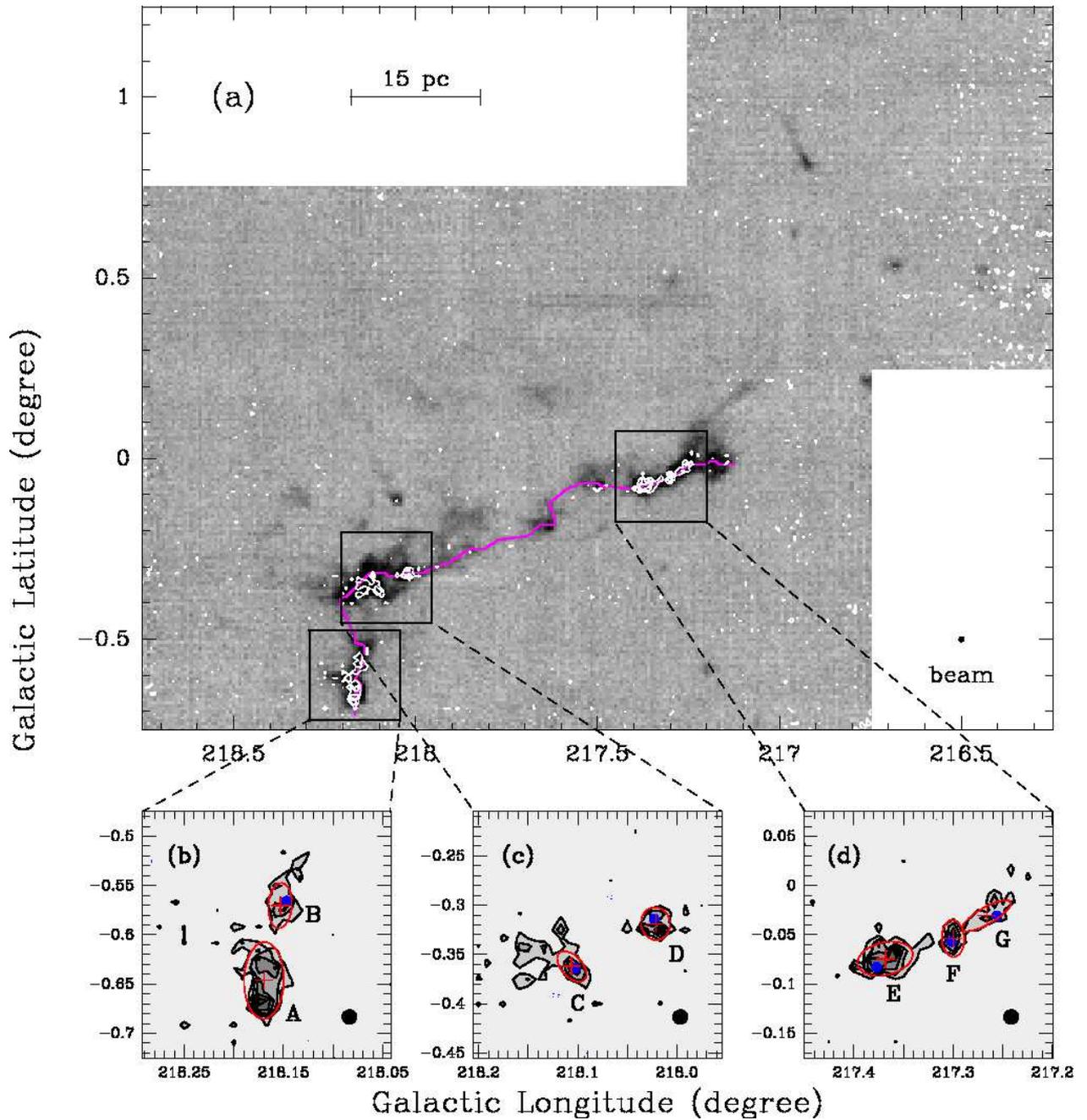}
\caption{{(a): C$^{18}$O intensity map (white contours) overlaid on the $^{13}$CO line image. Both lines have been integrated over the velocity range between 22 and 29~\kms. Contours start from 1.2~K~\kms (3~$\sigma$) to 3.6~K~\kms (9~$\sigma$) by 0.8~K~\kms (2~$\sigma$).  The horizontal bar represents 15~pc at the distance of 2.42~kpc. The purple solid line marks the highest H$_{2}$ column density line of the filament S287-main along its long axis. (b): Zoom in the region indicated in Fig.~\ref{Fig:clumps}a. Contours start from 1.2~K~\kms (3~$\sigma$) to 3.6~K~\kms (9~$\sigma$) by 0.6~K~\kms (1.5~$\sigma$). The clumps are marked with red crosses and ellipses, and their corresponding names are also given. The RMS MYSOs are marked with blue circles. The C$^{18}$O beam width is shown in the lower right corner of the panel. (c) Same as Fig.~\ref{Fig:clumps}b. (d) Same as Fig.~\ref{Fig:clumps}b.} \label{Fig:clumps}}
\end{figure*}

Clumps are fundamental units of star formation \citep[e.g.,][]{2007ARA&A..45..339B}. We searched for C$^{18}$O clumps because C$^{18}$O can trace higher density gas than $^{13}$CO because of its lower abundance. Moreover, the surveyed region is mapped in C$^{18}$O (1--0) for the first time. The observations reveal seven C$^{18}$O clumps that are all embedded in the filament S287-main (see Fig.~\ref{Fig:clumps}). The observed properties of the seven clumps are obtained with Gaussian fits to corresponding spectra by assuming one velocity component. These results are listed in Table~\ref{Tab:peakobs}. We find that the seven clumps are all within the 27 \kms\, cloud because of the coincidence on spatial distributions along the same line of sight as well as velocities, therefore we adopted a distance of 2.42~kpc for these clumps. All clumps lie along the H$_{2}$ highest column density line, except for clump C which is offset by 1.7 pc from the line. The size of each clump was fit using the C$^{18}$O (1--0) integrated intensity map with the task ``GAUSS\_2D'' in GILDAS, and the fit results are shown in Figs.~\ref{Fig:clumps}b--~\ref{Fig:clumps}d. The deconvolved radii of the clumps are defined as 
\begin{eqnarray}\label{f.rcl}
\theta_{\rm cl} &= & \nonumber (\frac{A}{\pi}-\frac{\theta_{\rm beam}^{2}}{4})^{1/2} \;, \\ 
R_{\rm cl} & = &\theta_{\rm cl} D \;,
\end{eqnarray}
 where $A$ is the area of the fit ellipse, $\theta_{\rm beam}$ is the FWHM beam width, and $D$ is the distance to the source. Following the LTE methods and Appendix~\ref{s.mass}, we calculated the excitation temperatures, opacities, column densities, LTE masses, and virial parameters of the clumps by assuming an isotopic ratio [$^{16}$O/$^{18}$O]=560 \citep{1994ARA&A..32..191W} and a constant [H$_{2}$/$^{12}$CO] abundance ratio of 1.1$\times 10^{4}$ \citep{1982ApJ...262..590F}. Since the opacities of $^{13}$CO (1--0) are higher than 0.4 in these clumps (see Table~\ref{Tab:phyclump}), the H$_{2}$ column densities derived from $^{13}$CO are underestimated by a factor of at least 1.2. Thus, we only discuss the properties derived from C$^{18}$O (1--0), which has opacities lower than 0.2. Assuming that the clumps are spheres with radius $R_{\rm cl}$, we obtain the average H$_{2}$ densities of the clumps. These physical parameters of C$^{18}$O clumps are tabulated in Table~\ref{Tab:phyclump}.

From Table~\ref{Tab:peakobs}, the observed C$^{18}$O line widths of the clumps are larger than those (0.3~\kms--2.1\kms) of low-mass protostars and pre-stellar cores \citep{2002A&A...389..908J}, but similar to those clumps in infrared dark clouds \citep[1.5~\kms--2.2~\kms,][]{2011ApJS..193...10Z}. From Table~\ref{Tab:phyclump}, the excitation temperatures derived from the peaks of $^{12}$CO (1-0) spectra are found to be 16$\sim$21 K with an average of 19~K, which is similar to the mean CO excitation temperature of ATLASGAL dust clumps dark at 8~$\mu$m \citep{2014A&A...570A..65G} and the mean kinetic temperature of ATLASGAL dust clumps derived from low-$J$ metastable ammonia inversion transitions \citep{2012A&A...544A.146W}. The LTE masses of the clumps are similar to those of ATLASGAL dust clumps \citep{2014A&A...570A..65G}. The densities are found to be around $\sim10^{4}$~cm$^{-3}$. The virial parameters are estimated to be lower than 1. Thus, we suggest that these clumps are all supercritical and will collapse or must be supported against self-gravity, for instance, by magnetic fields. Furthermore, Figs.~\ref{Fig:clumps}b--~\ref{Fig:clumps}d show that six of them are found to be associated with the Red MSX sources (RMSs)\footnote{They are G217.3771-00.0828, G217.3020-00.0567, G217.2558-00.0306, G217.6282-00.1827, G217.0441-00.0584, G218.0230-00.3139A, G218.0230-00.3139B, G218.1025-00.3638, and G218.1472-00.5656 from the RMS catalog \citep{2013ApJS..208...11L}. According to their spatial distribution and velocity information \citep{2008A&A...487..253U,2008ASPC..387..381U,2013ApJS..208...11L}, eight of them are found to be associated with the 27~\kms\,clouds while one of them is associated with the 50~\kms\,clouds.} which are believed to be massive young stellar object (MYSO) candidates \citep{2013ApJS..208...11L}. Clump A is the only C$^{18}$O clump that is not associated with MYSO candidates. This is probably because clump A is still at an early evolutionary stage or its embedded infrared sources are not luminous enough to be detected by MSX.

In the BGPS 1.1 mm source catalog \citep[Table~1 of][]{2010ApJS..188..123R}, we find 15 dust clumps in the surveyed region, 12 of which match the highest H$_{2}$ column density line of the filament S287-main very well (see Fig.~\ref{Fig:clumpsd}a). We made use of $^{13}$CO spectra at the corresponding pixels of dust clump central positions to obtain the velocities and line widths of the dust clumps that are listed in Table~\ref{Tab:dustclump}. As a result, 14 dust clumps are located in the 27~\kms\,cloud while the dust clump D2 is associated with the 50~\kms\,cloud. We therefore used a distance of 5.40 kpc for dust clump D2. Assuming a gas-to-dust ratio of 100, the H$_{2}$ column densities can be calculated from flux densities of the BGPS 1.1 mm dust emission using the following formula \citep[e.g.,][]{1983QJRAS..24..267H,2008A&A...487..993K}:
\begin{equation}\label{f.dustcol}
N({\rm H_{2}})=\frac{100F_{\nu}}{\mu m_{\rm H} \kappa_{\nu}B_{\nu}(T_{\rm d})\Omega_{\rm beam}}\;,
\end{equation}
where $F_{\nu}$ denotes the flux density, $\mu$ is the mean molecular weight per hydrogen molecule which is taken to be 2.8, and $m_{\rm H}$ is the mass of a hydrogen atom, $\kappa_{\nu}$ represents the dust absorption coefficient at 1.1 mm, taken to be 1.14~cm$^{2}$~g$^{-1}$ given by \citet{2010ApJ...717.1157D}, who logarithmically interpolated the dust opacities \citep[Table 1, Col. 6, dust with thin ice mantles coagulating for 10$^{5}$ yr at a gas density of 10$^{6}$~cm$^{-3}$,][]{1994A&A...291..943O} to the BGPS 1.1 mm band with the emissivity of the graybody spectrum applied, $B_{\nu}(T_{\rm d})$ is the Planck function for a dust temperature $T_{\rm D}$, $\Omega_{\rm beam}$ is the beam solid angle. The total dust clump mass is derived from the integrated flux density over the target:
\begin{equation}\label{f.dustmass}
M_{\rm dust}=\frac{100D^{2}S_{\rm int}}{\kappa_{\nu}B_{\nu}(T_{\rm d})}\;,
\end{equation}
where $D$ is the distance, and $S_{\rm int}$ is the integrated flux density. With a dust temperature of 20~K, the beam-averaged column density and the dust clump mass were derived. With the same method for C$^{18}$O clumps, we also obtained the number density and virial parameters for the dust clumps. These properties are tabulated in Table~\ref{Tab:dustclump}. Seven of the dust clumps have virial parameters larger than 2, which indicates that they are subcritical. This suggests that these dust clumps might be confined by external pressure. Comparing the dust clumps and the C$^{18}$O clumps, we find that six of the dust clumps are associated with C$^{18}$O clumps (see Fig.~\ref{Fig:clumpsd}b--d). For D7 and D15, which have high masses but no C$^{18}$O counterparts, this is most likely due to the C$^{18}$O depletion as observed in B68 \citep{2002ApJ...570L.101B}, since they show narrower line widths than dust clumps that have C$^{18}$O counterparts. The other seven have no C$^{18}$O counterparts, which mainly arises from the relatively poor sensitivity of C$^{18}$O.

\begin{figure*}[!htbp]
\centering
\includegraphics[width = 0.5 \textwidth]{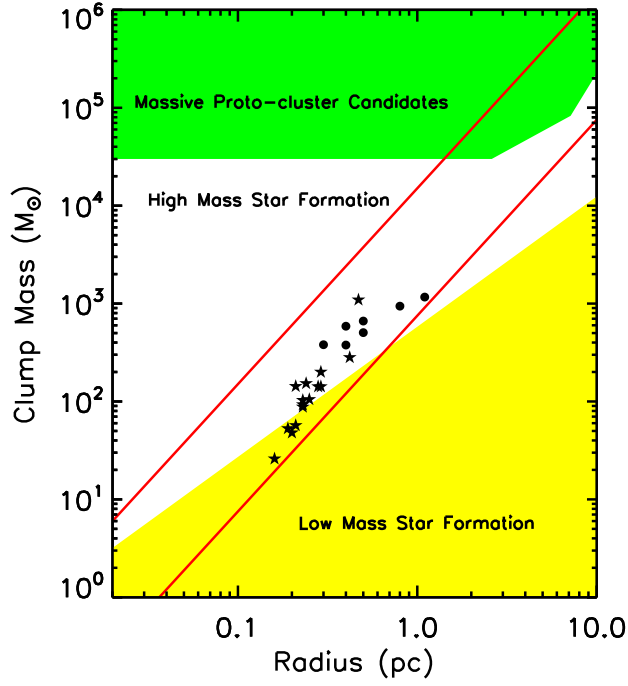}
\caption{{Mass-size relationship of the clumps, modified from Fig.~15 of \citet{2013MNRAS.431.1752U}. The upper and lower red lines represent surface densities of 1~g~cm$^{-2}$ \citep{2008Natur.451.1082K} and 0.05~g~cm$^{-2}$ \citep{2013MNRAS.431.1752U}, which are empirical upper and lower bounds for the clump surface densities required for massive star formation. The yellow shaded region demonstrates the clouds without high-mass star formation, according to the empirical relationship by \citet{2010ApJ...716..433K}. The green shaded region represents the parameter space where young massive cluster progenitors are expected \citep[e.g.,][]{2012ApJ...758L..28B}. The circles and pentagrams represent the C$^{18}$O clumps and the dust clumps, respectively.} \label{Fig:msf}}
\end{figure*}

Following previous studies \citep[e.g.,][]{2010ApJ...716..433K,2013MNRAS.431.1752U}, we also studied the mass-size relation for the C$^{18}$O clumps and the dust clumps, which is shown in Fig.~\ref{Fig:msf}. All of the C$^{18}$O clumps are found to have surface densities higher than the empirical lower bounds for massive star formation \citep{2010ApJ...716..433K,2013MNRAS.431.1752U}. This further supports that the C$^{18}$O clumps have the potential to form massive stars. The dust clumps all lie above the empirical lower bounds for massive star formation of \citet{2013MNRAS.431.1752U}, while four of them lie in the low-mass star formation shaded region by \citet{2010ApJ...716..433K}. Consequently, at least, eleven of the dust clumps have the potential to form massive stars. 

\renewcommand{\tabcolsep}{0.05cm}
\begin{table*}
\caption{Observed properties of the C$^{18}$O clumps.}\label{Tab:peakobs}
\tiny
\centering   
\begin{tabular}{ccccccccccccccccc}
\hline\hline                        
                &  &             & \multicolumn{4}{c}{$^{12}$CO} & & \multicolumn{4}{c}{$^{13}$CO} & &\multicolumn{4}{c}{C$^{18}$O} \\
\cline{4-7}  \cline{9-12}  \cline{14-17} \\
                &$l$ &$b$           &$T_{\rm mb}$ & $\int T_{\rm mb}$d$\upsilon$ & $\upsilon_{\rm lsr}$ & $ \Delta \upsilon$ & &$T_{\rm mb}$ & $\int T_{\rm mb}$d$\upsilon$ & $\upsilon_{\rm lsr}$ & $\Delta \upsilon$ & & $T_{\rm mb}$ & $\int T_{\rm mb}$d$\upsilon$ &$\upsilon_{\rm lsr}$ & $\Delta \upsilon$  \\  
name           &(\degree) &(\degree)             & (K)& (K~km~s$^{-1}$) & (km~s$^{-1}$) & (km~s$^{-1}$)    & & (K) & (K~km~s$^{-1}$) & (km~s$^{-1}$) & (km~s$^{-1}$)  &  & (K) & (K~km~s$^{-1}$) & (km~s$^{-1}$) & (km~s$^{-1}$) \\  
\hline
A & 218.167 & -0.667 & 14.8$\pm$0.5 & 46.9$\pm$0.5 & 25.6$\pm$0.1 & 3.0$\pm$0.1 &  & 8.0$\pm$0.3 & 16.9$\pm$0.3 & 25.6$\pm$0.1 & 2.0$\pm$0.1 &  & 2.1$\pm$0.3 & 2.7$\pm$0.2 & 25.3$\pm$0.1 & 1.2$\pm$0.1  \\ 
B & 218.150 & -0.575 & 15.2$\pm$0.5 & 44.6$\pm$0.5 & 25.5$\pm$0.1 & 2.8$\pm$0.1 &  & 7.3$\pm$0.3 & 11.8$\pm$0.2 & 25.4$\pm$0.1 & 1.5$\pm$0.1 &  & 1.6$\pm$0.3 & 1.4$\pm$0.2 & 25.2$\pm$0.1 & 0.8$\pm$0.1  \\ 
C & 218.108 & -0.367 & 15.2$\pm$0.5 & 70.7$\pm$0.6 & 26.8$\pm$0.1 & 4.4$\pm$0.1 &  & 5.7$\pm$0.3 & 17.5$\pm$0.3 & 26.3$\pm$0.1 & 2.9$\pm$0.1 &  & 1.0$\pm$0.3 & 2.1$\pm$0.2 & 25.7$\pm$0.1 & 1.9$\pm$0.2  \\ 
D & 218.017 & -0.317 & 18.1$\pm$0.6 & 66.9$\pm$0.7 & 28.0$\pm$0.1 & 3.5$\pm$0.1 &  & 7.2$\pm$0.3 & 18.4$\pm$0.3 & 27.5$\pm$0.1 & 2.4$\pm$0.1 &  & 1.0$\pm$0.3 & 1.8$\pm$0.2 & 26.9$\pm$0.1 & 1.6$\pm$0.3  \\ 
E & 217.358 & -0.067 & 16.4$\pm$0.5 & 77.6$\pm$0.6 & 26.4$\pm$0.1 & 4.5$\pm$0.1 &  & 8.1$\pm$0.3 & 22.3$\pm$0.3 & 26.1$\pm$0.1 & 2.6$\pm$0.1 &  & 1.5$\pm$0.3 & 3.5$\pm$0.3 & 25.9$\pm$0.1 & 2.2$\pm$0.2  \\ 
F & 217.300 & -0.050 & 17.7$\pm$0.4 & 86.9$\pm$0.6 & 26.6$\pm$0.1 & 4.6$\pm$0.1 &  & 9.4$\pm$0.2 & 22.3$\pm$0.2 & 26.5$\pm$0.1 & 2.2$\pm$0.1 &  & 1.5$\pm$0.3 & 3.0$\pm$0.3 & 26.5$\pm$0.1 & 1.8$\pm$0.2  \\ 
G & 217.258 & -0.017 & 12.7$\pm$0.4 & 55.4$\pm$0.5 & 27.7$\pm$0.1 & 4.1$\pm$0.1 &  & 5.7$\pm$0.2 & 16.0$\pm$0.2 & 27.6$\pm$0.1 & 2.6$\pm$0.1 &  & 0.8$\pm$0.3 & 1.8$\pm$0.3 & 27.5$\pm$0.1 & 2.0$\pm$0.4  \\ 

\hline                                   
\end{tabular}
\normalsize
\end{table*}

\renewcommand{\tabcolsep}{0.1cm}
\begin{table*}
\caption{Physical properties of the C$^{18}$O clumps.}\label{Tab:phyclump}
\scriptsize
\centering   
\begin{tabular}{ccccccccccccccc}
\hline\hline                        
                 &                      & & \multicolumn{2}{c}{$^{13}$CO} & &\multicolumn{7}{c}{C$^{18}$O} &  &\\
\cline{4-5}  \cline{7-13}   \\
name            &    $T_{\rm{ex}}$        & &  $\tau_{13}$ & $N({\rm H_{2}})$    & & $\theta_{\rm cl}$ & $R_{\rm cl}$ &  $\tau_{18}$ & $N({\rm H_{2}})$  & $M_{\rm LTE}$ & $\alpha_{\rm vir}$ & n$({\rm H_{2}})$&  & MYSO\tablefootmark{a}  \\  
                 & (K)                  & &  & ($\times 10^{22}$~cm$^{-2}$) & &  (\arcmin) &(pc)&    & ($\times 10^{22}$~cm$^{-2}$) & ($M_{\odot}$) &        &  (cm$^{-3}$)& & \\  
\hline
A &  18 &  &  0.77 & 1.6 & & 1.6 & 1.1 &0.15 & 1.9 &  1165 & 0.3 & $4.3\times 10^{3}$ &  & N \\ 
B &  19 &  &  0.65 & 1.1 & & 0.9 & 0.5 &0.11 & 1.0 &  663  & 0.1 & $1.9\times 10^{4}$ &  & Y \\ 
C &  19 &  &  0.47 & 1.6 & & 0.6 & 0.4 &0.07 & 1.5 &  587  & 0.5 & $5.3\times 10^{4}$ &  & Y \\ 
D &  22 &  &  0.51 & 1.9 & & 0.7 & 0.5 &0.06 & 1.5 &  505  & 0.5 & $2.4\times 10^{4}$ &  & Y \\ 
E &  20 &  &  0.68 & 2.2 & & 1.1 & 0.8 &0.10 & 2.7 &  941  & 0.8 & $9.2\times 10^{3}$ &  & Y \\ 
F &  21 &  &  0.75 & 2.3 & & 0.5 & 0.3 &0.09 & 2.4 &  380  & 0.7 & $4.2\times 10^{4}$ &  & Y \\ 
G &  16 &  &  0.59 & 1.3 & & 0.6 & 0.4 &0.07 & 1.1 &  377  & 0.8 & $3.4\times 10^{4}$ &  & Y \\ 

\hline                                   
\end{tabular}
\tablefoot{\\
\tablefoottext{a}{The $"$Y'' symbol indicates that the association with MYSO candidates has been observed while the $"$N'' symbol indicates that it has not been observed.}\\
 }
\normalsize
\end{table*}

\renewcommand{\tabcolsep}{0.1cm}
\begin{table*}
\caption{Properties of the dust clumps.}\label{Tab:dustclump}
\scriptsize
\centering   
\begin{tabular}{cccccccccccccccc}
\hline\hline                        
name\tablefootmark{a}   & R.A.(J2000) & DEC.(J2000) & $l$\tablefootmark{b}    & $b$\tablefootmark{b}     & $R_{cl}$\tablefootmark{b} & $\upsilon_{\rm lsr}$\tablefootmark{c} & $\Delta v$\tablefootmark{c} & $S_{\rm peak}$\tablefootmark{b,d} & $S_{\rm int}$\tablefootmark{b} & $T_{\rm ex}$\tablefootmark{e}  &N(H$_{2}$)    &  n(H$_{2}$)   & $M_{\rm dust}$ & $\alpha_{\rm vir}$ & C$^{18}$O clump  \\  
      & (hh:mm:ss.s) & (dd:mm:ss)       &(\degree)&(\degree) & (pc)    & (\kms)           & (\kms)    & (Jy/beam)                     & (Jy)    & (K)    & ($\times 10^{22}$cm$^{-3}$) &  ($\times10^{4}$~cm$^{-3}$)  &  ($M_{\odot}$) &                  & association \\  
\hline
 D1 & 06:58:43.1 & $-$04:56:23 & 218.157 & $-$0.636 & 0.21 & 26.0 & 1.8 & 0.26 & 0.73 & 21 & 0.4 & 2.1 &    57  &  2.6   & clump A \\
 D2 & 06:58:45.3 & $-$03:41:01 & 217.044 & $-$0.054 & 0.42 & 53.8 & 1.8 & 0.35 & 0.73 & 21 & 0.5 & 1.3 &   282  &  1.0   &         \\
 D3 & 06:59:06.7 & $-$04:52:21 & 218.142 & $-$0.518 & 0.20 & 25.2 & 1.4 & 0.23 & 0.62 & 16 & 0.3 & 2.0 &    48  &  1.7   &         \\
 D4 & 06:59:14.4 & $-$03:51:52 & 217.260 & $-$0.029 & 0.29 & 28.3 & 2.4 & 0.42 & 1.82 & 18 & 0.6 & 2.1 &   142  &  2.5   & clump G \\
 D5 & 06:59:14.8 & $-$03:54:39 & 217.302 & $-$0.049 & 0.29 & 26.5 & 2.2 & 0.51 & 2.56 & 22 & 0.7 & 3.0 &   200  &  1.5   & clump F \\
 D6 & 06:59:16.6 & $-$03:59:21 & 217.375 & $-$0.078 & 0.47 & 25.5 & 2.7 & 1.69 & 13.96& 29 & 2.4 & 3.6 &  1093  &  0.7   & clump E \\
 D7 & 06:59:21.6 & $-$04:15:58 & 217.631 & $-$0.186 & 0.28 & 25.7 & 1.4 & 0.46 & 1.82 & 19 & 0.6 & 2.2 &   142  &  0.8   &         \\ 
 D8 & 06:59:30.8 & $-$04:05:37 & 217.495 & $-$0.073 & 0.23 & 23.4 & 2.0 & 0.39 & 1.19 & 15 & 0.6 & 2.5 &    93  &  2.1   &         \\
 D9 & 06:59:35.3 & $-$04:45:59 & 218.102 & $-$0.364 & 0.21 & 26.1 & 2.6 & 0.79 & 1.83 & 19 & 1.1 & 5.5 &   143  &  2.0   & clump C \\
D10 & 06:59:36.5 & $-$04:40:23 & 218.021 & $-$0.317 & 0.23 & 27.5 & 2.7 & 0.35 & 1.32 & 24 & 0.5 & 2.8 &   103  &  3.4   & clump D \\
D11 & 06:59:38.8 & $-$04:41:38 & 218.044 & $-$0.318 & 0.25 & 28.1 & 2.3 & 0.33 & 1.35 & 24 & 0.5 & 2.3 &   105  &  2.6   &         \\ 
D12 & 06:59:41.0 & $-$04:51:36 & 218.196 & $-$0.386 & 0.23 & 30.0 & 1.9 & 0.37 & 1.14 & 25 & 0.5 & 2.5 &   88   &  1.9   &         \\ 
D13 & 06:59:42.4 & $-$04:49:56 & 218.174 & $-$0.368 & 0.16 & 28.3 & 3.9 & 0.09 & 0.34 & 23 & 0.1 & 2.3 &   26   &  18.5  &         \\ 
D14 & 06:59:44.7 & $-$04:48:58 & 218.164 & $-$0.352 & 0.19 & 28.9 & 3.7 & 0.28 & 0.68 & 24 & 0.4 & 2.6 &   53   &  10.0  &         \\ 
D15 & 07:00:23.0 & $-$04:36:26 & 218.051 & $-$0.115 & 0.24 & 26.6 & 1.6 & 0.71 & 1.96 & 16 & 1.0 & 3.7 &   153  &  0.9   &         \\ 

\hline                                   
\end{tabular}
\tablefoot{\\
\tablefoottext{a}{All dust clumps are located at a distance of 2.42 kpc except D2 which is located at 5.4 kpc.}\\
\tablefoottext{b}{These values are based on the catalog of \citet{2010ApJS..188..123R}.}\\
\tablefoottext{c}{The velocity information is based on the corresponding $^{13}$CO spectra.}\\
\tablefoottext{d}{The corresponding beamsize is 40\arcsec.}\\
\tablefoottext{e}{The CO excitation temperature is extracted toward the position of each dust clump.}\\
 }
\normalsize
\end{table*}

\subsection{Star formation in the surveyed region}\label{SF}
\subsubsection{Identification and classification of young stellar object candidates}
Low-mass YSOs can be assigned into Class I, Class II and Class III objects according to their spectral indices \citep[e.g.,][]{2009ApJS..181..321E}. They represent different evolutionary stages throughout the whole life of low-mass YSOs. Class I and II objects are referred to as disk-bearing YSOs. To select the disk-bearing YSO candidates in the studied region ($216.25$\degree $\le$ $l$ $\le$ $218.75$\degree, $-0.75$\degree$\le$ $b$ $\le$ $1.25$\degree)\footnote{The region used to classify YSO candidates covers 5~square degrees, which is larger than the area mapped in the three CO lines.}, we made use of the infrared data from the 2MASS and WISE surveys. The dusty circumstellar disks associated with the disk-bearing YSOs will create infrared excess, which causes their infrared colors to be different from the diskless Class III objects. On the other hand, it is impossible to distinguish the diskless YSOs and unrelated field objects only based on their infrared colors. Therefore, only the disk-bearing YSOs are investigated in this work.

\begin{figure*}[!htbp]
\centering
\includegraphics[width = 0.45 \textwidth]{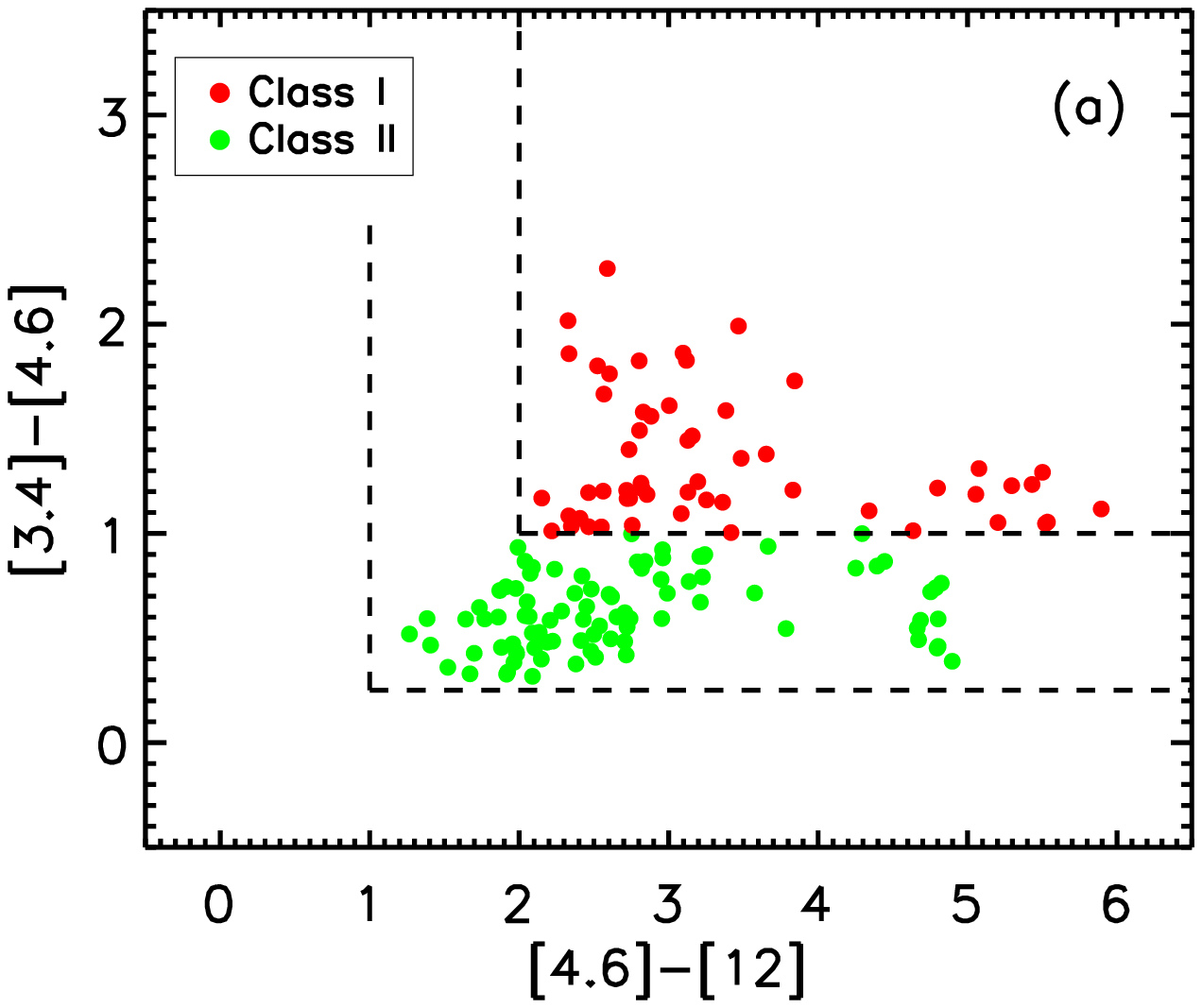}
\includegraphics[width = 0.45 \textwidth]{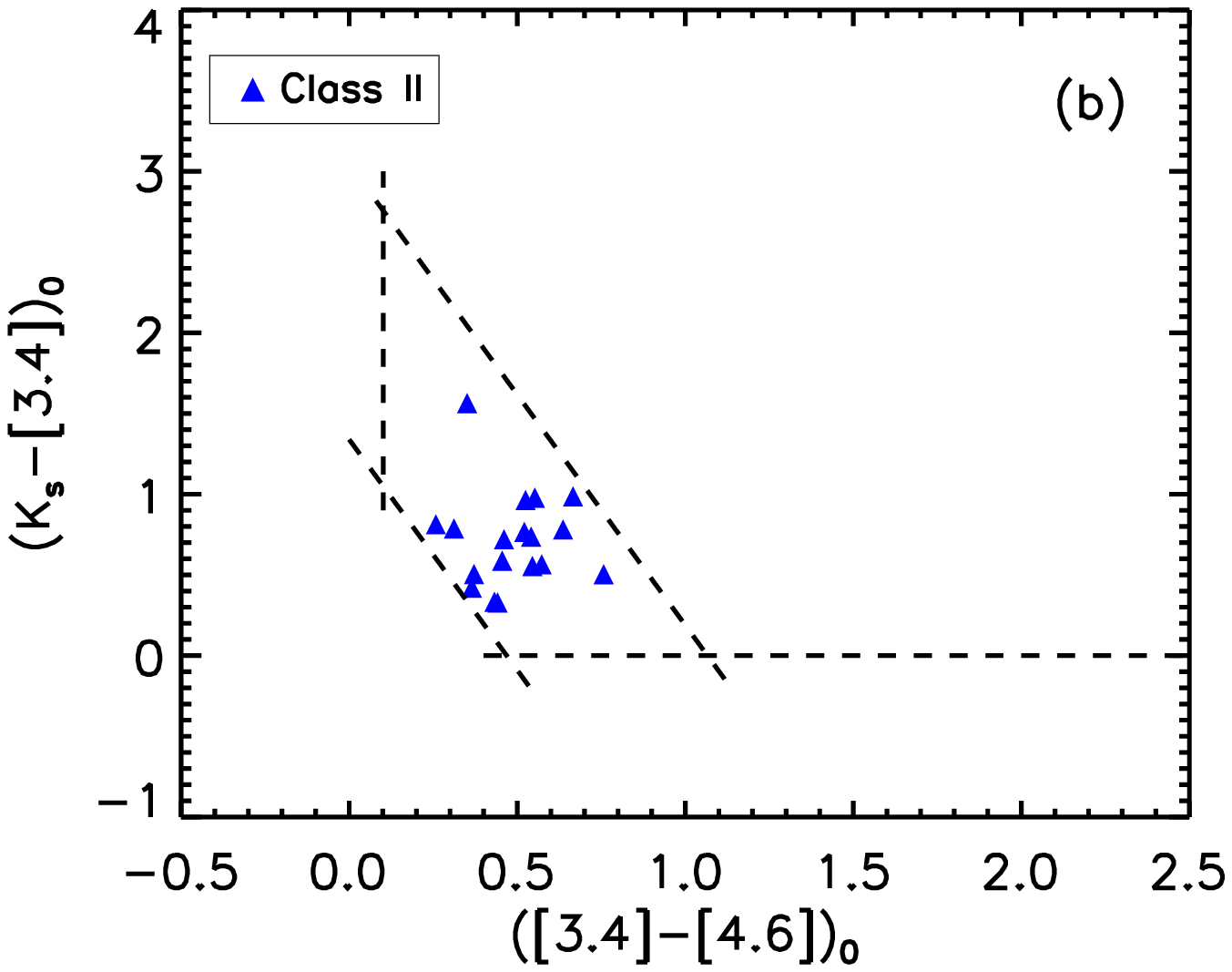}
\caption{{(a) WISE [3.4]$-$[4.6] vs. [4.6]$-$[12] color-color diagram showing the distribution of Class I (red circles) and Class II YSOs (green circles) selected with method 1. (b) Dereddened $K_{\rm s}-$[3.4], [3.4]$-$[4.6] color diagram. Only 18 additional Class II YSOs (blue triangles) are selected with method 2.} \label{Fig:cc}}
\end{figure*}

Two methods are employed to select YSO candidates in this work. The first is only based on the WISE data, and the details of criteria are given in \citet{2012ApJ...744..130K}. The selection is mainly based on the WISE [3.4]$-$[4.6] vs. [4.6]$-$[12] color-color diagram (hereafter the method 1, see Fig.~\ref{Fig:cc}a). The contaminations from extragalactic sources (star-forming galaxies and AGNs), shock emission blobs, and resolved PAH emission objects were removed based on their locations in the [3.4]$-$[4.6] vs. [4.6]$-$[12] color-color diagram and their WISE photometry \citep[see details in ][]{2012ApJ...744..130K}. For the sources that are not detected in the WISE [12] band, their dereddened photometry in the WISE [3.4] and [4.6] bands, in combination with the dereddened 2MASS $K_{\rm s}$ photometry, was used to construct the  ($K_{\rm s}$ $-$[3.4])$_{0}$ vs. ([3.4]$-$[4.6])$_{0}$ color-color diagram to find additional Class I and II YSO candidates (hereafter the method 2, see Fig.~\ref{Fig:cc}b). In method 2, the extinction used to deredden the photometry is estimated from its location in the $J-H$ vs. $H-K_{\rm s}$ color-color diagram as described in \citet{2013ApJS..207....5F}. The extinction law is based on the measurement of five nearby star-forming regions \citep{2007ApJ...663.1069F}. The criteria for this method are also described in \citet{2012ApJ...744..130K}. The targets were rejected when they were associated with 2MASS extended sources. As a result, we found 163 YSO candidates, of which 56 are Class I and 107 are Class II. All Class I YSO candidates are identified with method 1. Eighty-nine Class II YSO candidates are identified with method 1, and 18 with method 2. These identified disk-bearing YSO candidates are tabulated in Table~\ref{Tab:irpho}.

\subsubsection{Distribution of young stellar object candidates}
\begin{figure*}[!htbp]
\centering
\includegraphics[width = 0.9 \textwidth]{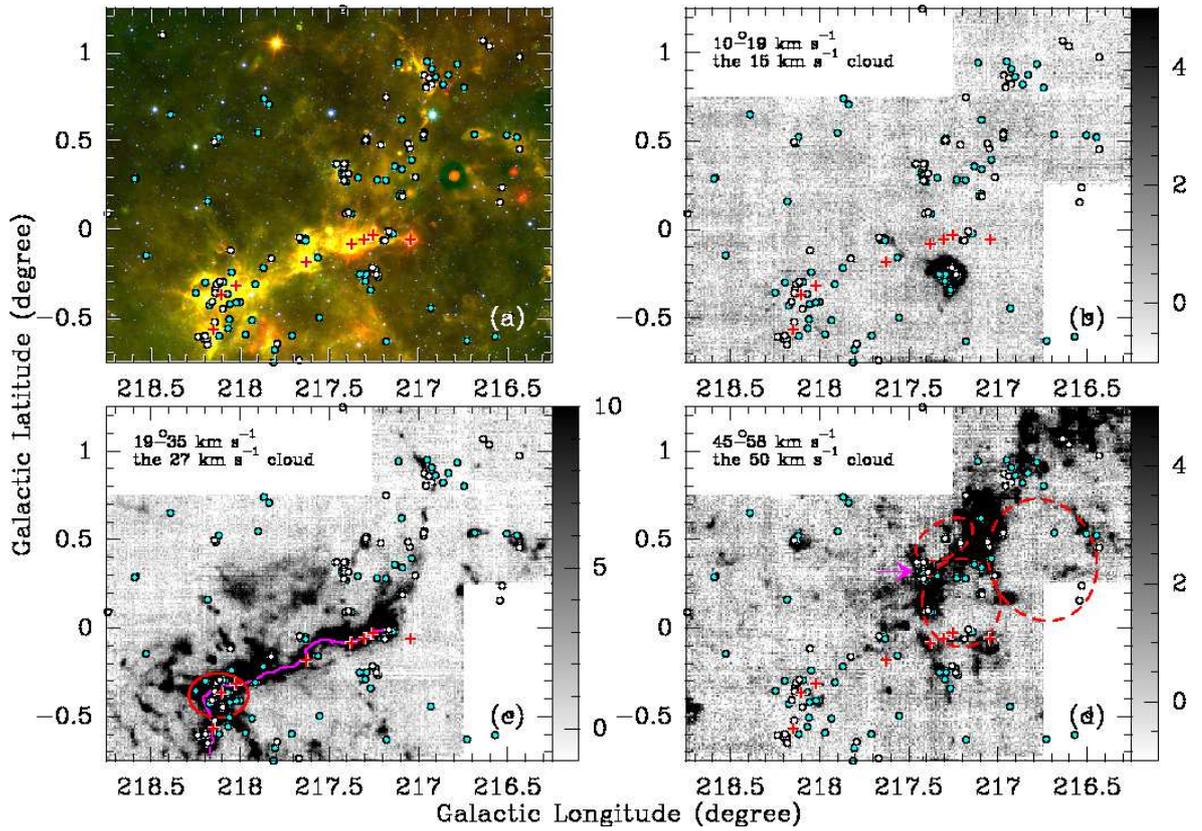}
\caption{{(a): Three-color composite image of the 22 $\mu$m (red), the 12 $\mu$m (green) and 4.6 $\mu$m emission (blue) from WISE overlaid with the YSO distribution. (b): Same as Fig.~\ref{Fig:allclouds}b, but overlaid with the YSO distribution. (c): Same as Fig.~\ref{Fig:allclouds}c, but overlaid with the YSO distribution. The solid red ellipse represents S287 while the purple solid line marks the highest H$_{2}$ column density line of the filament S287-main along its long axis. (d) Same as Fig.~\ref{Fig:allclouds}d, but overlaid with the YSO distribution. The dashed red ellipses represent shell A, shell B, and shell C. The purple arrow shows the overdensity of YSOs around the shells. In the four panels, the Class I and Class II YSOs are marked with white and cyan circles, while MSYOs are marked with red crosses.} \label{Fig:cloudyso}}
\end{figure*}

The spatial distribution of these YSO candidates is displayed in Fig.~\ref{Fig:cloudyso}, in which nine MYSO candidates from the RMS survey are also shown. Figures~\ref{Fig:cloudyso}b--d shows that most YSO candidates are distributed within molecular clouds, which is also observed in many other star-forming regions \citep[e.g.,][]{2012AJ....144..192M,2013ApJS..207....5F}. Comparing Fig.~\ref{Fig:cloudyso}a with Fig.~\ref{Fig:cloudyso}c, we find that most of bright thermal dust emission traced by 12~$\mu$m and 22~$\mu$m closely matches the morphology of the 27~\kms\,cloud, which suggests that most active star-forming activities occur in the 27~\kms\,cloud. No extended 12~$\mu$m and 22~$\mu$m emission is found toward the 15~\kms\,cloud, indicating that the dust temperature is colder in this region. Figure~\ref{Fig:cloudyso}b shows YSO candidates in island I. Figure~\ref{Fig:cloudyso}c shows that YSO candidates are found to lie in S287-main, tail, and knots, but are absent toward island II. This is presumably because island II is not dense enough to collapse and form stars. On the other hand, an overdensity of YSO candidates is found toward the H{\scriptsize II} region S287, indicating that triggered star formation may occur around this region. Significant YSO candidates coincide with the highest column density line of the filament. Class I YSO candidates are confined closer to the line than Class II YSO candidates. Particularly, eight MYSO candidates lie in the highest column density line. In the 50~\kms\,cloud, enhanced star formation is found around the intersection of shell B and shell C (indicated by the purple arrow in Fig.~\ref{Fig:cloudyso}d). Class I YSO candidates are found to be concentrated around the shells while the distribution of Class II YSO candidates is more random. Class I YSO candidates trace more recent star-forming activities, so that the concentration of Class I YSO candidates around the shells is indicative of sequential star formation. This is probably because the region is highly compressed by the expansion of shell B and shell C, probably triggering the formation of YSOs.

\section{Discussion}
\subsection{Star formation efficiencies and rates}\label{t:sfe}
Star formation efficiencies (SFEs) and rates (SFRs) play an vital role in theoretical descriptions of star formation and cloud evolution \citep[e.g.,][]{2012ARA&A..50..531K}. SFE is defined as below 
\begin{equation}\label{f.sfe}
{\rm SFE} = \frac{M_{*}}{M_{*}+M_{\rm cloud}} \;, 
\end{equation}
where $M_{*}$ is the total mass of YSOs and $M_{\rm cloud}$ is the total cloud mass. SFEs vary from less than 0.1\% to 50\% with a median value of 2\% among the molecular clouds in the inner Galaxy \citep{1986ApJ...301..398M}. 

 To calculate the total stellar mass, we made the following assumptions: (1) the mass distribution of YSOs follows the initial mass function (IMF) of \citet{2003PASP..115..763C}
\begin{eqnarray}\label{f.imf}
\nonumber \xi(m)&\sim& 1/m ~{\rm exp}[-({\rm log}(m)-{\rm log}(0.22))^{2}/(2\times 0.57^{2})], \; m<1 \;,\\
 \xi(m)&\sim& m^{-2.3}, \; m>1 \;, 
\end{eqnarray}
where $m=M/M_{\odot}$. (2) According to previous studies toward Taurus and L1641, the disk fractions, defined as $N$(II)/$N$(II+III), are found to be 50\%--60\% \citep{2010ApJS..186..111L,2013ApJS..207....5F}. Here, we assumed a disk fraction of 50\% for our case. (3) We assumed that the lower limit of the identified YSO mass is based on 14 mag in 2MASS $K_{\rm s}$ band, which corresponds to 3.3, 2.1, and 0.3 absolute magnitude in the 15~\kms\,cloud, the 27~\kms\,cloud, and the 50~\kms\,cloud. This results in lower mass limits of 0.2, 0.6, and 2.7 M$_{\odot}$ for the three clouds according to the model of \citet{2000A&A...358..593S}, assuming that their ages are all 1~Myr. On the other hand, the upper limit in the IMF will not affect the total mass significantly and was thus simply assumed to be 80~$M_{\odot}$. Given that we do not have spectroscopic information of YSOs, whether YSOs are associated with molecular clouds or not depends only on their spatial distribution. By cross-matching YSOs and the molecular clouds (see Fig.~\ref{Fig:cloudyso}), we find 2 Class I and 8 Class II YSO candidates associated with the 15\kms\,cloud, 26 Class I and 53 Class II candidates associated with the 27~\kms\,cloud, and 22 Class I and 37 Class II candidates associated with the 50~\kms\,cloud, respectively. Taking the MYSO candidates into account and assuming a disk fraction of 50\%, we find that the total number of YSOs is 18, 140 and 97 in the 15~\kms\,cloud, the 27~\kms\,cloud, and the 50~\kms\,cloud. With the IMF and the assumptions described above, the total stellar mass of the three clouds is estimated to be 10~$M_{\odot}$, 885~$M_{\odot}$ and 7288~$M_{\odot}$. We note that the number of YSOs in the 15~\kms\,cloud is too small and statistically unimportant, therefore the derived total stellar mass may have a large uncertainty. The total stellar mass of the 50~\kms\,cloud probably is significantly overestimated since the lower limit mass (2.7~$M_{\odot}$) in the IMF derived from the $K_{\rm s}$ band magnitude is too high and its total YSO number may be contaminated by foreground YSOs or circumstellar material such as post-AGB stars at such a far distance ($5.40$~kpc) \citep{2011A&A...526A..24V}. The discussions below therefore only focus on the 27~\kms\,cloud.

Since a part of molecular gas is detected in $^{12}$CO, but not in $^{13}$CO, the mass derived from $^{13}$CO is the lower limit of the total cloud mass. As pointed out by \citet{2010ApJ...723.1019H}, $^{13}$CO may underestimate the mass by a factor of about 4--5, which suggests that the total cloud mass of the 27~\kms\,cloud is about 1.0$\times10^{5}$~M$_{\odot}$ (a correction factor of 4 applied for the mass derived from $^{13}$CO). Along with the mass (8.3$\times10^{4}$~M$_{\odot}$) derived from $^{12}$CO, the SFEs of the 27~\kms\,cloud is thus estimated to be 1.1\% and 0.9\%. Therefore, the SFE of the 27~\kms\,cloud is lower than the SFEs (3.0\%--6.3\%) in nearby clouds \citep{2009ApJS..181..321E} and the median value (2\%) of the molecular clouds in the inner galaxy \citep{1986ApJ...301..398M}. This suggests that molecular clouds in the outer Galaxy may have lower SFEs.

We also calculated the SFR of the 27~\kms\,cloud with the total mass of YSOs by
\begin{equation}\label{f.sfr}
{\rm SFR}=\frac{M_{*}}{\tau} \;,
\end{equation}
where $\tau$ is the average age of YSOs assumed to be 2 Myr \citep[e.g.,][]{2009ApJS..181..321E}. This results in an SFR of 443~$M_{\odot}$~Myr$^{-1}$ for the 27~\kms\,cloud. The SFR for the 27~\kms\,cloud is found to be higher than individual nearby clouds such as the Perseus, Serpens, and Ophiuchus clouds. This is mainly due to its larger size, since the SFR surface density (0.28~$M_{\odot}{\rm yr^{-1}}{\rm kpc^{-2}}$) is similar to those of individual nearby clouds \citep{2010ApJ...723.1019H}.  

On the other hand, we can estimate the SFR surface density from the Kennicutt-Schmidt law employed for other galaxies \citep{1998ApJ...498..541K} with the formula
\begin{eqnarray}\label{f.sflaw}
  & &\Sigma({\rm SF})\,(M_{\odot}\;{\rm yr}^{-1}\;{\rm kpc}^{-2})\nonumber \\
&=&(2.5\pm0.7)\times 10^{-4} (\Sigma({\rm gas})/1M_{\odot}\;{\rm pc}^{-2})^{1.4\pm 0.15} \;,
\end{eqnarray}
 where $\Sigma({\rm SF})$ is the SFR surface density, $\Sigma({\rm gas})$ is the gas surface density. To be consistent with \citet{1998ApJ...498..541K}, we used the gas surface density of 53 $M_{\odot}\;{\rm pc}^{-2}$ derived from $^{12}$CO to calculate the expected SFR surface density for the 27~\kms\,cloud, and obtain a value of 0.065~$M_{\odot}\;{\rm Myr}^{-1}\;{\rm pc}^{-2}$ which is a factor of 4 lower than the observed value of 0.28~$M_{\odot}{\rm yr^{-1}}{\rm kpc^{-2}}$. This is consistent with the observations toward nearby clouds \citep{2010ApJ...723.1019H,2010ApJ...724..687L}.

\subsection{Fragmentation in the filament S287-main}
\begin{figure*}[!htbp]
\centering
\includegraphics[width = 0.9\textwidth]{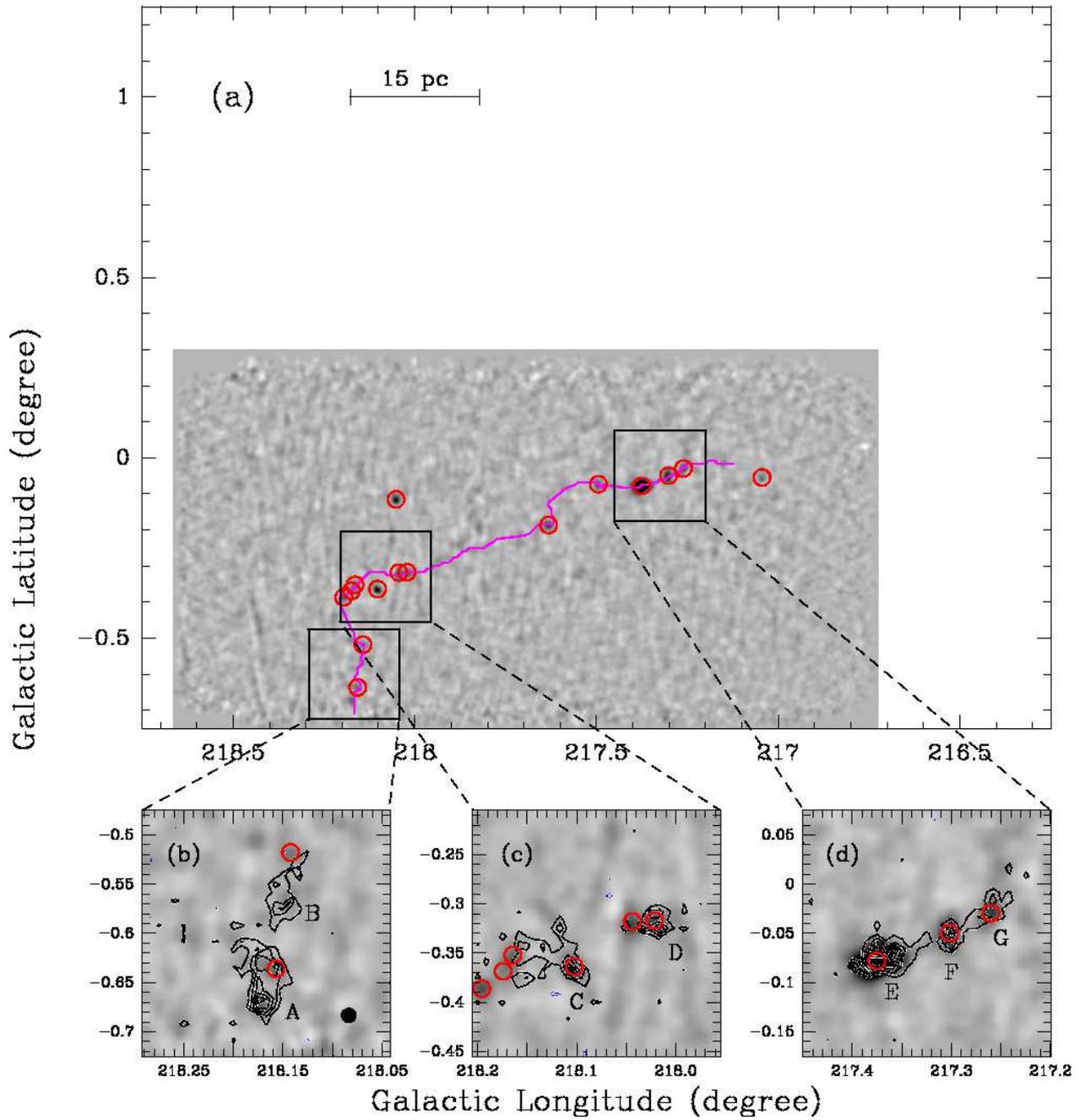}
\caption{{(a): Convolved BGPS 1.1 dust continuum map. The purple solid line marks the highest H$_{2}$ column density line of the filament S287-main along its long axis. (b): Zoom in the region indicated in Fig.~\ref{Fig:clumpsd}a overlaid with C$^{18}$O integrated intensity map. Contours start from 1.2~K~\kms (3~$\sigma$) to 3.6~K~\kms (9~$\sigma$) by 0.6~K~\kms (1.5~$\sigma$). (c) Same as Fig.~\ref{Fig:clumpsd}b. (d) Same as Fig.~\ref{Fig:clumpsd}b. In all panels, the open red circles represent the dust clumps from the catalog of \citet{2010ApJS..188..123R,2013ApJS..208...14G}.} \label{Fig:clumpsd}}
\end{figure*}

Filaments are found to be ubiquitous in molecular clouds and are thought to be going to fragment into prestellar cores \citep[e.g.,][]{2010A&A...518L.102A}. Some of them have the capability of forming massive stars and clusters \citep[e.g.,][]{2010A&A...518L..95H}. S287-main is such a filament, where the optical H{\scriptsize II} region S287 and the stellar association OB3 are located (see Fig.~\ref{Fig:allclouds}c and Fig.~\ref{Fig:cloud2-mom}a), indicating that these OB stars may form out of the filament. Moreover, there are massive clumps along the highest H$_{2}$ column density line of the filament (see Figs.~\ref{Fig:clumps}a and \ref{Fig:clumpsd}a), and these clumps are most likely to form massive stars (see Sect.~\ref{t:clumps}).

 Along the highest H$_{2}$ column density line of the filament, the mean excitation temperature is 17 K. If the CO population is fully thermalized, the kinetic temperature is equal to the excitation temperature. According to an unmagnetized isothermal filament model \citep{1964ApJ...140.1056O,1992ApJ...388..392I,1997ApJ...480..681I}, the critical line mass (mass per unit length) of the filament is  $M_{\rm line,crit}=2c_{\rm s}^{2}/{G}$, where $c_{\rm s}$ is the sound speed and $G$ is the gravitational constant. Thus, the critical line mass of S287-main is found to be 27~$M_{\odot}$~pc$^{-1}$. Following previous studies \citep[e.g.,][]{2010A&A...518L.102A,2011A&A...529L...6A,2015A&A...581A...5S}, we first investigated the radial profile perpendicular to the highest H$_{2}$ column density line, and then used a Gaussian fit to the radial profiles. We find a characteristic width of 1.8 pc. By integrating over the characteristic radial extent of S287-main, the filament is found to harbor a mass of 1.5$\times 10^{4}$~$M_{\odot}$, consist of 59\% mass of the 27~\kms\,cloud derived from $^{13}$CO (see Sect.~\ref{t:cloud}). Dividing the length (71 pc) of S287-main by the mass of filament, we obtain an observed line mass of 211~$M_{\odot}$~pc$^{-1}$, which is much higher than the critical line mass. This suggests that S287-main is supercritical. S287-main is thus unstable to perturbation and fragments into clumps. This is well consistent with the fact that clumps and YSOs lie along the highest H$_{2}$ column density line of S287-main (see Fig.~\ref{Fig:clumps}a,~\ref{Fig:clumpsd}a and \ref{Fig:cloudyso}c). 

To further investigate the fragmentation, we also compared observed parameters with theoretical models. In the turbulent fragmentation model of an isothermal gas cylinder \citep[e.g.,][]{1953ApJ...118..116C,1992ApJ...388..392I,1997ApJ...480..681I,2010ApJ...719L.185J,2014MNRAS.439.3275W}, the filament will fragment into clumps separated by a typical spacing of
\begin{equation}
\lambda_{\rm t} =22 H = 22\delta_{\rm v}(4\pi G \rho)^{-0.5}\;,\\
\end{equation}
where $\delta_{\rm v}$ is the turbulent velocity dispersion, and $\rho$ is the gas density at the center of the filament. From Table~\ref{Tab:peakobs} and \ref{Tab:dustclump}, we used the line widths of $^{13}$CO to obtain the turbulent velocity dispersion. The typical turbulent velocity dispersion is consequently about 1~\kms. Based on Table~\ref{Tab:phyclump} and \ref{Tab:dustclump}, we adopted 10$^{4}$ cm$^{-3}$ as the gas density at the center of the filament. This resulted in a theoretical spacing of 4.2 pc between clumps. This is similar to the observed separations (0.8--8 pc) of clumps in the filament. This agrees with previous studies \citep[e.g.,][]{2010ApJ...719L.185J,2015arXiv150807898S}. It suggests that turbulent fragmentation plays an important role in fragmenting the filament into clumps. In a model of gravitational fragmentation \citep{2002ApJ...578..914H}, the fragmentation length ($\lambda_{\rm c}$) and the corresponding collapse timescale ($\tau$) for the filament are
\begin{eqnarray}\label{f.frag}
\lambda_{\rm c}& =& 3.94\frac{c_{\rm s}^{2}}{G\Sigma}=1.5 T_{10}A_{\rm V}^{-1} \;{\rm pc}\;,\\
\tau &\sim& 3.7T_{10}^{1/2}A_{\rm V}^{-1} \;{\rm Myr}\;,
\end{eqnarray}
where $c_{\rm s}$ is the sound speed, $\Sigma$ is the surface density at the center of the filament, $A_{\rm v}$ is the visual extinction, and $T_{10}$ is the temperature in units of 10 K. Along the highest H$_{2}$ column density line of the filament, the mean column density is 9.2$\times 10^{21}$~cm$^{-2}$, which corresponds to a mean visual extinction of $\sim$10 mag \citep[according to $N({\rm H_{2}})=9.4\times 10^{20}\;{\rm cm}^{-2}\;(A_{\rm V}/{\rm mag})$,][]{1978ApJ...224..132B}. Hence, the fragmentation length is $\sim$0.3 pc, and the corresponding collapse timescale is 0.4 Myr. We explored the projected nearest neighbor separations of the Class I YSO candidates associated with the filaments, which ranges from 0.18 pc to 3.0 pc with a median of 1.05 pc. As pointed out by \citet{2002ApJ...578..914H}, the effective sound speed may have additional contributions from other complex motions which will increase the fragmentation length by a factor of a few. This could explain why significant portions of separations are larger than 0.3 pc. Furthermore, the predicted collapse timescale coincides with the lifetime of Class I YSOs \citep[$\sim$0.5 Myr][]{2009ApJS..181..321E,2013ApJS..207....5F}. On small-scales, our results support the hypothesis that the formation of YSOs roughly agrees with the model of gravitational fragmentation, consistent with previous studies toward other filaments \citep{2002ApJ...578..914H,2015A&A...581A...5S}. Therefore, we suggest that turbulence controls large-scale fragmentation in the filament, while gravitational fragmentation plays an important role in YSOs fragmentation on small scales.

\subsection{Comparison of star-forming molecular gas with non-star-forming molecular gas}
\begin{figure*}[!htbp]
\centering
\includegraphics[width = 0.45 \textwidth]{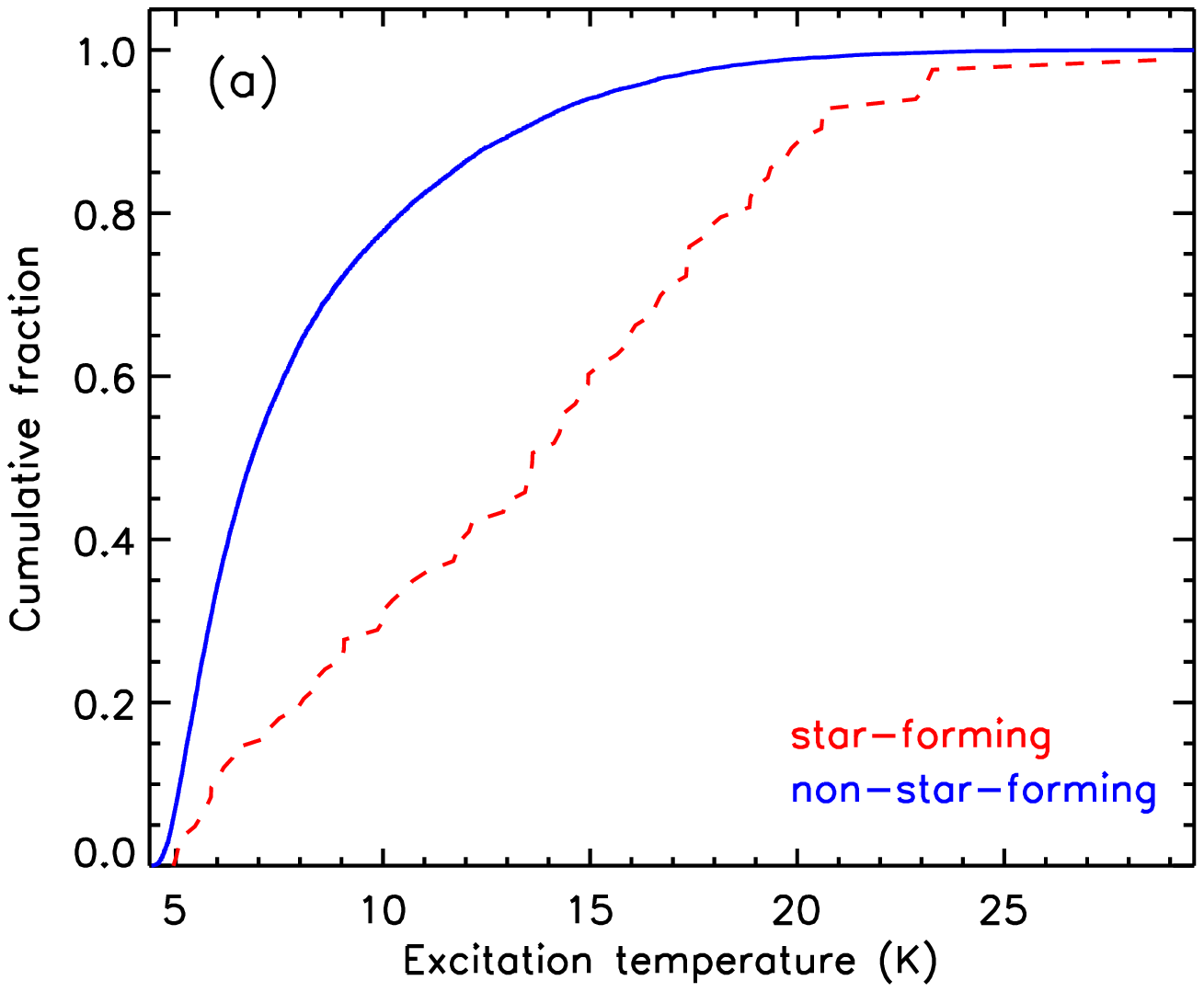}
\includegraphics[width = 0.45 \textwidth]{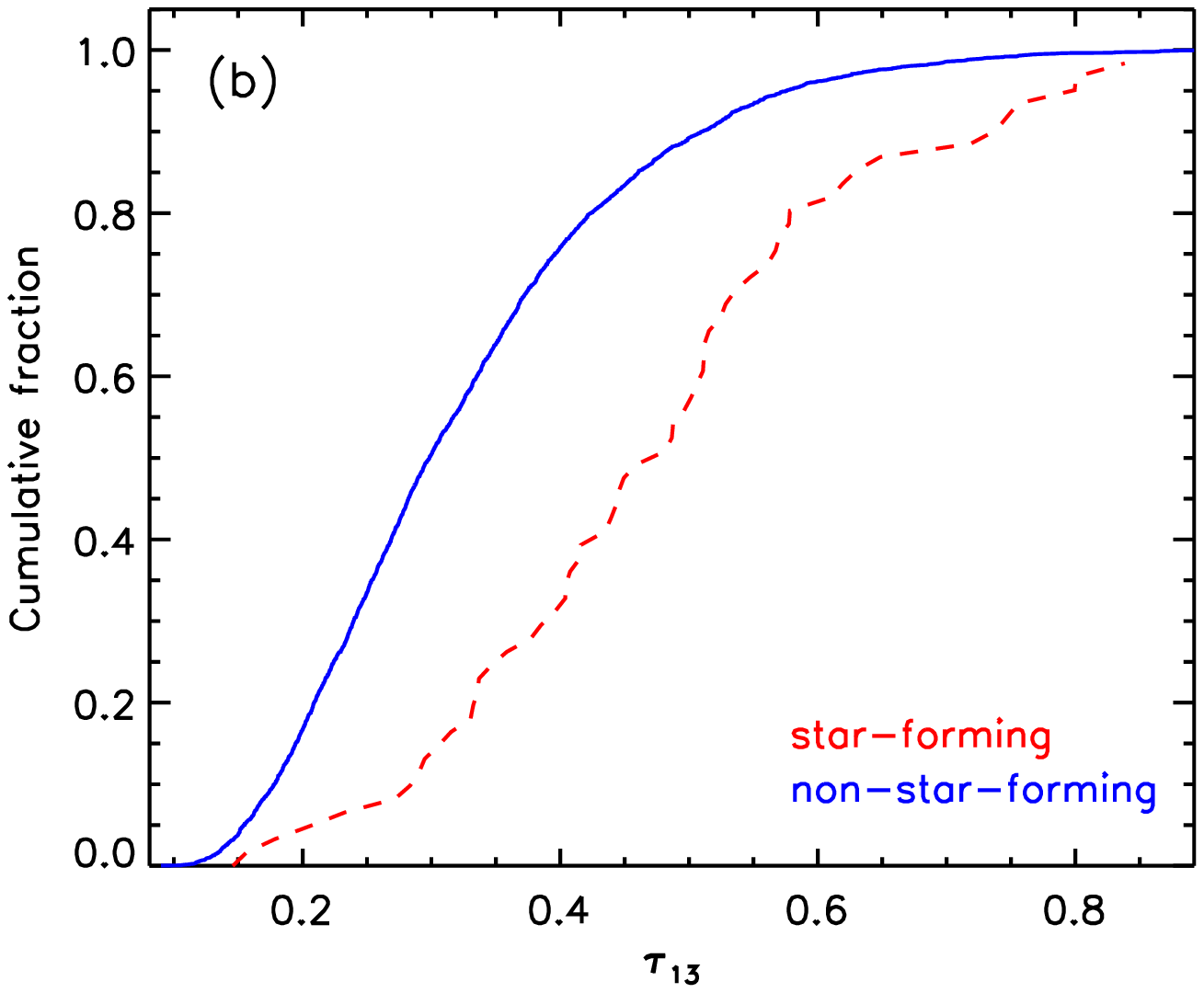}
\includegraphics[width = 0.45 \textwidth]{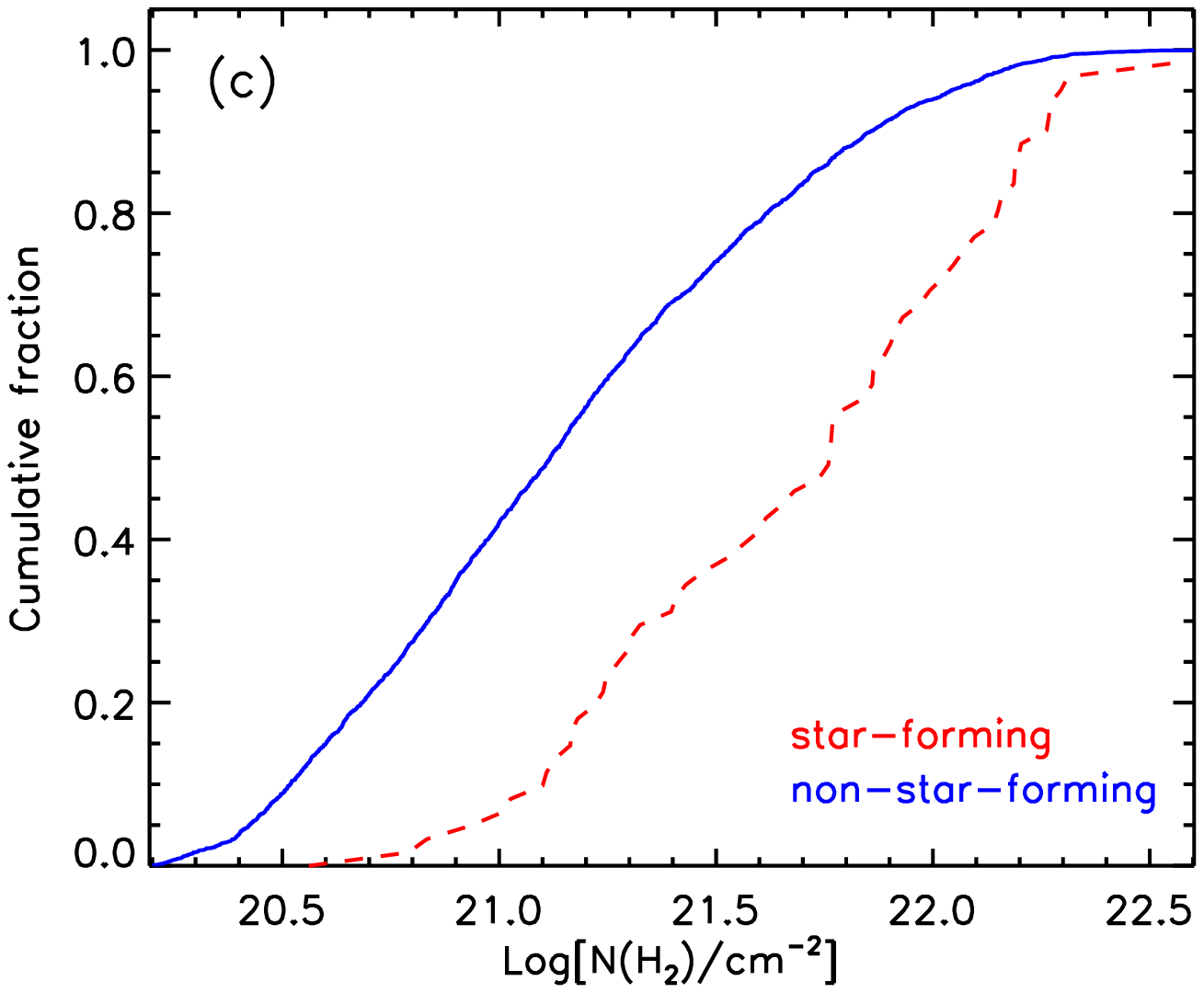}
\caption{{Cumulative distribution of excitation temperature (Fig.~\ref{Fig:ksmc}a), $^{13}$CO (1--0) opacities (Fig.~\ref{Fig:ksmc}b), and H$_{2}$ column densities (Fig.~\ref{Fig:ksmc}c) of star-forming (dashed red lines) and non-star-forming (solid blue lines) molecular gas for the 27~\kms\,cloud.} \label{Fig:ksmc}}
\end{figure*}

Here, we divided molecular gas into two types, star-forming and non-star-forming, to compare their properties. Molecular gas associated with YSO candidates was assigned to be star forming while molecular gas without YSO counterparts was assigned to be non-star forming. The two types of molecular gas are only identified by the spatial distribution correlation between YSO candidates and molecular clouds. To avoid the effects that the clouds in different environments (e.g., in different arms) may have different properties, we only carried out the comparison toward the 27 \kms\,cloud. We carried out a Kolmogorov-Smirnov test (K-S test) to investigate the difference between the two types of molecular gas. Based on their cumulative distributions of physical properties (Fig.~\ref{Fig:ksmc}), the K-S test confirms that the null hypothesis that the two distributions are the same can be rejected for excitation temperature, $^{13}$CO (1--0) opacity, and H$_{2}$ column density with greater than 3$\sigma$ confidence (all the probabilities of similarity are found to be $\ll$0.003, to reject the null hypothesis with $\ge$3$\sigma$ confidence the probability must be $<$0.003). Figure~\ref{Fig:ksmc}a shows that star-forming molecular gas tends to have a higher excitation temperature than non-star-forming molecular gas. This most likely attribute to the heating of YSOs. Figures~\ref{Fig:ksmc}b and \ref{Fig:ksmc}c demonstrate that star-forming gas has a higher $^{13}$CO (1--0) opacity and a higher H$_{2}$ column density than non-star-forming molecular gas, suggesting that star formation occurs in denser molecular gas. This is consistent with the fact that dense gas in molecular clouds regulates star formation from Galactic star-forming regions to external galaxies \citep{2004ApJ...606..271G,2005ApJ...635L.173W,2010ApJ...724..687L}.   

\subsection{Dense gas mass fraction in the 27~\kms\,cloud}\label{t:frac}
\begin{figure*}[!htbp]
\centering
\includegraphics[width = 0.9 \textwidth]{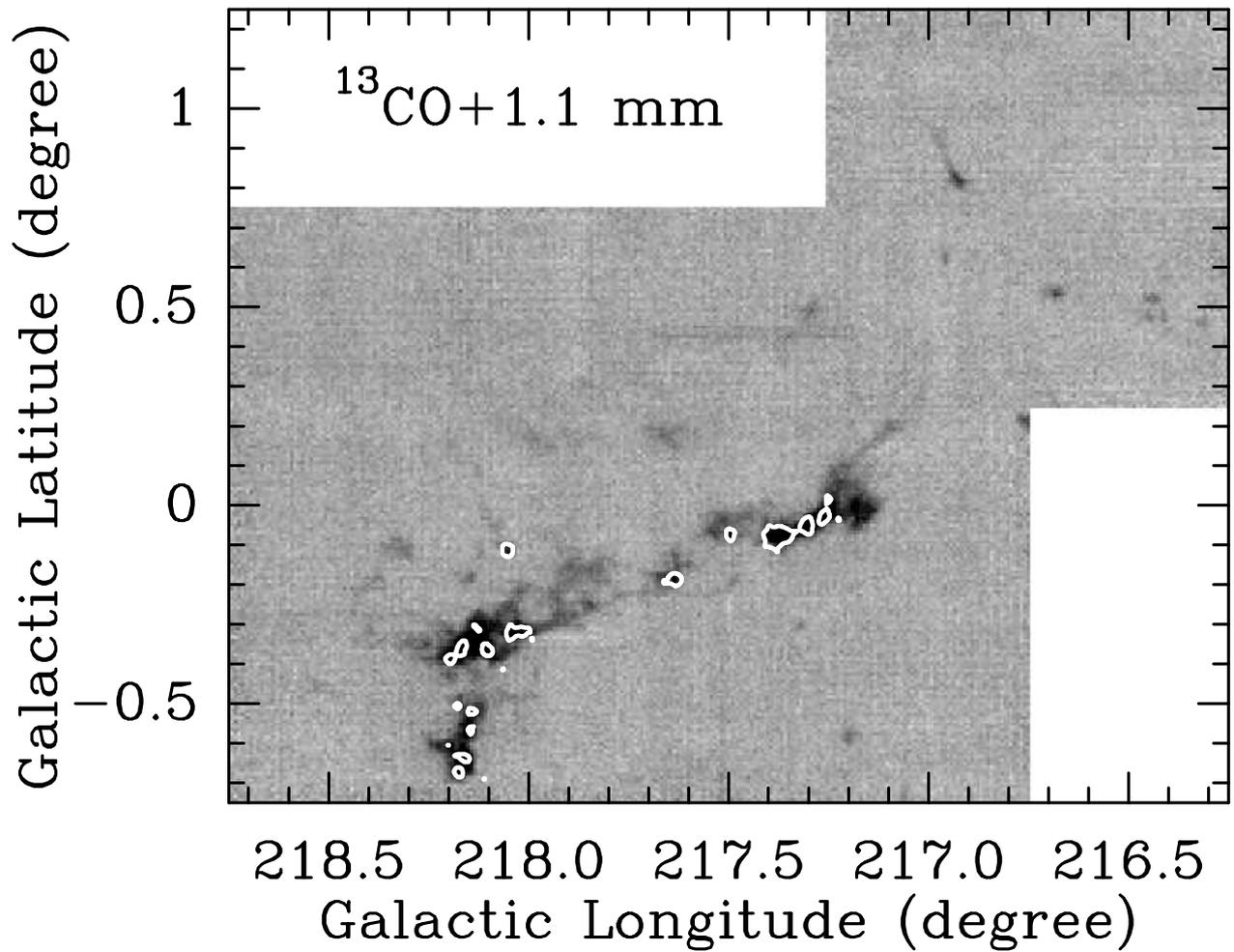}
\caption{{Convolved BGPS 1.1 dust continuum emission (white contours) overlaid on the $^{13}$CO (1--0) integrated intensity map. The $^{13}$CO (1--0) map has been integrated over the velocity range between 22 and 29~\kms. The contour for the BGPS 1.1 dust continuum emission corresponds to the threshold of flux densities, 0.09 Jy~beam$^{-1}$.} \label{Fig:dust}}
\end{figure*}

At present, star formation is believed to be regulated by dense gas in molecular clouds \citep[e.g.,][]{2004ApJ...606..271G,2005ApJ...635L.173W}. \citet{2010ApJ...724..687L,2012ApJ...745..190L} found that the SFR and molecular mass linearly correlate well with the same dense gas fraction ($f_{\rm DG}$) for both local Galactic clouds and a sample of external galaxies, and proposed $f_{\rm DG}$ as a fundamental parameter of star formation. Here, we used the BGPS 1.1 mm dust emission and our CO observations to investigate the dense gas fraction of the 27~\kms\,cloud since the dust emission mainly originates in the 27~\kms\,cloud (see Fig.~\ref{Fig:dust}). Although the coverage of the BGPS 1.1 mm dust emission is only about 45\%\,of our mapped region (see Fig.~\ref{Fig:clumpsd}a), the dust emission covers most of the prominent $^{13}$CO (1-0) emitting regions, and thus those regions not covered by the BGPS 1.1 mm dust emission are assumed to be free of dust emission. We took the dust emission above the threshold of a flux density of 0.09~Jy~beam$^{-1}$ (3$\sigma$ in the convolved image, indicated by the contour in Fig.~\ref{Fig:dust}) into account. The dust contours above the threshold were rechecked by eye, and the emission with a low fidelity (e.g., the dust contours without $^{13}$CO (1-0) counterparts) or not associated with the 27~\kms\,cloud was manually discarded.

We used Eqs.~(\ref{f.dustcol}) and (\ref{f.mass}) to estimate the total mass of dense gas in the cloud. From Eq.~(\ref{f.dustcol}), we see that the total mass of dense gas is subject to dust temperatures. Previous surveys showed that the dust temperature is about 10--20~K in the Milky Way \citep{2010ApJ...717.1157D,2011A&A...536A..25P}. Toward the dust clumps, the CO excitation temperature is found to be 15--29 K (see Table~\ref{Tab:dustclump}). Here, we assumed an average dust temperature from 10~K to 25~K for the dense gas in the 27~\kms\,cloud. Thus, the total mass of dense gas is estimated to be from 2700~$M_{\odot}$ to 694~$M_{\odot}$.To be consistent with the literature, we used the cloud mass derived from $^{13}$CO to calculate the dense gas fraction ($f_{\rm DG}$), which is estimated to be from 10.4\%\,($T_{\rm d}=10$~K) to 2.7\%\,($T_{\rm d}=25$~K). As a result, the $f_{\rm DG}$ of this cloud is in the range of 2.7\%--10.4\%. This is similar to the 6.5\%\,of the Gemini OB1 molecular cloud \citep{2010ApJ...717.1157D} and is roughly consistent with the statistical result ($7^{+13}_{-5}$)\%\,toward giant molecular clouds of the Galactic Ring Survey \citep{2014ApJ...780..173B} and the 2\%--12\% toward giant molecular filaments \citep{2014A&A...568A..73R}. 



\section{Summary}\label{Sec:sum}
Based on the Milky Way Imaging Scroll Painting (MWISP) project and the archival data, we have performed a study of molecular clouds and star formation toward the region within 216.25\degree $\le$ $l$ $\le$ 218.75\degree and $-$0.75\degree$\le$ $b$ $\le$ 1.25\degree. The main results are as follows.
\begin{itemize}
\item[1.] We decomposed the observed emission in this region into three molecular clouds, the 15~\kms\,cloud, the 27~\kms\,cloud, and the 50~\kms\,cloud, in the surveyed region. The 50~\kms\,cloud was resolved with an angular resolution of $\sim$1\arcmin\,for the first time. The properties of the three clouds were investigated and are tabulated in Table~\ref{Tab:cloudproperty}.
\item[2.]  By investigating the morphologies and velocity structures of the molecular clouds, we found that the 27~\kms\,cloud probably is affected by feedback from the stellar association Mon OB3, and the 50~\kms\,cloud is characterized by three large expanding molecular shells which are resolved for the first time.
\item[3.] This is the first time that the region was mapped in C$^{18}$O (1--0). In the 27~\kms\,cloud, we found seven C$^{18}$O clumps that are likely to form massive stars. Their masses range from 377~$M_{\odot}$ to 1165~$M_{\odot}$, and their virial parameters are all lower than unity. Based on the Bolocam Galactic Plane Survey (BGPS) 1.1 mm dust emission, we found 15 dust clumps, which have masses ranging from 26~$M_{\odot}$ to 1093 $M_{\odot}$. Six of these dust clumps are associated with C$^{18}$O clumps.
\item[4.] Based on 2MASS and WISE data, we identified 56 Class I and 107 Class II young stellar object (YSO) candidates with the color-color diagrams. The YSO distribution showed enhanced star formation around the H{\scriptsize II} region S287 and the intersection of shell B and shell C, probably indicative of triggering.
\item[5.] By counting the YSOs and assuming a universal initial mass function (IMF), the star formation efficiency (SFE) of the 27 \kms\,cloud was estimated to be lower than the SFE in the nearby clouds and the median value of the molecular clouds in the inner galaxy. Assuming the average age of YSOs to be 2~Myr, the star formation rate (SFR) of the 27 \kms\,cloud was found to be 443~$M_{\odot}\;{\rm Myr}^{-1}$, and its SFR surface density of 0.28~$M_{\odot}{\rm yr^{-1}}{\rm kpc^{-2}}$ is similar to those in individual nearby clouds.
\item[6.] From comparing the observed values of the filament S287-main with the models of fragmentation, we suggest that the turbulence controls the large-scale fragmentation in the filament while gravitational fragmentation plays an important role in the formation of YSOs on small scales.
\item[7.] From comparing star-forming molecular gas with non-star-forming molecular gas, we find that star-forming gas tend to have a higher excitation temperature, higher opacity, and higher column density than non-star-forming gas, indicating that star formation occurs in denser gas and star-forming gas is heated by YSOs.
\item[8.] Using the 1.1 mm dust emission to trace dense gas, we derived a dense gas fraction ($f_{\rm DG}$) of 2.7\%--10.4\% for the 27~\kms\,cloud.

\end{itemize}

\section*{ACKNOWLEDGMENTS}\label{sec.ack}
We thank all the staff of Qinghai Station of Purple Mountain Observatory for their assistance with our observations. We would like to thank the anonymous referee for the very helpful and constructive comments that led to improvements in this paper. We are grateful to Christian Henkel for his insightful comments on the manuscript. Y. Gong warmly thanks GuangXing Li for helpful discussions. We acknowledge support by the National Natural Science Foundation of China (NSFC) (grants nos. 11127903, 11233007, and 10973040) and the Strategic Priority Research Program of the Chinese Academy of Sciences (grant no. XDB09000000). M.F. acknowledges the NSFC under grant 11203081. This paper made use of information from the Red MSX Source survey database at www.ast.leeds.ac.uk/RMS which was constructed with support from the Science and Technology Facilities Council of the UK. This publication makes use of data products from the Two Micron All Sky Survey, which is a joint project of the University of Massachusetts and the Infrared Processing and Analysis Center/California Institute of Technology, funded by the National Aeronautics and Space Administration and the National Science Foundation. This research has made use of NASA's Astrophysics Data System.

\bibliographystyle{aa}
\bibliography{references}

\begin{thebibliography}{106}
\expandafter\ifx\csname natexlab\endcsname\relax\def\natexlab#1{#1}\fi

\bibitem[{{Aguirre} {et~al.}(2011){Aguirre}, {Ginsburg}, {Dunham}, {Drosback},
  {Bally}, {Battersby}, {Bradley}, {Cyganowski}, {Dowell}, {Evans}, {Glenn},
  {Harvey}, {Rosolowsky}, {Stringfellow}, {Walawender}, \&
  {Williams}}]{2011ApJS..192....4A}
{Aguirre}, J.~E., {Ginsburg}, A.~G., {Dunham}, M.~K., {et~al.} 2011, \apjs,
  192, 4

\bibitem[{{Andr{\'e}} {et~al.}(2010){Andr{\'e}}, {Men'shchikov}, {Bontemps},
  {K{\"o}nyves}, {Motte}, {Schneider}, {Didelon}, {Minier}, {Saraceno},
  {Ward-Thompson}, {di Francesco}, {White}, {Molinari}, {Testi}, {Abergel},
  {Griffin}, {Henning}, {Royer}, {Mer{\'{\i}}n}, {Vavrek}, {Attard},
  {Arzoumanian}, {Wilson}, {Ade}, {Aussel}, {Baluteau}, {Benedettini},
  {Bernard}, {Blommaert}, {Cambr{\'e}sy}, {Cox}, {di Giorgio}, {Hargrave},
  {Hennemann}, {Huang}, {Kirk}, {Krause}, {Launhardt}, {Leeks}, {Le Pennec},
  {Li}, {Martin}, {Maury}, {Olofsson}, {Omont}, {Peretto}, {Pezzuto}, {Prusti},
  {Roussel}, {Russeil}, {Sauvage}, {Sibthorpe}, {Sicilia-Aguilar}, {Spinoglio},
  {Waelkens}, {Woodcraft}, \& {Zavagno}}]{2010A&A...518L.102A}
{Andr{\'e}}, P., {Men'shchikov}, A., {Bontemps}, S., {et~al.} 2010, \aap, 518,
  L102

\bibitem[{{Arce} {et~al.}(2011){Arce}, {Borkin}, {Goodman}, {Pineda}, \&
  {Beaumont}}]{2011ApJ...742..105A}
{Arce}, H.~G., {Borkin}, M.~A., {Goodman}, A.~A., {Pineda}, J.~E., \&
  {Beaumont}, C.~N. 2011, \apj, 742, 105

\bibitem[{{Arzoumanian} {et~al.}(2011){Arzoumanian}, {Andr{\'e}}, {Didelon},
  {K{\"o}nyves}, {Schneider}, {Men'shchikov}, {Sousbie}, {Zavagno}, {Bontemps},
  {di Francesco}, {Griffin}, {Hennemann}, {Hill}, {Kirk}, {Martin}, {Minier},
  {Molinari}, {Motte}, {Peretto}, {Pezzuto}, {Spinoglio}, {Ward-Thompson},
  {White}, \& {Wilson}}]{2011A&A...529L...6A}
{Arzoumanian}, D., {Andr{\'e}}, P., {Didelon}, P., {et~al.} 2011, \aap, 529, L6

\bibitem[{{Avedisova} \& {Kondratenko}(1984)}]{1984NInfo..56...59A}
{Avedisova}, V.~S. \& {Kondratenko}, G.~I. 1984, Nauchnye Informatsii, 56, 59

\bibitem[{{Battisti} \& {Heyer}(2014)}]{2014ApJ...780..173B}
{Battisti}, A.~J. \& {Heyer}, M.~H. 2014, \apj, 780, 173

\bibitem[{{Beaumont} \& {Williams}(2010)}]{2010ApJ...709..791B}
{Beaumont}, C.~N. \& {Williams}, J.~P. 2010, \apj, 709, 791

\bibitem[{{Bergin} {et~al.}(2002){Bergin}, {Alves}, {Huard}, \&
  {Lada}}]{2002ApJ...570L.101B}
{Bergin}, E.~A., {Alves}, J., {Huard}, T., \& {Lada}, C.~J. 2002, \apjl, 570,
  L101

\bibitem[{{Bergin} \& {Tafalla}(2007)}]{2007ARA&A..45..339B}
{Bergin}, E.~A. \& {Tafalla}, M. 2007, \araa, 45, 339

\bibitem[{{Bertoldi} \& {McKee}(1992)}]{1992ApJ...395..140B}
{Bertoldi}, F. \& {McKee}, C.~F. 1992, \apj, 395, 140

\bibitem[{{Binney} {et~al.}(1991){Binney}, {Gerhard}, {Stark}, {Bally}, \&
  {Uchida}}]{1991MNRAS.252..210B}
{Binney}, J., {Gerhard}, O.~E., {Stark}, A.~A., {Bally}, J., \& {Uchida}, K.~I.
  1991, \mnras, 252, 210

\bibitem[{{Blitz} {et~al.}(1982){Blitz}, {Fich}, \&
  {Stark}}]{1982ApJS...49..183B}
{Blitz}, L., {Fich}, M., \& {Stark}, A.~A. 1982, \apjs, 49, 183

\bibitem[{{Bohlin} {et~al.}(1978){Bohlin}, {Savage}, \&
  {Drake}}]{1978ApJ...224..132B}
{Bohlin}, R.~C., {Savage}, B.~D., \& {Drake}, J.~F. 1978, \apj, 224, 132

\bibitem[{{Bonnor}(1956)}]{1956MNRAS.116..351B}
{Bonnor}, W.~B. 1956, \mnras, 116, 351

\bibitem[{{Bressert} {et~al.}(2012){Bressert}, {Ginsburg}, {Bally},
  {Battersby}, {Longmore}, \& {Testi}}]{2012ApJ...758L..28B}
{Bressert}, E., {Ginsburg}, A., {Bally}, J., {et~al.} 2012, \apjl, 758, L28

\bibitem[{{Buckner} \& {Froebrich}(2013)}]{2013MNRAS.436.1465B}
{Buckner}, A.~S.~M. \& {Froebrich}, D. 2013, \mnras, 436, 1465

\bibitem[{{Chabrier}(2003)}]{2003PASP..115..763C}
{Chabrier}, G. 2003, \pasp, 115, 763

\bibitem[{{Chandrasekhar} \& {Fermi}(1953)}]{1953ApJ...118..116C}
{Chandrasekhar}, S. \& {Fermi}, E. 1953, \apj, 118, 116

\bibitem[{{Dame} {et~al.}(2001){Dame}, {Hartmann}, \&
  {Thaddeus}}]{2001ApJ...547..792D}
{Dame}, T.~M., {Hartmann}, D., \& {Thaddeus}, P. 2001, \apj, 547, 792

\bibitem[{{Dunham} {et~al.}(2010){Dunham}, {Rosolowsky}, {Evans}, {Cyganowski},
  {Aguirre}, {Bally}, {Battersby}, {Bradley}, {Dowell}, {Drosback}, {Ginsburg},
  {Glenn}, {Harvey}, {Merello}, {Schlingman}, {Shirley}, {Stringfellow},
  {Walawender}, \& {Williams}}]{2010ApJ...717.1157D}
{Dunham}, M.~K., {Rosolowsky}, E., {Evans}, II, N.~J., {et~al.} 2010, \apj,
  717, 1157

\bibitem[{{Ebert}(1955)}]{1955ZA.....37..217E}
{Ebert}, R. 1955, \zap, 37, 217

\bibitem[{{Eden} {et~al.}(2013){Eden}, {Moore}, {Morgan}, {Thompson}, \&
  {Urquhart}}]{2013MNRAS.431.1587E}
{Eden}, D.~J., {Moore}, T.~J.~T., {Morgan}, L.~K., {Thompson}, M.~A., \&
  {Urquhart}, J.~S. 2013, \mnras, 431, 1587

\bibitem[{{Evans} {et~al.}(2009){Evans}, {Dunham}, {J{\o}rgensen}, {Enoch},
  {Mer{\'{\i}}n}, {van Dishoeck}, {Alcal{\'a}}, {Myers}, {Stapelfeldt},
  {Huard}, {Allen}, {Harvey}, {van Kempen}, {Blake}, {Koerner}, {Mundy},
  {Padgett}, \& {Sargent}}]{2009ApJS..181..321E}
{Evans}, II, N.~J., {Dunham}, M.~M., {J{\o}rgensen}, J.~K., {et~al.} 2009,
  \apjs, 181, 321

\bibitem[{{Fang} {et~al.}(2013){Fang}, {Kim}, {van Boekel}, {Sicilia-Aguilar},
  {Henning}, \& {Flaherty}}]{2013ApJS..207....5F}
{Fang}, M., {Kim}, J.~S., {van Boekel}, R., {et~al.} 2013, \apjs, 207, 5

\bibitem[{{Flaherty} {et~al.}(2007){Flaherty}, {Pipher}, {Megeath}, {Winston},
  {Gutermuth}, {Muzerolle}, {Allen}, \& {Fazio}}]{2007ApJ...663.1069F}
{Flaherty}, K.~M., {Pipher}, J.~L., {Megeath}, S.~T., {et~al.} 2007, \apj, 663,
  1069

\bibitem[{{Frerking} {et~al.}(1982){Frerking}, {Langer}, \&
  {Wilson}}]{1982ApJ...262..590F}
{Frerking}, M.~A., {Langer}, W.~D., \& {Wilson}, R.~W. 1982, \apj, 262, 590

\bibitem[{{Fukui}(1989)}]{1989ESOC...33...95F}
{Fukui}, Y. 1989, in European Southern Observatory Conference and Workshop
  Proceedings, Vol.~33, European Southern Observatory Conference and Workshop
  Proceedings, ed. B.~{Reipurth}, 95--117

\bibitem[{{Gao} \& {Solomon}(2004)}]{2004ApJ...606..271G}
{Gao}, Y. \& {Solomon}, P.~M. 2004, \apj, 606, 271

\bibitem[{{Giannetti} {et~al.}(2014){Giannetti}, {Wyrowski}, {Brand},
  {Csengeri}, {Fontani}, {Walmsley}, {Nguyen Luong}, {Beuther}, {Schuller},
  {G{\"u}sten}, \& {Menten}}]{2014A&A...570A..65G}
{Giannetti}, A., {Wyrowski}, F., {Brand}, J., {et~al.} 2014, \aap, 570, A65

\bibitem[{{Ginsburg} {et~al.}(2013){Ginsburg}, {Glenn}, {Rosolowsky},
  {Ellsworth-Bowers}, {Battersby}, {Dunham}, {Merello}, {Shirley}, {Bally},
  {Evans}, {Stringfellow}, \& {Aguirre}}]{2013ApJS..208...14G}
{Ginsburg}, A., {Glenn}, J., {Rosolowsky}, E., {et~al.} 2013, \apjs, 208, 14

\bibitem[{{Green}(2014)}]{2014BASI...42...47G}
{Green}, D.~A. 2014, Bulletin of the Astronomical Society of India, 42, 47

\bibitem[{{Hartmann}(2002)}]{2002ApJ...578..914H}
{Hartmann}, L. 2002, \apj, 578, 914

\bibitem[{{Hartmann}(2009)}]{2009apsf.book.....H}
{Hartmann}, L. 2009, {Accretion Processes in Star Formation: Second Edition}
  (Cambridge University Press)

\bibitem[{{Heiderman} {et~al.}(2010){Heiderman}, {Evans}, {Allen}, {Huard}, \&
  {Heyer}}]{2010ApJ...723.1019H}
{Heiderman}, A., {Evans}, II, N.~J., {Allen}, L.~E., {Huard}, T., \& {Heyer},
  M. 2010, \apj, 723, 1019

\bibitem[{{Henning} {et~al.}(2010){Henning}, {Linz}, {Krause}, {Ragan},
  {Beuther}, {Launhardt}, {Nielbock}, \& {Vasyunina}}]{2010A&A...518L..95H}
{Henning}, T., {Linz}, H., {Krause}, O., {et~al.} 2010, \aap, 518, L95

\bibitem[{{Heyer} \& {Dame}(2015)}]{2015ARA&A..53..583H}
{Heyer}, M. \& {Dame}, T.~M. 2015, \araa, 53, 583

\bibitem[{{Heyer} {et~al.}(2009){Heyer}, {Krawczyk}, {Duval}, \&
  {Jackson}}]{2009ApJ...699.1092H}
{Heyer}, M., {Krawczyk}, C., {Duval}, J., \& {Jackson}, J.~M. 2009, \apj, 699,
  1092

\bibitem[{{Hildebrand}(1983)}]{1983QJRAS..24..267H}
{Hildebrand}, R.~H. 1983, \qjras, 24, 267

\bibitem[{{Howard} {et~al.}(1998){Howard}, {Pipher}, \&
  {Forrest}}]{1998ApJ...509..749H}
{Howard}, E.~M., {Pipher}, J.~L., \& {Forrest}, W.~J. 1998, \apj, 509, 749

\bibitem[{{Inutsuka} \& {Miyama}(1992)}]{1992ApJ...388..392I}
{Inutsuka}, S.-I. \& {Miyama}, S.~M. 1992, \apj, 388, 392

\bibitem[{{Inutsuka} \& {Miyama}(1997)}]{1997ApJ...480..681I}
{Inutsuka}, S.-i. \& {Miyama}, S.~M. 1997, \apj, 480, 681

\bibitem[{{Jackson} {et~al.}(2010){Jackson}, {Finn}, {Chambers}, {Rathborne},
  \& {Simon}}]{2010ApJ...719L.185J}
{Jackson}, J.~M., {Finn}, S.~C., {Chambers}, E.~T., {Rathborne}, J.~M., \&
  {Simon}, R. 2010, \apjl, 719, L185

\bibitem[{{J{\o}rgensen} {et~al.}(2002){J{\o}rgensen}, {Sch{\"o}ier}, \& {van
  Dishoeck}}]{2002A&A...389..908J}
{J{\o}rgensen}, J.~K., {Sch{\"o}ier}, F.~L., \& {van Dishoeck}, E.~F. 2002,
  \aap, 389, 908

\bibitem[{{Kauffmann} {et~al.}(2008){Kauffmann}, {Bertoldi}, {Bourke}, {Evans},
  \& {Lee}}]{2008A&A...487..993K}
{Kauffmann}, J., {Bertoldi}, F., {Bourke}, T.~L., {Evans}, II, N.~J., \& {Lee},
  C.~W. 2008, \aap, 487, 993

\bibitem[{{Kauffmann} {et~al.}(2013){Kauffmann}, {Pillai}, \&
  {Goldsmith}}]{2013ApJ...779..185K}
{Kauffmann}, J., {Pillai}, T., \& {Goldsmith}, P.~F. 2013, \apj, 779, 185

\bibitem[{{Kauffmann} {et~al.}(2010){Kauffmann}, {Pillai}, {Shetty}, {Myers},
  \& {Goodman}}]{2010ApJ...716..433K}
{Kauffmann}, J., {Pillai}, T., {Shetty}, R., {Myers}, P.~C., \& {Goodman},
  A.~A. 2010, \apj, 716, 433

\bibitem[{{Kennicutt} \& {Evans}(2012)}]{2012ARA&A..50..531K}
{Kennicutt}, R.~C. \& {Evans}, N.~J. 2012, \araa, 50, 531

\bibitem[{{Kennicutt}(1998)}]{1998ApJ...498..541K}
{Kennicutt}, Jr., R.~C. 1998, \apj, 498, 541

\bibitem[{{Kharchenko} {et~al.}(2005){Kharchenko}, {Piskunov}, {R{\"o}ser},
  {Schilbach}, \& {Scholz}}]{2005A&A...438.1163K}
{Kharchenko}, N.~V., {Piskunov}, A.~E., {R{\"o}ser}, S., {Schilbach}, E., \&
  {Scholz}, R.-D. 2005, \aap, 438, 1163

\bibitem[{{Kim} {et~al.}(2004){Kim}, {Kawamura}, {Yonekura}, \&
  {Fukui}}]{2004PASJ...56..313K}
{Kim}, B.~G., {Kawamura}, A., {Yonekura}, Y., \& {Fukui}, Y. 2004, \pasj, 56,
  313

\bibitem[{{Koenig} {et~al.}(2012){Koenig}, {Leisawitz}, {Benford}, {Rebull},
  {Padgett}, \& {Assef}}]{2012ApJ...744..130K}
{Koenig}, X.~P., {Leisawitz}, D.~T., {Benford}, D.~J., {et~al.} 2012, \apj,
  744, 130

\bibitem[{{Krumholz} \& {McKee}(2008)}]{2008Natur.451.1082K}
{Krumholz}, M.~R. \& {McKee}, C.~F. 2008, \nat, 451, 1082

\bibitem[{{Lada} {et~al.}(2012){Lada}, {Forbrich}, {Lombardi}, \&
  {Alves}}]{2012ApJ...745..190L}
{Lada}, C.~J., {Forbrich}, J., {Lombardi}, M., \& {Alves}, J.~F. 2012, \apj,
  745, 190

\bibitem[{{Lada} {et~al.}(2010){Lada}, {Lombardi}, \&
  {Alves}}]{2010ApJ...724..687L}
{Lada}, C.~J., {Lombardi}, M., \& {Alves}, J.~F. 2010, \apj, 724, 687

\bibitem[{{Lee}(1994)}]{1994JKAS...27..147L}
{Lee}, Y. 1994, Journal of Korean Astronomical Society, 27, 147

\bibitem[{{Li} {et~al.}(2013){Li}, {Wyrowski}, {Menten}, \&
  {Belloche}}]{2013A&A...559A..34L}
{Li}, G.-X., {Wyrowski}, F., {Menten}, K., \& {Belloche}, A. 2013, \aap, 559,
  A34

\bibitem[{{Li} {et~al.}(2015){Li}, {Yuen}, {Otto}, {Leung}, {Sridharan},
  {Zhang}, {Liu}, {Tang}, \& {Qiu}}]{2015Natur.520..518L}
{Li}, H.-B., {Yuen}, K.~H., {Otto}, F., {et~al.} 2015, \nat, 520, 518

\bibitem[{{Luhman} {et~al.}(2010){Luhman}, {Allen}, {Espaillat}, {Hartmann}, \&
  {Calvet}}]{2010ApJS..186..111L}
{Luhman}, K.~L., {Allen}, P.~R., {Espaillat}, C., {Hartmann}, L., \& {Calvet},
  N. 2010, \apjs, 186, 111

\bibitem[{{Lumsden} {et~al.}(2013){Lumsden}, {Hoare}, {Urquhart}, {Oudmaijer},
  {Davies}, {Mottram}, {Cooper}, \& {Moore}}]{2013ApJS..208...11L}
{Lumsden}, S.~L., {Hoare}, M.~G., {Urquhart}, J.~S., {et~al.} 2013, \apjs, 208,
  11

\bibitem[{{Maddalena} \& {Thaddeus}(1985)}]{1985ApJ...294..231M}
{Maddalena}, R.~J. \& {Thaddeus}, P. 1985, \apj, 294, 231

\bibitem[{{Megeath} {et~al.}(2012){Megeath}, {Gutermuth}, {Muzerolle},
  {Kryukova}, {Flaherty}, {Hora}, {Allen}, {Hartmann}, {Myers}, {Pipher},
  {Stauffer}, {Young}, \& {Fazio}}]{2012AJ....144..192M}
{Megeath}, S.~T., {Gutermuth}, R., {Muzerolle}, J., {et~al.} 2012, \aj, 144,
  192

\bibitem[{{Mel'Nik} \& {Dambis}(2009)}]{2009MNRAS.400..518M}
{Mel'Nik}, A.~M. \& {Dambis}, A.~K. 2009, \mnras, 400, 518

\bibitem[{{Mel'Nik} \& {Efremov}(1995)}]{1995AstL...21...10M}
{Mel'Nik}, A.~M. \& {Efremov}, Y.~N. 1995, Astronomy Letters, 21, 10

\bibitem[{{Milam} {et~al.}(2005){Milam}, {Savage}, {Brewster}, {Ziurys}, \&
  {Wyckoff}}]{2005ApJ...634.1126M}
{Milam}, S.~N., {Savage}, C., {Brewster}, M.~A., {Ziurys}, L.~M., \& {Wyckoff},
  S. 2005, \apj, 634, 1126

\bibitem[{{Myers} {et~al.}(1986){Myers}, {Dame}, {Thaddeus}, {Cohen},
  {Silverberg}, {Dwek}, \& {Hauser}}]{1986ApJ...301..398M}
{Myers}, P.~C., {Dame}, T.~M., {Thaddeus}, P., {et~al.} 1986, \apj, 301, 398

\bibitem[{{Neckel} \& {Staude}(1984)}]{1984A&A...131..200N}
{Neckel}, T. \& {Staude}, H.~J. 1984, \aap, 131, 200

\bibitem[{{Ossenkopf} \& {Henning}(1994)}]{1994A&A...291..943O}
{Ossenkopf}, V. \& {Henning}, T. 1994, \aap, 291, 943

\bibitem[{{Ostriker}(1964)}]{1964ApJ...140.1056O}
{Ostriker}, J. 1964, \apj, 140, 1056

\bibitem[{{Planck Collaboration} {et~al.}(2011){Planck Collaboration},
  {Abergel}, {Ade}, {Aghanim}, {Arnaud}, {Ashdown}, {Aumont}, {Baccigalupi},
  {Balbi}, {Banday}, {Barreiro}, {Bartlett}, {Battaner}, {Benabed},
  {Beno{\^i}t}, {Bernard}, {Bersanelli}, {Bhatia}, {Bock}, {Bonaldi}, {Bond},
  {Borrill}, {Bouchet}, {Boulanger}, {Bucher}, {Burigana}, {Cabella},
  {Cardoso}, {Catalano}, {Cay{\'o}n}, {Challinor}, {Chamballu}, {Chiang},
  {Chiang}, {Christensen}, {Clements}, {Colombi}, {Couchot}, {Coulais},
  {Crill}, {Cuttaia}, {Danese}, {Davies}, {Davis}, {de Bernardis}, {de
  Gasperis}, {de Rosa}, {de Zotti}, {Delabrouille}, {Delouis}, {D{\'e}sert},
  {Dickinson}, {Dobashi}, {Donzelli}, {Dor{\'e}}, {D{\"o}rl}, {Douspis},
  {Dupac}, {Efstathiou}, {En{\ss}lin}, {Eriksen}, {Finelli}, {Forni},
  {Frailis}, {Franceschi}, {Galeotta}, {Ganga}, {Giard}, {Giardino},
  {Giraud-H{\'e}raud}, {Gonz{\'a}lez-Nuevo}, {G{\'o}rski}, {Gratton},
  {Gregorio}, {Gruppuso}, {Guillet}, {Hansen}, {Harrison},
  {Henrot-Versill{\'e}}, {Herranz}, {Hildebrandt}, {Hivon}, {Hobson}, {Holmes},
  {Hovest}, {Hoyland}, {Huffenberger}, {Jaffe}, {Jones}, {Jones}, {Juvela},
  {Keih{\"a}nen}, {Keskitalo}, {Kisner}, {Kneissl}, {Knox}, {Kurki-Suonio},
  {Lagache}, {Lamarre}, {Lasenby}, {Laureijs}, {Lawrence}, {Leach}, {Leonardi},
  {Leroy}, {Linden-V{\o}rnle}, {L{\'o}pez-Caniego}, {Lubin},
  {Mac{\'{\i}}as-P{\'e}rez}, {MacTavish}, {Maffei}, {Mandolesi}, {Mann},
  {Maris}, {Marshall}, {Martin}, {Mart{\'{\i}}nez-Gonz{\'a}lez}, {Masi},
  {Matarrese}, {Matthai}, {Mazzotta}, {McGehee}, {Meinhold}, {Melchiorri},
  {Mendes}, {Mennella}, {Mitra}, {Miville-Desch{\^e}nes}, {Moneti}, {Montier},
  {Morgante}, {Mortlock}, {Munshi}, {Murphy}, {Naselsky}, {Natoli},
  {Netterfield}, {N{\o}rgaard-Nielsen}, {Noviello}, {Novikov}, {Novikov},
  {Osborne}, {Pajot}, {Paladini}, {Pasian}, {Patanchon}, {Perdereau},
  {Perotto}, {Perrotta}, {Piacentini}, {Piat}, {Plaszczynski}, {Pointecouteau},
  {Polenta}, {Ponthieu}, {Poutanen}, {Pr{\'e}zeau}, {Prunet}, {Puget}, {Reach},
  {Rebolo}, {Reinecke}, {Renault}, {Ricciardi}, {Riller}, {Ristorcelli},
  {Rocha}, {Rosset}, {Rubi{\~n}o-Mart{\'{\i}}n}, {Rusholme}, {Sandri},
  {Santos}, {Savini}, {Scott}, {Seiffert}, {Shellard}, {Smoot}, {Starck},
  {Stivoli}, {Stolyarov}, {Sudiwala}, {Sygnet}, {Tauber}, {Terenzi},
  {Toffolatti}, {Tomasi}, {Torre}, {Tristram}, {Tuovinen}, {Umana},
  {Valenziano}, {Verstraete}, {Vielva}, {Villa}, {Vittorio}, {Wade}, {Wandelt},
  {Yvon}, {Zacchei}, \& {Zonca}}]{2011A&A...536A..25P}
{Planck Collaboration}, {Abergel}, A., {Ade}, P.~A.~R., {et~al.} 2011, \aap,
  536, A25

\bibitem[{{Ragan} {et~al.}(2014){Ragan}, {Henning}, {Tackenberg}, {Beuther},
  {Johnston}, {Kainulainen}, \& {Linz}}]{2014A&A...568A..73R}
{Ragan}, S.~E., {Henning}, T., {Tackenberg}, J., {et~al.} 2014, \aap, 568, A73

\bibitem[{{Rathborne} {et~al.}(2009){Rathborne}, {Johnson}, {Jackson}, {Shah},
  \& {Simon}}]{2009ApJS..182..131R}
{Rathborne}, J.~M., {Johnson}, A.~M., {Jackson}, J.~M., {Shah}, R.~Y., \&
  {Simon}, R. 2009, \apjs, 182, 131

\bibitem[{{Reid} {et~al.}(2014){Reid}, {Menten}, {Brunthaler}, {Zheng}, {Dame},
  {Xu}, {Wu}, {Zhang}, {Sanna}, {Sato}, {Hachisuka}, {Choi}, {Immer},
  {Moscadelli}, {Rygl}, \& {Bartkiewicz}}]{2014ApJ...783..130R}
{Reid}, M.~J., {Menten}, K.~M., {Brunthaler}, A., {et~al.} 2014, \apj, 783, 130

\bibitem[{{Rosolowsky} {et~al.}(2010){Rosolowsky}, {Dunham}, {Ginsburg},
  {Bradley}, {Aguirre}, {Bally}, {Battersby}, {Cyganowski}, {Dowell},
  {Drosback}, {Evans}, {Glenn}, {Harvey}, {Stringfellow}, {Walawender}, \&
  {Williams}}]{2010ApJS..188..123R}
{Rosolowsky}, E., {Dunham}, M.~K., {Ginsburg}, A., {et~al.} 2010, \apjs, 188,
  123

\bibitem[{{Samal} {et~al.}(2015){Samal}, {Ojha}, {Jose}, {Zavagno},
  {Takahashi}, {Neichel}, {Kim}, {Chauhan}, {Pandey}, {Zinchenko}, {Tamura}, \&
  {Ghosh}}]{2015A&A...581A...5S}
{Samal}, M.~R., {Ojha}, D.~K., {Jose}, J., {et~al.} 2015, \aap, 581, A5

\bibitem[{{Schinnerer} {et~al.}(2013){Schinnerer}, {Meidt}, {Pety}, {Hughes},
  {Colombo}, {Garc{\'{\i}}a-Burillo}, {Schuster}, {Dumas}, {Dobbs}, {Leroy},
  {Kramer}, {Thompson}, \& {Regan}}]{2013ApJ...779...42S}
{Schinnerer}, E., {Meidt}, S.~E., {Pety}, J., {et~al.} 2013, \apj, 779, 42

\bibitem[{{Shan} {et~al.}(2012){Shan}, {Yang}, {Shi}, {Yao}, {Zuo}, {Lin},
  {Chen}, {Zhang}, {Duan}, {Cao}, {Li}, {Li}, {Liu}, \& {Zhong}}]{pmo2}
{Shan}, W.~L., {Yang}, J., {Shi}, S.~C., {et~al.} 2012, IEEE Transactions on
  Terahertz Science and Technology, 2, 593

\bibitem[{{Sharpless}(1959)}]{1959ApJS....4..257S}
{Sharpless}, S. 1959, \apjs, 4, 257

\bibitem[{{Shu} {et~al.}(1987){Shu}, {Adams}, \&
  {Lizano}}]{1987ARA&A..25...23S}
{Shu}, F.~H., {Adams}, F.~C., \& {Lizano}, S. 1987, \araa, 25, 23

\bibitem[{{Siess} {et~al.}(2000){Siess}, {Dufour}, \&
  {Forestini}}]{2000A&A...358..593S}
{Siess}, L., {Dufour}, E., \& {Forestini}, M. 2000, \aap, 358, 593

\bibitem[{{Skrutskie} {et~al.}(2006){Skrutskie}, {Cutri}, {Stiening},
  {Weinberg}, {Schneider}, {Carpenter}, {Beichman}, {Capps}, {Chester},
  {Elias}, {Huchra}, {Liebert}, {Lonsdale}, {Monet}, {Price}, {Seitzer},
  {Jarrett}, {Kirkpatrick}, {Gizis}, {Howard}, {Evans}, {Fowler}, {Fullmer},
  {Hurt}, {Light}, {Kopan}, {Marsh}, {McCallon}, {Tam}, {Van Dyk}, \&
  {Wheelock}}]{2006AJ....131.1163S}
{Skrutskie}, M.~F., {Cutri}, R.~M., {Stiening}, R., {et~al.} 2006, \aj, 131,
  1163

\bibitem[{{Solomon} {et~al.}(1987){Solomon}, {Rivolo}, {Barrett}, \&
  {Yahil}}]{1987ApJ...319..730S}
{Solomon}, P.~M., {Rivolo}, A.~R., {Barrett}, J., \& {Yahil}, A. 1987, \apj,
  319, 730

\bibitem[{{Su} {et~al.}(2014){Su}, {Fang}, {Yang}, {Zhou}, \&
  {Chen}}]{2014ApJ...788..122S}
{Su}, Y., {Fang}, M., {Yang}, J., {Zhou}, P., \& {Chen}, Y. 2014, \apj, 788,
  122

\bibitem[{{Su} {et~al.}(2015){Su}, {Zhang}, {Shao}, \&
  {Yang}}]{2015arXiv150807898S}
{Su}, Y., {Zhang}, S., {Shao}, X., \& {Yang}, J. 2015, ArXiv e-prints

\bibitem[{{Sun} {et~al.}(2015){Sun}, {Xu}, {Yang}, {Li}, {Du}, {Zhang}, \&
  {Zhou}}]{2015ApJ...798L..27S}
{Sun}, Y., {Xu}, Y., {Yang}, J., {et~al.} 2015, \apjl, 798, L27

\bibitem[{{Sz{\H u}cs} {et~al.}(2014){Sz{\H u}cs}, {Glover}, \&
  {Klessen}}]{2014MNRAS.445.4055S}
{Sz{\H u}cs}, L., {Glover}, S.~C.~O., \& {Klessen}, R.~S. 2014, \mnras, 445,
  4055

\bibitem[{{Tielens}(2005)}]{2005pcim.book.....T}
{Tielens}, A.~G.~G.~M. 2005, {The Physics and Chemistry of the Interstellar
  Medium}

\bibitem[{{Ulich} \& {Haas}(1976)}]{1976ApJS...30..247U}
{Ulich}, B.~L. \& {Haas}, R.~W. 1976, \apjs, 30, 247

\bibitem[{{Ungerechts} {et~al.}(2000){Ungerechts}, {Umbanhowar}, \&
  {Thaddeus}}]{2000ApJ...537..221U}
{Ungerechts}, H., {Umbanhowar}, P., \& {Thaddeus}, P. 2000, \apj, 537, 221

\bibitem[{{Urquhart} {et~al.}(2008{\natexlab{a}}){Urquhart}, {Busfield},
  {Hoare}, {Lumsden}, {Oudmaijer}, {Moore}, {Gibb}, {Purcell}, {Burton},
  {Mar{\'e}chal}, {Jiang}, \& {Wang}}]{2008A&A...487..253U}
{Urquhart}, J.~S., {Busfield}, A.~L., {Hoare}, M.~G., {et~al.}
  2008{\natexlab{a}}, \aap, 487, 253

\bibitem[{{Urquhart} {et~al.}(2008{\natexlab{b}}){Urquhart}, {Hoare},
  {Lumsden}, {Oudmaijer}, \& {Moore}}]{2008ASPC..387..381U}
{Urquhart}, J.~S., {Hoare}, M.~G., {Lumsden}, S.~L., {Oudmaijer}, R.~D., \&
  {Moore}, T.~J.~T. 2008{\natexlab{b}}, in Astronomical Society of the Pacific
  Conference Series, Vol. 387, Massive Star Formation: Observations Confront
  Theory, ed. H.~{Beuther}, H.~{Linz}, \& T.~{Henning}, 381

\bibitem[{{Urquhart} {et~al.}(2013){Urquhart}, {Moore}, {Schuller}, {Wyrowski},
  {Menten}, {Thompson}, {Csengeri}, {Walmsley}, {Bronfman}, \&
  {K{\"o}nig}}]{2013MNRAS.431.1752U}
{Urquhart}, J.~S., {Moore}, T.~J.~T., {Schuller}, F., {et~al.} 2013, \mnras,
  431, 1752

\bibitem[{{van Dishoeck} \& {Black}(1988)}]{1988ApJ...334..771V}
{van Dishoeck}, E.~F. \& {Black}, J.~H. 1988, \apj, 334, 771

\bibitem[{{Vieira} {et~al.}(2011){Vieira}, {Gregorio-Hetem}, {Hetem},
  {Stasi{\'n}ska}, \& {Szczerba}}]{2011A&A...526A..24V}
{Vieira}, R.~G., {Gregorio-Hetem}, J., {Hetem}, A., {Stasi{\'n}ska}, G., \&
  {Szczerba}, R. 2011, \aap, 526, A24

\bibitem[{{Wang} {et~al.}(2015){Wang}, {Testi}, {Ginsburg}, {Walmsley},
  {Molinari}, \& {Schisano}}]{2015MNRAS.450.4043W}
{Wang}, K., {Testi}, L., {Ginsburg}, A., {et~al.} 2015, \mnras, 450, 4043

\bibitem[{{Wang} {et~al.}(2014){Wang}, {Zhang}, {Testi}, {van der Tak}, {Wu},
  {Zhang}, {Pillai}, {Wyrowski}, {Carey}, {Ragan}, \&
  {Henning}}]{2014MNRAS.439.3275W}
{Wang}, K., {Zhang}, Q., {Testi}, L., {et~al.} 2014, \mnras, 439, 3275

\bibitem[{{Watson} {et~al.}(1976){Watson}, {Anicich}, \&
  {Huntress}}]{1976ApJ...205L.165W}
{Watson}, W.~D., {Anicich}, V.~G., \& {Huntress}, Jr., W.~T. 1976, \apjl, 205,
  L165

\bibitem[{{Wienen} {et~al.}(2012){Wienen}, {Wyrowski}, {Schuller}, {Menten},
  {Walmsley}, {Bronfman}, \& {Motte}}]{2012A&A...544A.146W}
{Wienen}, M., {Wyrowski}, F., {Schuller}, F., {et~al.} 2012, \aap, 544, A146

\bibitem[{{Williams} \& {Maddalena}(1996)}]{1996ApJ...464..247W}
{Williams}, J.~P. \& {Maddalena}, R.~J. 1996, \apj, 464, 247

\bibitem[{{Wilson} \& {Rood}(1994)}]{1994ARA&A..32..191W}
{Wilson}, T.~L. \& {Rood}, R. 1994, \araa, 32, 191

\bibitem[{{Wright} {et~al.}(2010){Wright}, {Eisenhardt}, {Mainzer}, {Ressler},
  {Cutri}, {Jarrett}, {Kirkpatrick}, {Padgett}, {McMillan}, {Skrutskie},
  {Stanford}, {Cohen}, {Walker}, {Mather}, {Leisawitz}, {Gautier}, {McLean},
  {Benford}, {Lonsdale}, {Blain}, {Mendez}, {Irace}, {Duval}, {Liu}, {Royer},
  {Heinrichsen}, {Howard}, {Shannon}, {Kendall}, {Walsh}, {Larsen}, {Cardon},
  {Schick}, {Schwalm}, {Abid}, {Fabinsky}, {Naes}, \&
  {Tsai}}]{2010AJ....140.1868W}
{Wright}, E.~L., {Eisenhardt}, P.~R.~M., {Mainzer}, A.~K., {et~al.} 2010, \aj,
  140, 1868

\bibitem[{{Wu} {et~al.}(2005){Wu}, {Evans}, {Gao}, {Solomon}, {Shirley}, \&
  {Vanden Bout}}]{2005ApJ...635L.173W}
{Wu}, J., {Evans}, II, N.~J., {Gao}, Y., {et~al.} 2005, \apjl, 635, L173

\bibitem[{{Yang} {et~al.}(2002){Yang}, {Jiang}, {Wang}, {Ju}, \&
  {Wang}}]{2002ApJS..141..157Y}
{Yang}, J., {Jiang}, Z., {Wang}, M., {Ju}, B., \& {Wang}, H. 2002, \apjs, 141,
  157

\bibitem[{{Zeng} {et~al.}(2006){Zeng}, {Mao}, \& {Pei}}]{2006bookZ}
{Zeng}, Q., {Mao}, R.~Q., \& {Pei}, C.~C. 2006, {Microwave spectrum of
  astrophysics diagnosis (Chinese edition)}

\bibitem[{{Zhang} {et~al.}(2014){Zhang}, {Xu}, \& {Yang}}]{2014AJ....147...46Z}
{Zhang}, S., {Xu}, Y., \& {Yang}, J. 2014, \aj, 147, 46

\bibitem[{{Zhang} {et~al.}(2011){Zhang}, {Yang}, {Xu}, {Pandian}, {Menten}, \&
  {Henkel}}]{2011ApJS..193...10Z}
{Zhang}, S.~B., {Yang}, J., {Xu}, Y., {et~al.} 2011, \apjs, 193, 10

\bibitem[{{Zhou} {et~al.}(2014){Zhou}, {Yang}, {Fang}, \&
  {Su}}]{2014ApJ...791..109Z}
{Zhou}, X., {Yang}, J., {Fang}, M., \& {Su}, Y. 2014, \apj, 791, 109

\end{thebibliography}
\clearpage


\clearpage

\begin{appendix}
\section{LTE mass, virial mass, and the virial parameter}\label{s.mass}
From the H$_{2}$ column densities under the LTE condition, the total mass of an object can be derived from the H$_{2}$ column densities by 
\begin{equation}\label{f.mass}
 M = \mu m_{\rm H} D^{2}  \int N(\rm{H}_{2}){\rm d}\Omega\;,
\end{equation}
where $\mu$ is the mean molecular weight per hydrogen molecule which is assumed to be 2.8, $m_{\rm H}$ is the mass of the atomic hydrogen, $D$ is the distance to the object, {\rm d}$\Omega$ is the solid angle element and $N({\rm H_{2}})$ is the column density of the molecular hydrogen.

According to the definition by \citet{1992ApJ...395..140B}, the virial parameter represents the ratio between the kinetic and half gravitational potential energy, and can be estimated by
\begin{equation}\label{f.virial}
\alpha_{\rm vir}=\frac{5\delta_{\rm v}^{2}R}{GM} \;,
\end{equation}
 where $\delta_{\rm v}$ is the one-dimensional velocity dispersion, $R$ is the radius, $G$ is the gravitational constant ($\sim\frac{1}{232} M_{\odot}^{-1}$\,~(\kms)$^{2}$~pc), and $M$ is the total mass. In analyses, the virial parameters are used to evaluate whether objects are subcritical or supercritical. According to the discussion about non-magnetized spheres \citep{2013ApJ...779..185K}, the critical virial parameter is found to be 2, which is based on the isothermal hydrostatic equilibrium spheres model \citep{1955ZA.....37..217E,1956MNRAS.116..351B}. However, the magnetic field is not negligible in molecular clouds \citep[e.g.,][]{2015Natur.520..518L}, which will result in a lower critical virial parameter. Thus $\alpha_{\rm vir}=2$ is the upper limit of the critical virial parameter \citep{2013ApJ...779..185K}. To simplify Eq.~(\ref{f.virial}), the virial parameter can be written as $\alpha_{\rm vir}=\frac{M_{\rm vir}}{M}$ \citep[e.g.,][]{2013ApJ...779..185K}. Thus, the virial mass can be expressed as 
\begin{equation}\label{f.vmass}
M_{\rm vir}=\frac{5\delta_{\rm v}^{2}R}{G}\approx209(\frac{R}{\rm pc}) (\frac{\Delta v}{\rm{km~s^{-1}}})^{2} M_{\odot}\;,
\end{equation}
where $\Delta v$ is the FWHM line width, corresponding to $\sqrt{8{\rm ln 2}}\delta_{\rm v}$.

\section{Excitation temperature, opacity, and column density toward the surveyed region}\label{s.statistics}
\begin{figure*}[!htbp]
\centering
\includegraphics[width = 0.3 \textwidth]{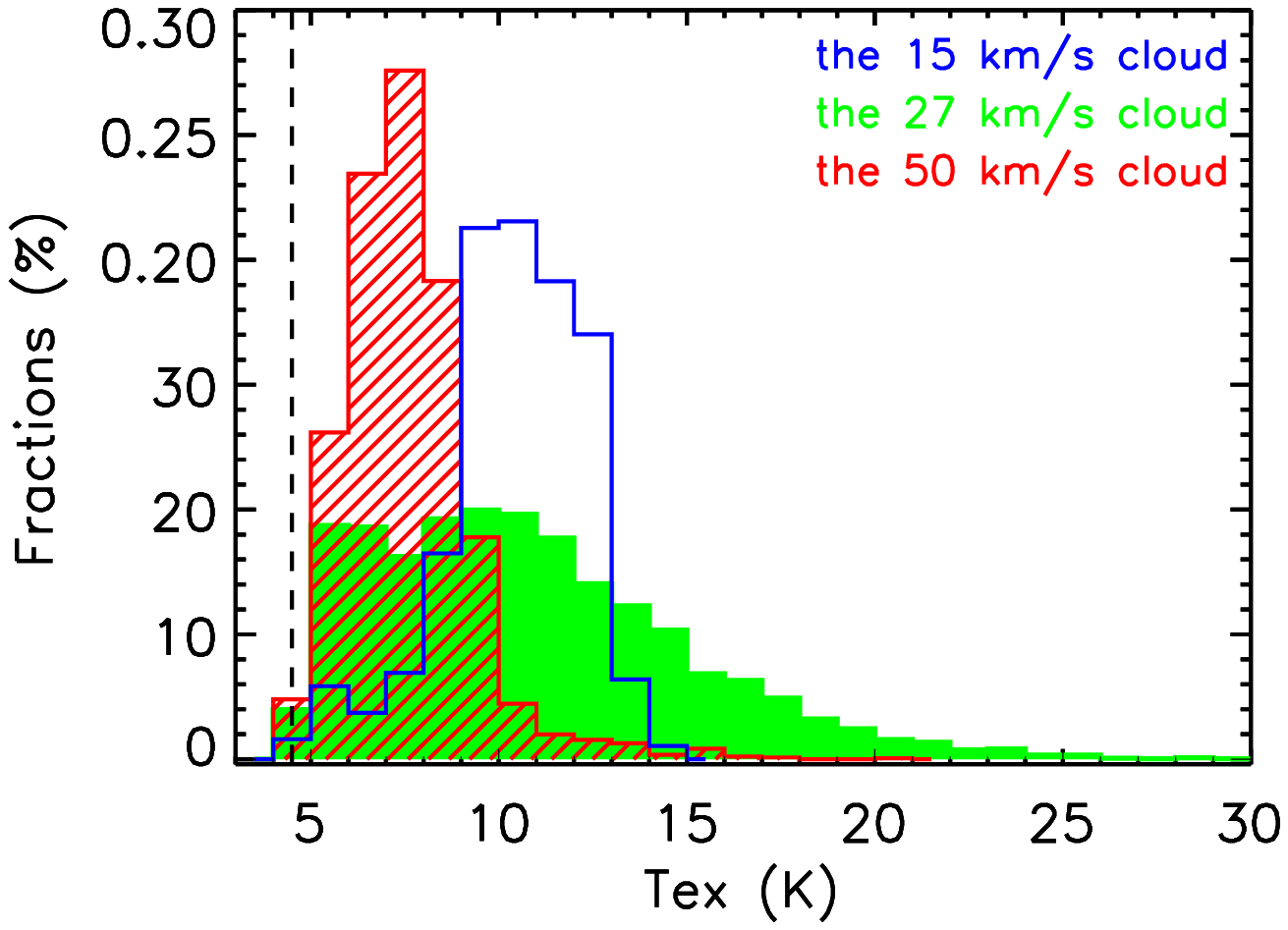}
\includegraphics[width = 0.3 \textwidth]{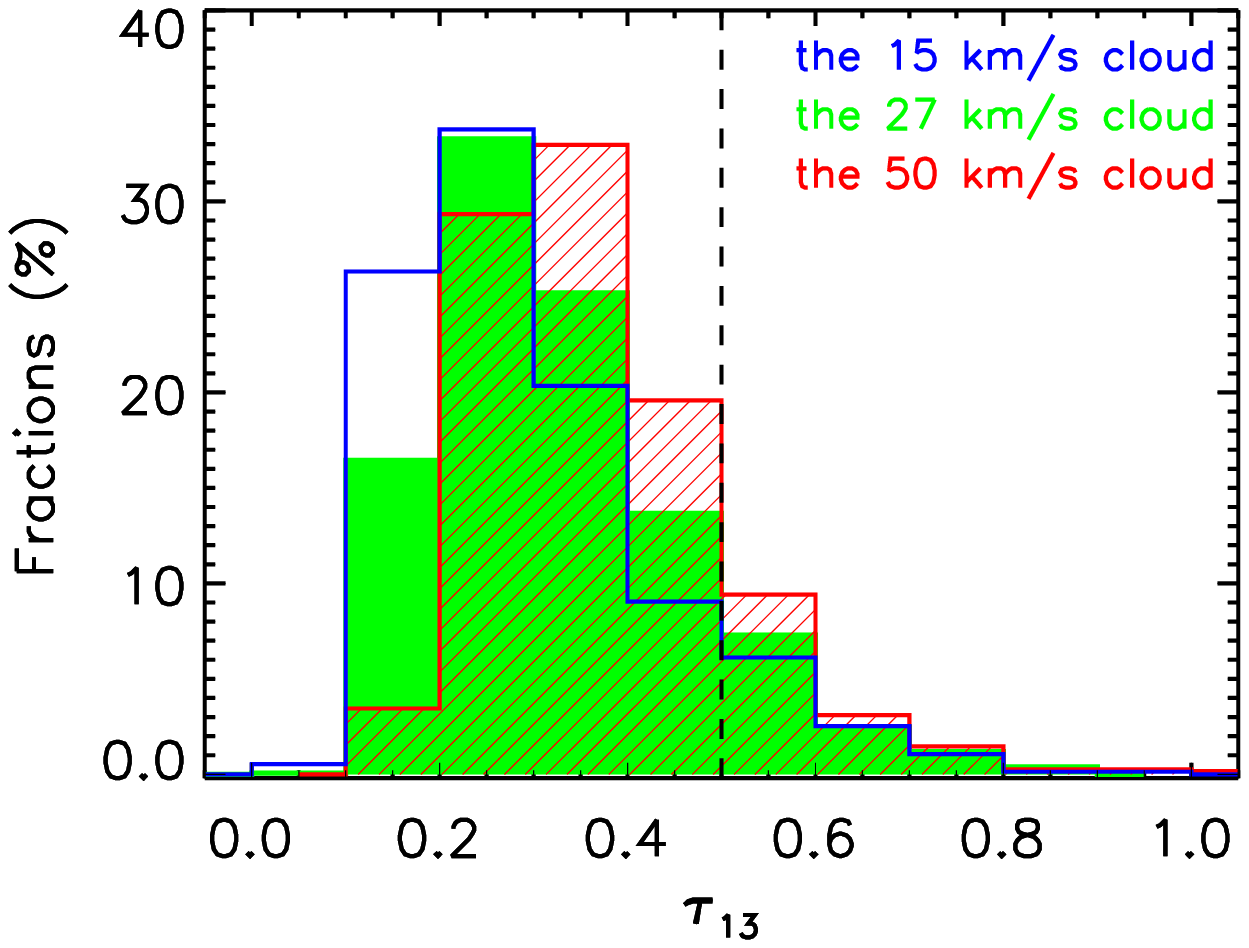}
\includegraphics[width = 0.3 \textwidth]{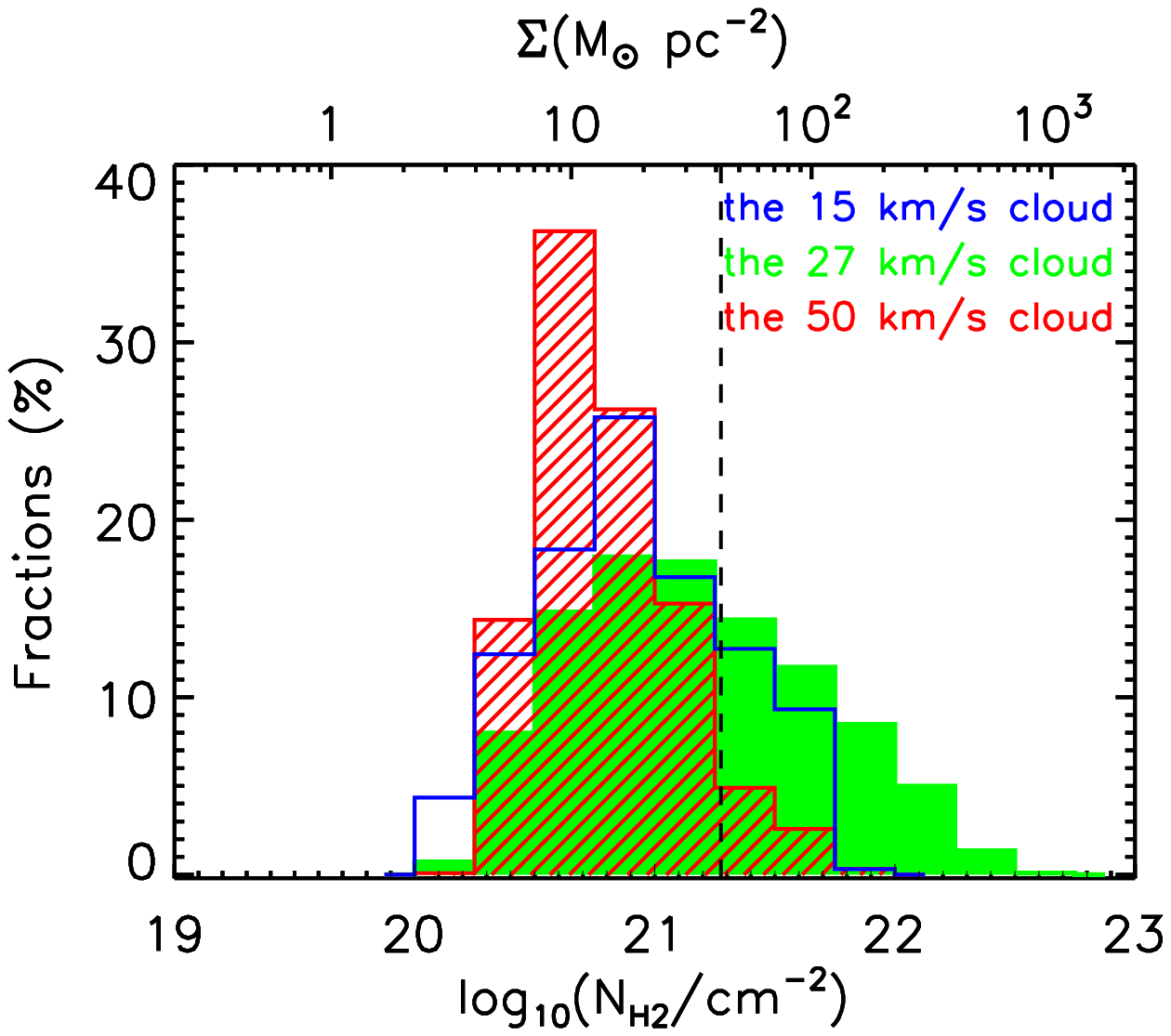}
\caption{{Statistics of the excitation temperature (\textit{left}), $^{13}$CO (1--0) opacity (\textit{middle}), H$_{2}$ column density and surface density derived from $^{13}$CO (1--0) and labeled at the bottom and the top side (\textit{right}) for the three molecular clouds: the 15~\kms\,cloud (open blue histogram), the 27~\kms\,cloud (filled green histogram), and the 50~\kms\,cloud (line-filled red histogram). In the left panel, the black dashed line marks T$_{\rm ex}=~4.5~$K, which corresponds to a $^{12}$CO (1--0) peak temperature of 1.5~K ($\sim$3~$\sigma$, see Sect.~\ref{Sec:obs}). In the middle panel, the black dashed line marks $\tau_{13}=0.5$. In the right panel, the black dashed line represents the median mass surface density of molecular clouds in the Galactic Ring Survey \citep{2009ApJ...699.1092H}.} \label{Fig:property}}
\end{figure*}

\begin{figure*}[!htbp]
\centering
\includegraphics[width = 0.9 \textwidth]{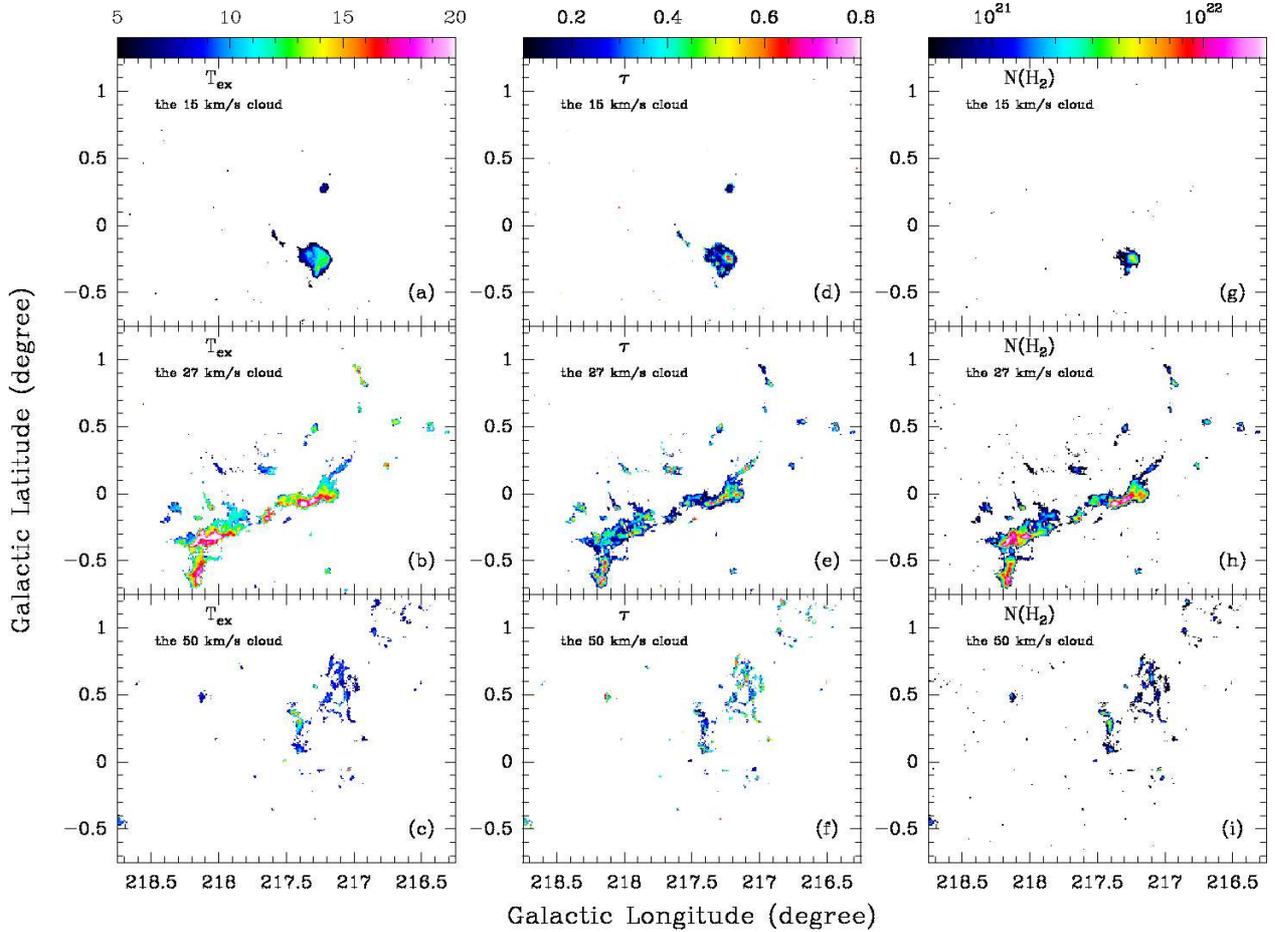}
\caption{{Distribution of the excitation temperature derived from $^{12}$CO (1--0) (Figs.~\ref{Fig:cloudphy}a--\ref{Fig:cloudphy}c), $^{13}$CO (1--0) opacity (Figs.~\ref{Fig:cloudphy}d--\ref{Fig:cloudphy}f), H$_{2}$ column density ((Figs.~\ref{Fig:cloudphy}g--\ref{Fig:cloudphy}i)) derived from $^{13}$CO (1--0) of the three clouds. Corresponding molecular clouds and physical properties are given in the upper left corner of each panel. The color bar of the first column represents the excitation temperature in units of K. The color bar of the second column represents the $^{13}$CO (1--0) opacity. The color bar of the third column represents the H$_{2}$ column density in units of cm$^{-2}$.} \label{Fig:cloudphy}}
\end{figure*}

Following the LTE method, we made a statistical study toward the three clouds. We only took $^{12}$CO (1--0) and $^{13}$CO (1--0) emission detected above $>3\sigma$ ($1\sigma=0.5$~K for $^{12}$CO (1--0) and $1\sigma=0.3$~K for $^{13}$CO (1--0) at a velocity resolution of 0.17~\kms) into account. Figure~\ref{Fig:property} shows the statistical results of excitation temperature derived from optically thick $^{12}$CO (1--0), $^{13}$CO (1--0) opacity, and H$_{2}$ column density toward the three clouds, while Fig.~\ref{Fig:cloudphy} displays their distributions. We find that excitation temperatures range between 4.5~K and 14.2~K with a median temperature of 10.4~K for the 15 \kms\,cloud, 4.5~K and 29.6~K with a median temperature of 10.1~K for the 27~\kms\,cloud, 4.5~K and 20.8~K with a median temperature of 7.4~K for the 50~\kms\,cloud. Parts of the three clouds present excitation temperatures lower than typical kinetic temperatures ($\sim 10$~K) of molecular clouds. As pointed out by \citet{2009ApJ...699.1092H}, this is likely because cloud densities are lower than the critical density ($\sim 10^{3}$~cm$^{-3}$ at 10~K) of $^{12}$CO (1--0) or the filling factor of $^{12}$CO (1--0) emission is lower than unity. $^{13}$CO (1--0) emission is found to be almost optically thin everywhere in the survey. The highest $^{13}$CO (1--0) opacities are 0.97, 0.89 and 1.03 for the 15~\kms\,cloud, the 27~\kms\,cloud, and the 50~\kms\,cloud, respectively. The histogram of $\tau(^{13}{\rm CO})$ shows 48\% of all values below 0.3, 88\% below 0.5, and only lower than 1\% above 0.8. We also calculated surface densities by multiplying H$_{2}$ column densities by the mass of molecular hydrogen. We find that the surface densities of the three clouds (see Table~\ref{Tab:cloudproperty}) are similar to those found in \citet{2009ApJ...699.1092H}.

\section{$^{12}$CO/$^{13}$CO line ratios}\label{t:ratio}
\begin{figure*}[!htbp]
\centering
\includegraphics[width = 0.9 \textwidth]{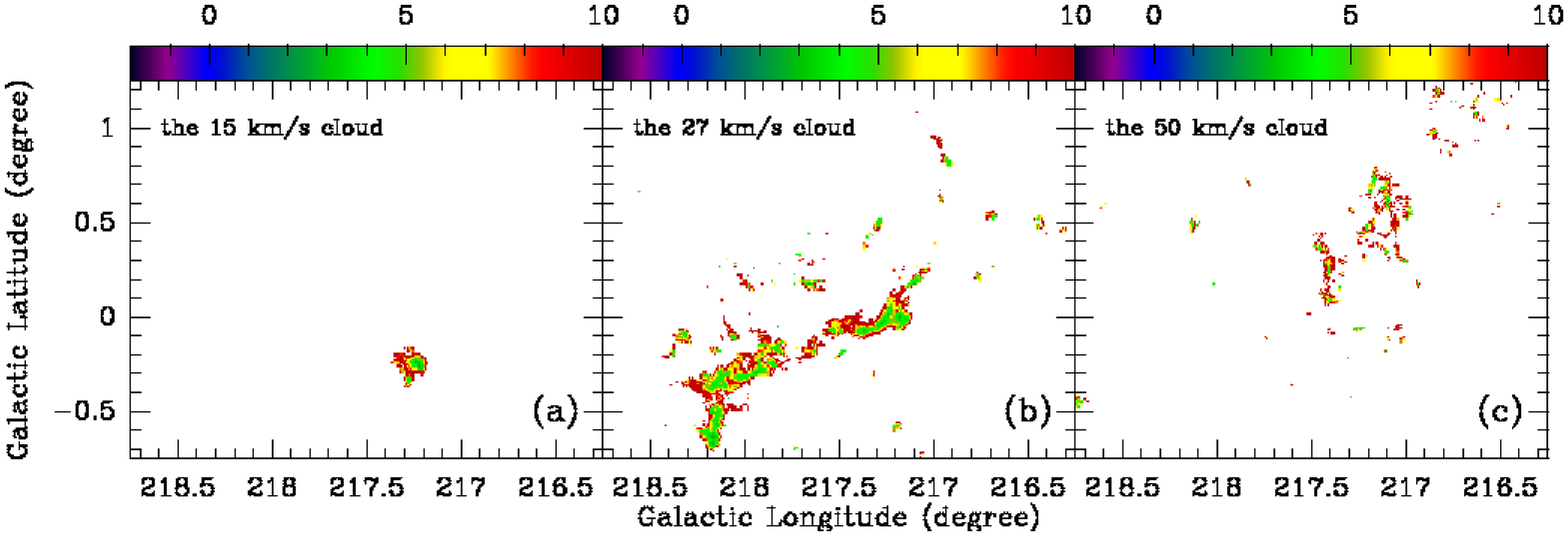}
\caption{{Integrated intensity ratio $^{12}$CO/$^{13}$CO maps of the 15~\kms\,cloud (a), the 27~\kms\,cloud (b), and the 50~\kms\,cloud (c).} \label{Fig:codismap}}
\end{figure*}

Figure~\ref{Fig:codismap} displays the $^{12}$CO/$^{13}$CO integrated intensity ratio maps of the three clouds, where the $^{13}$CO (1--0) emission is detected above 3$\sigma$. We find that the $^{12}$CO/$^{13}$CO integrated intensity ratios at the surface of molecular clouds have much higher values ($>$10) than those ($\sim$3) of their inner parts. This might be attributed to opacity effects. Taking line broadening by opacity into account, the opacities of $^{12}$CO and $^{13}$CO can be estimated with the formula below \citep[Eq. 4.20;][]{2006bookZ},
\begin{eqnarray}\label{f:tau}
\frac{\int T_{\rm mb}(^{12}{\rm CO}){\rm d}v}{\int T_{\rm mb}(^{13}{\rm CO}){\rm d}v}&=&\nonumber  \frac{1-{\rm exp}(-\tau_{12})}{1-{\rm exp}(-\tau_{12}/r)} \times \\
             & & \sqrt{\frac{{\rm ln}\{ \frac{\tau_{12}}{{\rm ln} [2/(1+{\rm exp}(-\tau_{12}))]} \}}{{\rm ln}\{ \frac{\tau_{12}/r}{{\rm ln} [2/(1+{\rm exp}(-\tau_{12}/r))]} \}}}\;,
\end{eqnarray}
where $\tau_{12}$ is the opacity of $^{12}$CO and $r$ is the isotopic ratio [$^{12}$C/$^{13}$C]. Adopting [$^{12}$C/$^{13}$C] isotopic ratios calculated from formula~(\ref{f.ratio}), we find that the opacities of $^{12}$CO vary from lower than 18 to 135, and those of $^{13}$CO vary from lower than 0.1 to 1.3, respectively. Consequently, the correction factors $\tau/[1-{\rm exp(-\tau)}]$ range from lower than 18 to 100  for $^{12}$CO, and vary within a factor of 2 for $^{13}$CO. Therefore, the integrated intensity ratio gradient from the surface of molecular clouds to their inner parts is mainly due to the opacity difference of $^{12}$CO. The derived opacities also support the assumptions used in Sect.~\ref{t:cloud} that $^{12}$CO (1--0) is optically thick and $^{13}$CO (1--0) is almost optically thin for the clouds. However, the isotopic ratio [$^{12}$C/$^{13}$C] may vary considerably, which contributes to the uncertainties of the estimated opacities. Selective photodissociation by UV photons may take effect at the very edge of the molecular clouds due to the difference in self-shielding between $^{12}$CO and $^{13}$CO \citep[e.g,][]{1988ApJ...334..771V}, which will result in a higher isotopic ratio [$^{12}$C/$^{13}$C]. On the other hand, farther inside of molecular clouds (2~mag $<A_{\rm V}<$~5 mag), when $^{12}$CO and $^{13}$CO are both protected from UV photons, the charge exchange reaction with $^{13}$C$^{+}$ will enrich $^{13}$CO, leading to a lower isotopic ratio [$^{12}$C/$^{13}$C] \citep{1976ApJ...205L.165W,2014MNRAS.445.4055S}.  

\clearpage
\section{Identified young stellar objects within the surveyed region}
\renewcommand{\tabcolsep}{0.15 cm}
\normalsize \longtab{1}{
\begin{landscape}
\begin{longtable}{cccccccccccc}
\caption{The infrared photometric magnitudes of identified YSO candidates in the surveyed region.}\label{Tab:irpho}\\
\hline\hline
name                &      $l$    &$b$       &      $J$        &      $H$        & $K_{\rm{s}}$    &    W1         &   W2          & W3           &  W4            & $A_{\rm{v}}$ & Class \\
                   &(\degree)  &(\degree)&  (mag)       &  (mag)        & (mag)         &  (mag)        & (mag)         & (mag)        & (mag)          & (mag)      &       \\
\hline
\endfirsthead
\caption{continued.}\\
\hline\hline
name                &      $l$    &$b$       &      $J$        &      $H$        & $K_{\rm{s}}$    &    W1         &   W2          & W3           &  W4            & $A_{\rm{v}}$ & Class \\
                   &(\degree)  &(\degree)&  (mag)       &  (mag)        & (mag)         &  (mag)        & (mag)         & (mag)        & (mag)          & (mag)      &       \\
\hline
\endhead
\hline
\endfoot

J065851.61$-$030533.6 & 216.530 & 0.239  & 16.97        & 15.64$\pm$0.14 & 15.33$\pm$0.20 & 13.85$\pm$0.03 &  12.24$\pm$0.02  & 9.23$\pm$0.03  & 6.84$\pm$0.09   &   & I  \\ 
J065942.29$-$033643.6 & 217.089 & 0.189  & 15.26$\pm$0.06 & 14.39$\pm$0.05 & 13.67$\pm$0.04 & 12.07$\pm$0.02 &  11.04$\pm$0.02  & 8.49$\pm$0.02  & 6.07$\pm$0.05 &   & I  \\ 
J065835.14$-$030827.2 & 216.542 & 0.156  & 15.17$\pm$0.05 & 13.77$\pm$0.04 & 12.39$\pm$0.03 & 10.73$\pm$0.02 &  9.69$\pm$0.02   & 6.94$\pm$0.02  & 4.91$\pm$0.03 &  1.2 & I  \\ 
J065957.46$-$032958.7 & 217.017 & 0.297  & \nodata    & \nodata    & \nodata    & 13.86$\pm$0.03 &  11.59$\pm$0.02  & 9.00$\pm$0.03  & 6.27$\pm$0.05             &   & I  \\ 
J065908.52$-$034612.8 & 217.165 & -0.008 & 16.66        & 15.24$\pm$0.12 & 13.57$\pm$0.05 & 11.74$\pm$0.02 &  10.25$\pm$0.02  & 7.45$\pm$0.02  & 5.08$\pm$0.03   &   & I  \\ 
J065834.39$-$035645.5 & 217.257 & -0.214 & 13.73$\pm$0.03 & 12.40$\pm$0.02 & 11.38$\pm$0.02 & 10.98$\pm$0.02 &  9.54$\pm$0.02   & 6.41$\pm$0.02  & 4.03$\pm$0.02 &  3.2 & I  \\ 
J065940.30$-$044745.6 & 218.138 & -0.359 & \nodata    & \nodata    & \nodata    & 15.34$\pm$0.05 &  14.32$\pm$0.06  & 9.69$\pm$0.10  & 7.09$\pm$0.13             &   & I  \\ 
J065944.99$-$044716.6 & 218.140 & -0.338 & 17.01        & 15.32$\pm$0.11 & 14.02$\pm$0.06 & 12.93$\pm$0.03 &  11.90$\pm$0.03  & 9.43$\pm$0.06  & 7.43$\pm$0.24   &   & I  \\ 
J065850.06$-$045716.7 & 218.184 & -0.617 & 17.56        & 15.95        & 14.92$\pm$0.14 & 13.60$\pm$0.04 &  12.59$\pm$0.04  & 9.18$\pm$0.06  & 7.30$\pm$0.33     &   & I  \\ 
J065801.41$-$043704.5 & 217.792 & -0.643 & \nodata    & \nodata    & \nodata    & 16 58$\pm$0.19 &   15.35$\pm$0.17 & 9.92$\pm$0.09  & 8.64$\pm$0.43             &   & I  \\ 
J065950.67$-$044603.5 & 218.132 & -0.308 & 17.53        & 15.89        & 14.59$\pm$0.08 & 12.67$\pm$0.05 &  11.52$\pm$0.05  & 8.16$\pm$0.05  & 5.33$\pm$0.06     &   & I  \\ 
J070038.50$-$035139.9 & 217.417 & 0.283  & 16.40$\pm$0.12 & 14.49$\pm$0.05 & 13.45$\pm$0.04 & 12.03$\pm$0.02 &  10.82$\pm$0.02  & 8.11$\pm$0.04  & 5.45$\pm$0.04 &  12.2& I  \\ 
J070023.01$-$043635.7 & 218.053 & -0.116 & 16.69        & 13.98        & 11.63$\pm$0.04 & 9.06$\pm$0.02  &  7.04$\pm$0.02   & 4.71$\pm$0.01  & 1.05$\pm$0.01     &   & I  \\ 
J065941.09$-$044754.4 & 218.141 & -0.357 & 18.50        & 16.09        & 15.06$\pm$0.14 & 13.79$\pm$0.03 &  11.93$\pm$0.03  & 9.60$\pm$0.07  & 6.18$\pm$0.06     &   & I  \\ 
J065858.39$-$045944.1 & 218.236 & -0.605 & 15.66$\pm$0.07 & 14.28$\pm$0.05 & 13.27$\pm$0.04 & 12.61$\pm$0.03 &  11.60$\pm$0.02  & 9.39$\pm$0.06  & 6.50$\pm$0.07 &  3.9 & I  \\ 
J065916.85$-$044757.8 & 218.096 & -0.447 & \nodata    & \nodata    & \nodata    & 14.83$\pm$0.05 &  13.73$\pm$0.05  & 9.38$\pm$0.05  & 7.69$\pm$0.21             &   & I  \\ 
J065855.19$-$045724.1 & 218.195 & -0.599 & \nodata    & \nodata    & \nodata    & 13.57$\pm$0.04 &  11.77$\pm$0.03  & 9.24$\pm$0.07  & 5.88$\pm$0.14             &   & I  \\ 
J065823.79$-$035637.8 & 217.235 & -0.253 & 16.45        & 15.14        & 14.15$\pm$0.07 & 11.80$\pm$0.02 &  10.13$\pm$0.02  & 7.56$\pm$0.02  & 4.82$\pm$0.05     &   & I  \\ 
J065956.46$-$035619.0 & 217.406 & 0.093  & 16.76        & 15.92$\pm$0.17 & 14.47$\pm$0.08 & 12.73$\pm$0.03 &  11.00$\pm$0.02  & 7.16$\pm$0.02  & 4.94$\pm$0.03   &   & I  \\ 
J065956.30$-$035628.9 & 217.408 & 0.091  & 15.02$\pm$0.04 & 13.99$\pm$0.04 & 13.05$\pm$0.03 & 11.84$\pm$0.03 &  10.59$\pm$0.02  & 7.39$\pm$0.02  & 4.99$\pm$0.04 &   & I  \\ 
J065955.89$-$035522.1 & 217.391 & 0.098  & 17.49        & 16.29        & 14.84$\pm$0.09 & 11.94$\pm$0.03 &  10.35$\pm$0.02  & 6.97$\pm$0.02  & 4.17$\pm$0.02     &   & I  \\ 
J065947.96$-$042534.9 & 217.824 & -0.162 & 15.27$\pm$0.05 & 14.60$\pm$0.06 & 14.33$\pm$0.05 & 13.27$\pm$0.03 &  12.10$\pm$0.03  & 9.38$\pm$0.04  & 7.73$\pm$0.16 &  0.4 & I  \\ 
J065859.69$-$034851.4 & 217.187 & -0.061 & \nodata    & \nodata    & \nodata    & 13.74$\pm$0.03 &  12.58$\pm$0.03  & 9.33$\pm$0.08  & 7.13$\pm$0.15             &   & I  \\ 
J070043.05$-$034924.1 & 217.392 & 0.317  & \nodata    & \nodata    & \nodata    & 15.86$\pm$0.09 &  14.80$\pm$0.09  & 9.60$\pm$0.05  & 8.41$\pm$0.38             &   & I  \\ 
J065951.95$-$044435.5 & 218.113 & -0.292 & 16.50        & 14.91$\pm$0.09 & 13.31$\pm$0.04 & 11.16$\pm$0.02 &  9.30$\pm$0.02   & 6.20$\pm$0.02  & 3.43$\pm$0.03   &   & I  \\ 
J065843.07$-$045802.6 & 218.182 & -0.649 & 15.33$\pm$0.05 & 13.68$\pm$0.04 & 12.81$\pm$0.03 & 11.59$\pm$0.02 &  10.40$\pm$0.02  & 7.27$\pm$0.02  & 5.06$\pm$0.04 &  9.8 & I  \\ 
J065906.15$-$045226.2 & 218.142 & -0.521 & 17.06        & 15.37        & 14.58$\pm$0.11 & 13.52$\pm$0.03 &  11.69$\pm$0.03  & 8.57$\pm$0.05  & 5.27$\pm$0.05     &   & I  \\ 
J065932.01$-$045001.5 & 218.156 & -0.407 & 12.74$\pm$0.02 & 11.80$\pm$0.03 & 11.06$\pm$0.02 & 9.72$\pm$0.02  &  8.64$\pm$0.02   & 6.30$\pm$0.03  & 4.09$\pm$0.04 &   & I  \\ 
J065728.00$-$043304.4 & 217.669 & -0.736 & \nodata    & \nodata    & \nodata    & 13.64$\pm$0.03 &  12.28$\pm$0.03  & 8.79$\pm$0.03  & 5.90$\pm$0.05             &   & I  \\ 
J070037.45$-$035148.2 & 217.417 & 0.278  & 17.85        & 17.52        & 15.38$\pm$0.14 & 13.67$\pm$0.03 &  12.20$\pm$0.02  & 9.05$\pm$0.06  & 7.00$\pm$0.12     &   & I  \\ 
J070102.09$-$035120.2 & 217.457 & 0.373  & \nodata    & \nodata    & \nodata    & 15.82$\pm$0.09 &  14.60$\pm$0.08  & 9.80$\pm$0.07  & 7.73$\pm$0.22             &   & I  \\ 
J070053.56$-$034944.9 & 217.417 & 0.354  & 18.16        & 15.79        & 15.05$\pm$0.10 & 12.85$\pm$0.03 &  11.65$\pm$0.03  & 7.82$\pm$0.03  & 5.49$\pm$0.07     &   & I  \\ 
J070056.71$-$034839.4 & 217.407 & 0.374  & \nodata    & \nodata    & \nodata    & 16.88$\pm$0.16 &  15.57$\pm$0.16  & 10.49$\pm$0.10 & 8.63$\pm$0.00             &   & I  \\ 
J070057.46$-$034911.7 & 217.416 & 0.372  & \nodata    & \nodata    & \nodata    & 16.74$\pm$0.16 &  15.45$\pm$0.17  & 9.95$\pm$0.07  & 7.32$\pm$0.12             &   & I  \\ 
J065955.64$-$041359.1 & 217.666 & -0.045 & \nodata    & \nodata    & \nodata    & 16.27$\pm$0.10 &  15.23$\pm$0.13  & 9.71$\pm$0.06  & 7.06$\pm$0.09             &   & I  \\ 
J070035.37$-$032706.3 & 217.047 & 0.459  & 17.01$\pm$0.19 & 15.60$\pm$0.12 & 15.08$\pm$0.12 & 13.59$\pm$0.03 &  12.39$\pm$0.03  & 9.82$\pm$0.05  & 7.12$\pm$0.08 &  7.6 & I  \\ 
J070139.76$-$031036.7 & 216.925 & 0.823  & \nodata    & \nodata    & \nodata    & 13.24$\pm$0.03 &  11.47$\pm$0.02  & 8.87$\pm$0.05  & 6.16$\pm$0.06             &   & I  \\ 
J070057.83$-$033514.3 & 217.210 & 0.480  & 11.03$\pm$0.02 & 9.83$\pm$0.03  & 8.92$\pm$0.03  &  7.69$\pm$0.02 &  6.66$\pm$0.02   & 4.31$\pm$0.01  & 2.36$\pm$0.02 &  2.1 & I  \\ 
J070154.93$-$031128.2 & 216.967 & 0.873  & \nodata    & \nodata    & \nodata    & 16.69$\pm$0.18 &  15.58$\pm$0.16  & 9.68$\pm$0.06  & 7.68$\pm$0.22             &   & I  \\ 
J070042.50$-$032656.2 & 217.058 & 0.487  & \nodata    & \nodata    & \nodata    & 13.99$\pm$0.03 &  12.41$\pm$0.03  & 9.58$\pm$0.04  & 6.61$\pm$0.06             &   & I  \\ 
J070151.97$-$032613.7 & 217.180 & 0.749  & 18.23        & 17.58        & 15.08$\pm$0.10 & 12.30$\pm$0.03 &  10.48$\pm$0.02  & 7.67$\pm$0.02  & 5.41$\pm$0.03     &   & I  \\ 
J070140.27$-$031235.8 & 216.955 & 0.810  & \nodata    & \nodata    & \nodata    & 16.49$\pm$0.20 &  15.44$\pm$0.15  & 9.90$\pm$0.10  & 7.26$\pm$0.14             &   & I  \\ 
J070045.77$-$032023.4 & 216.967 & 0.549  & 15.03$\pm$0.05 & 14.34$\pm$0.07 & 13.05        & 12.10$\pm$0.03 &  10.53$\pm$0.02  & 7.65$\pm$0.02  & 5.13$\pm$0.03   &   & I  \\ 
J070043.75$-$032051.8 & 216.970 & 0.538  & 17.25        & 15.29$\pm$0.09 & 13.95$\pm$0.05 & 12.58$\pm$0.03 &  11.37$\pm$0.02  & 8.54$\pm$0.02  & 6.13$\pm$0.04   &   & I  \\ 
J070139.07$-$031252.0 & 216.957 & 0.803  & \nodata    & \nodata    & \nodata    & 14.65$\pm$0.05 &  12.66$\pm$0.03  & 9.19$\pm$0.08  & 5.75$\pm$0.04             &   & I  \\ 
J070148.59$-$031026.5 & 216.939 & 0.857  & \nodata    & \nodata    & \nodata    & 16.65$\pm$0.17 &  15.47$\pm$0.14  & 10.41$\pm$0.17 & 7.98$\pm$0.27             &   & I  \\ 
J070112.61$-$033904.3 & 217.295 & 0.506  & 18.68        & 17.26        & 15.42$\pm$0.14 & 13.93$\pm$0.03 &  12.76$\pm$0.03  & 10.61$\pm$0.08 & 8.16$\pm$0.24     &   & I  \\ 
J070405.41$-$032530.5 & 217.423 & 1.248  & 16.61$\pm$0.13 & 15.63$\pm$0.12 & 14.68$\pm$0.10 & 12.82$\pm$0.03 &  11.75$\pm$0.02  & 9.34$\pm$0.05  & 7.13$\pm$0.09 &   & I  \\ 
J065611.11$-$032307.0 & 216.486 & -0.488 & 17.12$\pm$0.21 & 15.72$\pm$0.12 & 14.69$\pm$0.11 & 13.63$\pm$0.03 &  12.44$\pm$0.03  & 9.59$\pm$0.05  & 7.18$\pm$0.15 &  4.1 & I  \\ 
J070243.99$-$042432.2 & 218.143 & 0.496  & 15.29$\pm$0.06 & 13.49$\pm$0.04 & 12.12$\pm$0.03 & 10.35$\pm$0.02 &  9.16$\pm$0.02   & 6.70$\pm$0.01  & 4.60$\pm$0.02 &  6.8 & I  \\ 
J070223.13$-$050733.9 & 218.741 & 0.091  & 18.74        & 16.38        & 15.19$\pm$0.10 & 13.35$\pm$0.03 &  12.11$\pm$0.02  & 9.30$\pm$0.04  & 6.96$\pm$0.09     &   & I  \\ 
J070524.99$-$042324.3 & 218.433 & 1.099  & \nodata    & \nodata    & \nodata    & 13.42$\pm$0.03 &  12.25$\pm$0.03  & 9.51$\pm$0.03  & 7.28$\pm$0.09             &   & I  \\ 
J065926.91$-$025424.4 & 216.432 & 0.455  & \nodata    & \nodata    & \nodata    & 16.80$\pm$0.17 &  15.57$\pm$0.15  & 10.28$\pm$0.10 & 8.26$\pm$0.30             &   & I  \\ 
J070118.07$-$024008.5 & 216.432 & 0.975  & 16.85        & 15.45        & 15.26$\pm$0.13 & 14.26$\pm$0.03 &  12.88$\pm$0.03  & 9.22$\pm$0.04  & 7.10$\pm$0.11     &   & I  \\ 
J070200.42$-$024833.5 & 216.637 & 1.068  & 17.88        & 15.88$\pm$0.16 & 14.87$\pm$0.09 & 13.97$\pm$0.03 &  12.87$\pm$0.03  & 9.79$\pm$0.04  & 6.19$\pm$0.04   &   & I  \\ 
J070149.95$-$024730.1 & 216.602 & 1.037  & 16.37$\pm$0.11 & 14.60$\pm$0.06 & 13.15$\pm$0.03 & 10.88$\pm$0.02 &  9.48$\pm$0.02   & 6.75$\pm$0.02  & 3.75$\pm$0.02 &  5.9 & I  \\ 
J065945.95$-$033631.8 & 217.093 & 0.204  & 14.91$\pm$0.04 & 13.78$\pm$0.05 & 12.98$\pm$0.04 & 11.88$\pm$0.02 &  11.14$\pm$0.02  & 9.16$\pm$0.04  & 7.72$\pm$0.21 &  2.2 & II \\ 
J065904.10$-$034606.3 & 217.155 & -0.023 & 14.52$\pm$0.04 & 14.07$\pm$0.05 & 13.84$\pm$0.06 & 13.37$\pm$0.03 &  12.78$\pm$0.03  & 10.03$\pm$0.10 & 6.62$\pm$0.08 &   & II \\ 
J065902.84$-$034520.9 & 217.141 & -0.022 & 15.52$\pm$0.09 & 14.36$\pm$0.07 & 13.81$\pm$0.07 & 12.94$\pm$0.03 &  12.33$\pm$0.03  & 10.30$\pm$0.13 & 7.17$\pm$0.12 &  5.0 & II \\ 
J070015.25$-$033239.2 & 217.091 & 0.342  & 14.52$\pm$0.03 & 13.76$\pm$0.04 & 13.15$\pm$0.03 & 12.24$\pm$0.02 &  11.65$\pm$0.02  & 9.88$\pm$0.05  & 7.30$\pm$0.10 &   & II \\ 
J065944.70$-$033730.2 & 217.105 & 0.192  & 15.48$\pm$0.07 & 14.11$\pm$0.04 & 13.44$\pm$0.04 & 12.93$\pm$0.03 &  12.34$\pm$0.03  & 10.70$\pm$0.09 & 8.21$\pm$0.28 &  6.9 & II \\ 
J070017.70$-$034146.3 & 217.231 & 0.282  & 14.18$\pm$0.04 & 13.52$\pm$0.06 & 13.00$\pm$0.04 & 12.14$\pm$0.02 &  11.75$\pm$0.02  & 9.79$\pm$0.05  & 7.25$\pm$0.11 &   & II \\ 
J070023.03$-$033425.7 & 217.132 & 0.358  & 15.61        & 15.98$\pm$0.15 & 15.21$\pm$0.12 & 13.69$\pm$0.03 &  13.32$\pm$0.03  & 10.94$\pm$0.12 & 8.31$\pm$0.23   &   & II \\ 
J065709.54$-$034534.2 & 216.930 & -0.443 & 14.57$\pm$0.04 & 13.84$\pm$0.03 & 13.28$\pm$0.04 & 12.63$\pm$0.02 &  12.30$\pm$0.03  & 10.63$\pm$0.09 & 7.80$\pm$0.19 &   & II \\ 
J070020.43$-$032821.4 & 217.037 & 0.394  & 15.96$\pm$0.09 & 14.82$\pm$0.08 & 14.11$\pm$0.05 & 13.23$\pm$0.03 &  12.62$\pm$0.03  & 10.56$\pm$0.09 & 7.83$\pm$0.15 &  2.9 & II \\ 
J065851.37$-$045807.5 & 218.199 & -0.619 & 17.82        & 15.48$\pm$0.14 & 14.72$\pm$0.11 & 13.44$\pm$0.03 &  12.56$\pm$0.03  & 9.60$\pm$0.06  & 7.05$\pm$0.10   &   & II \\ 
J070012.32$-$030533.2 & 216.684 & 0.538  & 15.06        & 14.05$\pm$0.07 & 13.24$\pm$0.05 & 12.09$\pm$0.02 &  11.26$\pm$0.02  & 9.03$\pm$0.03  & 7.15$\pm$0.12   &   & II \\ 
J070011.97$-$033912.7 & 217.182 & 0.280  & 14.66$\pm$0.04 & 13.89$\pm$0.05 & 13.26$\pm$0.03 & 12.33$\pm$0.02 &  11.66$\pm$0.02  & 9.60$\pm$0.05  & 7.34$\pm$0.12 &   & II \\ 
J070034.44$-$035123.6 & 217.405 & 0.270  & 14.01$\pm$0.03 & 13.72$\pm$0.04 & 13.69$\pm$0.04 & 13.26$\pm$0.03 &  12.86$\pm$0.03  & 10.71$\pm$0.13 & 8.55$\pm$0.00 &   & II \\ 
J070052.40$-$035025.2 & 217.425 & 0.344  & 15.75$\pm$0.10 & 14.21$\pm$0.07 & 12.96$\pm$0.05 & 10.79$\pm$0.02 &  9.95$\pm$0.02   & 5.70$\pm$0.02  & 3.83$\pm$0.03 &  4.2 & II \\ 
J070043.81$-$035102.0 & 217.418 & 0.308  & 17.05$\pm$0.20 & 15.79$\pm$0.15 & 15.07$\pm$0.10 & 13.91$\pm$0.05 &  13.36$\pm$0.07  & 8.70$\pm$0.06  & 6.48$\pm$0.16 &  4.5 & II \\ 
J065822.65$-$040057.9 & 217.297 & -0.290 & 14.12$\pm$0.03 & 13.33$\pm$0.03 & 12.82$\pm$0.03 & 11.92$\pm$0.02 &  11.44$\pm$0.02  & 9.02$\pm$0.04  & 7.32$\pm$0.12 &   & II \\ 
J065948.47$-$041233.6 & 217.632 & -0.061 & \nodata    & \nodata    & \nodata    & 14.03$\pm$0.05 &  13.44$\pm$0.07  & 8.76$\pm$0.10  & 7.91$\pm$0.40             &   & II \\ 
J070039.28$-$035156.3 & 217.422 & 0.284  & 17.66        & 15.39$\pm$0.11 & 14.49$\pm$0.06 & 13.14$\pm$0.03 &  12.43$\pm$0.03  & 9.44$\pm$0.08  & 8.06$\pm$0.44   &   & II \\ 
J065949.74$-$044527.0 & 218.121 & -0.307 & 15.92$\pm$0.09 & 14.39$\pm$0.05 & 13.67$\pm$0.04 & 13.02$\pm$0.04 &  12.13$\pm$0.03  & 8.92$\pm$0.05  & 7.01$\pm$0.19 &  8.2 & II \\ 
J065955.74$-$043938.2 & 218.047 & -0.240 & 15.14$\pm$0.05 & 14.15$\pm$0.05 & 13.38$\pm$0.03 & 12.63$\pm$0.03 &  12.03$\pm$0.03  & 10.17$\pm$0.09 & 8.45$\pm$0.35 &  0.3 & II \\ 
J065900.28$-$044735.1 & 218.059 & -0.506 & 15.63$\pm$0.07 & 14.70$\pm$0.07 & 14.29$\pm$0.08 & 13.57$\pm$0.03 &  12.92$\pm$0.03  & 10.46$\pm$0.09 & 7.31$\pm$0.10 &  2.8 & II \\ 
J065951.29$-$041346.6 & 217.655 & -0.060 & 16.09$\pm$0.09 & 15.41$\pm$0.10 & 15.37$\pm$0.12 & 13.81$\pm$0.06 &  13.09$\pm$0.06  & 8.33$\pm$0.05  & 7.06$\pm$0.12 &  0.7 & II \\ 
J065834.46$-$040138.0 & 217.329 & -0.251 & 14.85$\pm$0.04 & 13.89$\pm$0.05 & 13.38$\pm$0.04 & 12.82$\pm$0.03 &  12.37$\pm$0.03  & 10.49$\pm$0.10 & 8.26$\pm$0.00 &  3.3 & II \\ 
J065952.31$-$041245.9 & 217.642 & -0.048 & \nodata    & \nodata    & \nodata    & 13.37$\pm$0.04 &  12.37$\pm$0.04  & 8.08$\pm$0.04  & 7.78$\pm$0.27             &   & II \\ 
J065819.92$-$035629.2 & 217.225 & -0.266 & 15.34$\pm$0.06 & 14.38$\pm$0.05 & 13.73$\pm$0.05 & 13.03$\pm$0.02 &  12.67$\pm$0.03  & 11.14$\pm$0.19 & 8.64$\pm$0.48 &  0.8 & II \\ 
J065950.00$-$041218.4 & 217.631 & -0.053 & \nodata    & \nodata    & \nodata    & 13.68$\pm$0.04 &  12.94$\pm$0.06  & 8.15$\pm$0.08  & 7.06$\pm$0.16             &   & II \\ 
J065952.75$-$041308.3 & 217.648 & -0.049 & \nodata    & \nodata    & \nodata    & 13.54$\pm$0.04 &  12.67$\pm$0.06  & 8.23$\pm$0.04  & 7.23$\pm$0.22             &   & II \\ 
J070056.68$-$034954.2 & 217.425 & 0.364  & 16.36$\pm$0.11 & 15.32$\pm$0.09 & 14.75$\pm$0.08 & 13.44$\pm$0.04 &  12.95$\pm$0.04  & 8.28$\pm$0.03  & 5.98$\pm$0.05 &  3.8 & II \\ 
J070047.48$-$035123.2 & 217.430 & 0.319  & 16.82$\pm$0.17 & 15.63$\pm$0.14 & 14.81$\pm$0.09 & 12.85$\pm$0.04 &  12.26$\pm$0.03  & 7.45$\pm$0.05  & 5.81$\pm$0.17 &  2.8 & II \\ 
J065952.32$-$035440.4 & 217.374 & 0.090  & 15.44$\pm$0.06 & 13.65$\pm$0.04 & 12.44$\pm$0.03 & 11.47$\pm$0.02 &  10.58$\pm$0.02  & 7.34$\pm$0.02  & 4.86$\pm$0.03 &  8.1 & II \\ 
J065850.17$-$044924.1 & 218.067 & -0.557 & 15.14$\pm$0.05 & 14.56$\pm$0.06 & 14.34$\pm$0.08 & 13.30$\pm$0.03 &  12.38$\pm$0.03  & 9.42$\pm$0.05  & 6.90$\pm$0.09 &   & II \\ 
J065830.44$-$035933.8 & 217.291 & -0.250 & 14.12$\pm$0.06 & 13.30$\pm$0.06 & 12.98$\pm$0.05 & 12.40$\pm$0.04 &  12.00$\pm$0.04  & 9.48$\pm$0.07  & 7.23$\pm$0.24 &  1.6 & II \\ 
J065953.27$-$045332.5 & 218.248 & -0.355 & 15.42$\pm$0.05 & 14.43$\pm$0.05 & 13.79$\pm$0.04 & 12.67$\pm$0.03 &  11.95$\pm$0.03  & 8.38$\pm$0.04  & 5.86$\pm$0.05 &  1.4 & II \\ 
J065845.17$-$044101.4 & 217.933 & -0.511 & 16.68$\pm$0.15 & 15.66$\pm$0.16 & 14.66$\pm$0.13 & 13.78$\pm$0.04 &  13.23$\pm$0.04  & 10.51$\pm$0.16 & 7.68$\pm$0.14 &   & II \\ 
J065741.02$-$044104.5 & 217.812 & -0.749 & 15.91$\pm$0.09 & 14.93$\pm$0.07 & 14.40$\pm$0.08 & 13.71$\pm$0.03 &  13.09$\pm$0.04  & 10.38$\pm$0.08 & 8.47$\pm$0.31 &  3.2 & II \\ 
J065914.39$-$044141.9 & 217.999 & -0.409 & 15.56$\pm$0.06 & 14.48$\pm$0.06 & 13.78$\pm$0.06 & 12.99$\pm$0.03 &  12.39$\pm$0.03  & 9.74$\pm$0.08  & 7.44$\pm$0.20 &  2.3 & II \\ 
J065926.90$-$043423.2 & 217.914 & -0.307 & 16.07$\pm$0.09 & 15.21$\pm$0.10 & 15.05$\pm$0.15 & 14.80$\pm$0.05 &  14.34$\pm$0.07  & 9.54$\pm$0.08  & 6.78$\pm$0.09 &  2.4 & II \\ 
J065656.73$-$040359.8 & 217.179 & -0.631 & 15.71$\pm$0.07 & 14.86$\pm$0.07 & 14.44$\pm$0.09 & 13.84$\pm$0.03 &  13.31$\pm$0.04  & 11.18$\pm$0.14 & 8.21$\pm$0.20 &  2.4 & II \\ 
J065806.71$-$042023.1 & 217.554 & -0.496 & 15.17$\pm$0.05 & 14.35$\pm$0.07 & 13.67$\pm$0.05 & 13.05$\pm$0.03 &  12.63$\pm$0.03  & 10.65$\pm$0.11 & 8.58$\pm$0.00 &   & II \\ 
J070039.46$-$035134.7 & 217.417 & 0.288  & 15.77$\pm$0.08 & 14.87$\pm$0.07 & 13.94$\pm$0.05 & 11.76$\pm$0.03 &  10.86$\pm$0.02  & 7.62$\pm$0.02  & 4.80$\pm$0.03 &   & II \\ 
J065916.15$-$044253.3 & 218.020 & -0.411 & 13.67$\pm$0.04 & 12.87$\pm$0.04 & 12.36$\pm$0.04 & 11.54$\pm$0.02 &  11.10$\pm$0.02  & 8.62$\pm$0.03  & 6.79$\pm$0.07 &   & II \\ 
J070032.09$-$034713.9 & 217.339 & 0.294  & 17.05        & 16.07$\pm$0.19 & 15.24$\pm$0.13 & 13.89$\pm$0.03 &  13.02$\pm$0.03  & 10.98$\pm$0.16 & 7.75$\pm$0.15   &   & II \\ 
J070102.37$-$035144.3 & 217.463 & 0.371  & 15.33$\pm$0.05 & 14.83$\pm$0.06 & 14.82$\pm$0.09 & 14.58$\pm$0.04 &  14.13$\pm$0.05  & 9.34$\pm$0.06  & 6.73$\pm$0.08 &   & II \\ 
J065901.08$-$034920.4 & 217.197 & -0.059 & 16.32$\pm$0.11 & 14.45$\pm$0.06 & 13.50$\pm$0.04 & 12.52$\pm$0.03 &  11.66$\pm$0.02  & 8.87$\pm$0.05  & 6.58$\pm$0.07 &  11.6& II \\ 
J065941.32$-$042928.3 & 217.869 & -0.216 & 14.16$\pm$0.02 & 13.50$\pm$0.02 & 13.07$\pm$0.03 & 12.36$\pm$0.02 &  11.65$\pm$0.02  & 9.05$\pm$0.03  & 6.75$\pm$0.08 &   & II \\ 
J070036.49$-$035141.2 & 217.413 & 0.276  & 15.58$\pm$0.07 & 13.89        & 13.15        & 12.11$\pm$0.03 &  11.31$\pm$0.02  & 8.89$\pm$0.05  & 6.76$\pm$0.09     &   & II \\ 
J065918.52$-$044500.0 & 218.056 & -0.419 & 15.96$\pm$0.08 & 14.93$\pm$0.07 & 14.17$\pm$0.07 & 13.09$\pm$0.03 &  12.50$\pm$0.03  & 10.07$\pm$0.17 & 8.23$\pm$0.34 &  0.8 & II \\ 
J065846.65$-$045314.1 & 218.117 & -0.599 & 13.85$\pm$0.03 & 13.00$\pm$0.03 & 12.22$\pm$0.03 & 11.18$\pm$0.02 &  10.51$\pm$0.02  & 7.30$\pm$0.02  & 5.17$\pm$0.04 &   & II \\ 
J070136.17$-$043548.8 & 218.181 & 0.160  & 15.20$\pm$0.04 & 14.10$\pm$0.05 & 13.57$\pm$0.04 & 13.03$\pm$0.02 &  12.38$\pm$0.03  & 10.65$\pm$0.09 & 8.67$\pm$0.35 &  4.5 & II \\  
J065929.39$-$045128.3 & 218.172 & -0.428 & \nodata    & \nodata    & \nodata    & 14.86$\pm$0.06 &  14.48$\pm$0.09  & 9.58$\pm$0.08  & 7.36$\pm$0.28             &   & II \\ 
J065931.67$-$044409.0 & 218.068 & -0.363 & \nodata    & \nodata    & \nodata    & 15.21$\pm$0.06 &  14.45$\pm$0.09  & 9.62$\pm$0.08  & 8.28$\pm$0.35             &   & II \\ 
J065959.66$-$044906.2 & 218.194 & -0.298 & 15.28$\pm$0.05 & 14.49$\pm$0.06 & 13.97$\pm$0.05 & 13.26$\pm$0.03 &  12.78$\pm$0.03  & 10.60$\pm$0.14 & 8.60$\pm$0.38 &   & II \\ 
J065755.38$-$043846.4 & 217.805 & -0.678 & 15.92$\pm$0.10 & 14.76$\pm$0.10 & 14.10$\pm$0.08 & 13.36$\pm$0.04 &  12.87$\pm$0.04  & 10.26$\pm$0.11 & 7.78$\pm$0.28 &  4.8 & II \\ 
J065829.95$-$035607.6 & 217.239 & -0.226 & 14.09$\pm$0.03 & 13.12$\pm$0.04 & 12.55$\pm$0.03 & 11.76$\pm$0.02 &  11.28$\pm$0.02  & 9.33$\pm$0.05  & 7.11$\pm$0.12 &  1.7 & II \\ 
J065808.65$-$040044.7 & 217.267 & -0.340 & 11.72$\pm$0.02 & 10.73$\pm$0.02 & 9.88$\pm$0.02  & 8.80$\pm$0.02  &  8.07$\pm$0.02   & 5.59$\pm$0.01  & 3.27$\pm$0.02 &   & II \\ 
J065801.98$-$043127.0 & 217.709 & -0.598 & 14.28$\pm$0.03 & 13.60$\pm$0.04 & 13.25$\pm$0.04 & 12.82$\pm$0.03 &  12.30$\pm$0.03  & 11.03$\pm$0.14 & 8.35$\pm$0.00 &  0.4 & II \\ 
J065949.07$-$044337.8 & 218.093 & -0.295 & 12.58        & 11.73        & 11.26$\pm$0.03 & 10.36$\pm$0.03 &  9.90$\pm$0.02   & 8.49$\pm$0.13  & 6.45$\pm$0.00     &   & II \\ 
J065920.23$-$041132.6 & 217.563 & -0.157 & 16.07$\pm$0.09 & 14.58$\pm$0.06 & 13.42$\pm$0.03 & 12.46$\pm$0.03 &  11.72$\pm$0.03  & 9.80$\pm$0.08  & 7.32$\pm$0.11 &  4.2 & II \\ 
J065832.08$-$044505.3 & 217.969 & -0.591 & 16.73$\pm$0.15 & 15.29$\pm$0.12 & 14.80$\pm$0.11 & 13.73$\pm$0.03 &  13.14$\pm$0.04  & 10.18$\pm$0.10 & 7.68$\pm$0.15 &  7.8 & II \\ 
J070043.71$-$035014.5 & 217.406 & 0.313  & 16.57$\pm$0.15 & 14.74$\pm$0.05 & 13.65$\pm$0.04 & 12.41$\pm$0.02 &  11.87$\pm$0.02  & 8.08$\pm$0.03  & 5.57$\pm$0.06 &  9.6 & II \\ 
J070040.60$-$035124.9 & 217.417 & 0.293  & \nodata    & \nodata    & \nodata    & 14.61$\pm$0.05 &  13.76$\pm$0.06  & 9.37$\pm$0.06  & 7.86$\pm$0.28             &   & II \\ 
J070039.00$-$035202.2 & 217.423 & 0.282  & 16.85        & 15.55$\pm$0.14 & 14.65$\pm$0.08 & 13.40$\pm$0.04 &  12.63$\pm$0.03  & 9.50$\pm$0.08  & 7.55$\pm$0.28 &   & II \\ 
J070157.90$-$030814.6 & 216.924 & 0.908  & 16.76$\pm$0.15 & 15.78$\pm$0.14 & 14.98$\pm$0.10 & 13.63$\pm$0.03 &  12.77$\pm$0.03  & 9.92$\pm$0.05  & 7.31$\pm$0.10 &   & II \\ 
J070112.57$-$033931.3 & 217.302 & 0.502  & 16.28$\pm$0.11 & 15.11$\pm$0.09 & 14.66$\pm$0.08 & 13.76$\pm$0.03 &  13.24$\pm$0.04  & 11.15$\pm$0.13 & 8.49$\pm$0.32 &  4.5 & II \\ 
J070209.70$-$030827.3 & 216.950 & 0.950  & 15.40$\pm$0.05 & 14.64$\pm$0.06 & 14.28$\pm$0.07 & 13.43$\pm$0.03 &  13.12$\pm$0.03  & 11.03$\pm$0.11 & 8.87$\pm$0.36 &  1.5 & II \\ 
J070145.96$-$030819.0 & 216.903 & 0.863  & 15.99$\pm$0.10 & 14.90$\pm$0.09 & 14.12$\pm$0.05 & 13.22$\pm$0.02 &  12.70$\pm$0.03  & 10.20$\pm$0.06 & 7.09$\pm$0.09 &  1.8 & II \\ 
J070140.79$-$031043.8 & 216.929 & 0.826  & 13.88$\pm$0.03 & 12.75$\pm$0.04 & 12.06$\pm$0.03 & 11.06$\pm$0.02 &  10.22$\pm$0.02  & 7.40$\pm$0.02  & 4.61$\pm$0.03 &  3.0 & II \\ 
J070225.79$-$031715.0 & 217.111 & 0.943  & 14.65$\pm$0.03 & 13.59$\pm$0.04 & 12.73$\pm$0.03 & 11.52$\pm$0.02 &  10.89$\pm$0.02  & 8.61$\pm$0.02  & 6.68$\pm$0.06 &  0.5 & II \\ 
J070140.44$-$030409.6 & 216.831 & 0.875  & 14.78$\pm$0.03 & 13.92$\pm$0.04 & 13.41$\pm$0.03 & 12.56$\pm$0.02 &  12.13$\pm$0.03  & 10.43$\pm$0.06 & 8.55$\pm$0.28 &  0.7 & II \\ 
J070115.43$-$030126.4 & 216.743 & 0.803  & 12.88$\pm$0.02 & 12.30$\pm$0.03 & 11.86$\pm$0.02 & 11.04$\pm$0.02 &  10.62$\pm$0.02  & 7.91$\pm$0.02  & 5.75$\pm$0.04 &   & II \\ 
J070148.23$-$025950.6 & 216.781 & 0.936  & 16.67        & 15.67$\pm$0.12 & 14.77$\pm$0.09 & 13.83$\pm$0.03 &  13.10$\pm$0.03  & 11.23$\pm$0.14 & 8.32$\pm$0.23   &   & II \\ 
J070040.32$-$032122.5 & 216.971 & 0.521  & 15.35$\pm$0.05 & 14.51$\pm$0.05 & 13.97$\pm$0.04 & 13.47$\pm$0.05 &  13.15$\pm$0.05  & 11.23$\pm$0.17 & 8.50$\pm$0.41 &  0.1 & II \\ 
J070111.16$-$033850.6 & 217.289 & 0.502  & 14.12$\pm$0.03 & 13.34$\pm$0.03 & 12.80$\pm$0.03 & 12.22$\pm$0.02 &  11.77$\pm$0.02  & 9.66$\pm$0.04  & 8.21$\pm$0.22 &   & II \\ 
J070115.03$-$032509.3 & 217.093 & 0.621  & 16.10$\pm$0.08 & 15.28$\pm$0.09 & 14.86$\pm$0.08 & 13.96$\pm$0.03 &  13.17$\pm$0.04  & 9.94$\pm$0.05  & 8.23$\pm$0.20 &  1.9 & II \\ 
J070132.29$-$030713.3 & 216.861 & 0.821  & 15.30$\pm$0.05 & 14.46$\pm$0.05 & 13.98$\pm$0.04 & 13.06$\pm$0.03 &  12.51$\pm$0.03  & 9.97$\pm$0.05  & 7.57$\pm$0.13 &  0.5 & II \\ 
J070114.34$-$033829.5 & 217.290 & 0.516  & 13.33$\pm$0.03 & 12.48$\pm$0.03 & 11.90$\pm$0.03 & 10.84$\pm$0.02 &  10.41$\pm$0.02  & 8.43$\pm$0.02  & 6.44$\pm$0.05 &   & II \\ 
J070227.87$-$041007.9 & 217.899 & 0.547  & 14.18$\pm$0.03 & 13.70$\pm$0.04 & 13.28$\pm$0.03 & 12.73$\pm$0.02 &  12.24$\pm$0.03  & 10.02$\pm$0.06 & 7.11$\pm$0.10 &   & II \\ 
J070305.74$-$040306.0 & 217.866 & 0.740  & 15.88$\pm$0.10 & 14.85$\pm$0.08 & 13.97$\pm$0.05 & 12.72$\pm$0.03 &  12.02$\pm$0.03  & 9.41$\pm$0.04  & 7.49$\pm$0.14 &   & II \\ 
J070255.66$-$040226.3 & 217.837 & 0.708  & 15.13$\pm$0.05 & 14.18$\pm$0.05 & 13.61$\pm$0.04 & 12.60$\pm$0.02 &  12.02$\pm$0.02  & 9.81$\pm$0.04  & 7.89$\pm$0.17 &  1.4 & II \\ 
J065555.51$-$033043.6 & 216.569 & -0.604 & 14.77$\pm$0.04 & 14.31$\pm$0.04 & 14.00$\pm$0.05 & 13.05$\pm$0.02 &  12.72$\pm$0.03  & 10.79$\pm$0.10 & 8.28$\pm$0.25 &   & II \\ 
J065607.68$-$033949.5 & 216.727 & -0.628 & 15.96$\pm$0.09 & 15.29$\pm$0.09 & 14.56$\pm$0.09 & 13.39$\pm$0.03 &  12.45$\pm$0.03  & 8.78$\pm$0.03  & 6.45$\pm$0.06 &   & II \\ 
J070109.06$-$050234.4 & 218.526 & -0.144 & \nodata    & \nodata    & \nodata    & 13.00$\pm$0.03 &  12.16$\pm$0.03  & 10.07$\pm$0.05 & 7.52$\pm$0.12             &   & II \\ 
J070247.06$-$042233.2 & 218.119 & 0.523  & 14.49$\pm$0.03 & 13.77$\pm$0.04 & 13.17$\pm$0.03 & 12.62$\pm$0.02 &  12.13$\pm$0.02  & 9.43$\pm$0.03  & 6.99$\pm$0.08 &   & II \\ 
J070249.16$-$045340.9 & 218.584 & 0.293  & 16.28$\pm$0.10 & 15.33$\pm$0.09 & 14.58$\pm$0.07 & 13.66$\pm$0.03 &  12.85$\pm$0.03  & 10.78$\pm$0.10 & 8.65$\pm$0.00 &   & II \\ 
J070344.36$-$043329.6 & 218.390 & 0.651  & 16.33$\pm$0.10 & 15.10$\pm$0.08 & 14.15$\pm$0.07 & 13.28$\pm$0.03 &  12.69$\pm$0.03  & 11.30$\pm$0.17 & 8.47$\pm$0.00 &  2.2 & II \\ 
J070240.89$-$042427.0 & 218.135 & 0.486  & 18.14        & 16.38$\pm$0.22 & 14.99$\pm$0.09 & 13.93$\pm$0.03 &  12.94$\pm$0.03  & 10.18$\pm$0.06 & 7.40$\pm$0.11   &   & II \\ 
J070248.80$-$045410.1 & 218.591 & 0.288  & 16.83$\pm$0.17 & 15.90$\pm$0.15 & 15.05$\pm$0.09 & 13.82$\pm$0.03 &  13.10$\pm$0.04  & 10.73$\pm$0.10 & 8.47$\pm$0.32 &   & II \\ 
J065943.07$-$025321.4 & 216.447 & 0.523  & 10.86$\pm$0.03 & 10.11$\pm$0.03 & 9.63$\pm$0.02  & 8.43$\pm$0.02  &  7.49$\pm$0.02   & 5.50$\pm$0.01  & 4.43$\pm$0.03 &   & II \\ 
J065951.79$-$025604.8 & 216.504 & 0.534  & 17.05$\pm$0.18 & 15.80$\pm$0.14 & 14.87$\pm$0.08 & 13.96$\pm$0.03 &  13.18$\pm$0.03  & 10.23$\pm$0.07 & 7.60$\pm$0.14 &  2.7 & II \\ 
J065857.91$-$033007.1 & 216.906 &0.076   & 16.44$\pm$0.16 & 15.36$\pm$0.17 & 14.80$\pm$0.13 & 13.81$\pm$0.03 & 13.50$\pm$0.04   & 11.94$\pm$0.44 & 8.42$\pm$0.00 & 4.5  & II \\
J065945.11$-$044431.0 & 218.099 &$-$0.317& 15.96$\pm$0.09 & 14.80$\pm$0.07 & 13.90$\pm$0.07 & 13.06$\pm$0.04 & 12.52$\pm$0.04   & 9.84           & 7.12$\pm$0.37 & 1.8  & II \\ 
J065844.91$-$045644.3 & 218.166 &$-$0.632& 16.94$\pm$0.20 & 15.34$\pm$0.10 & 14.49$\pm$0.10 & 13.60$\pm$0.04 & 12.74$\pm$0.03   & 10.32          & 7.86$\pm$0.46 & 9.4  & II \\ 
J065926.67$-$041447.1 & 217.623 &$-$0.158& 15.63$\pm$0.06 & 15.15$\pm$0.09 & 14.68$\pm$0.12 & 13.11$\pm$0.05 & 12.76$\pm$0.05   & 8.68$\pm$0.51  & 5.21$\pm$0.14 & 0.0  & II \\ 
J065749.62$-$043208.3 & 217.696 &$-$0.649& 16.04$\pm$0.09 & 15.01$\pm$0.11 & 14.84$\pm$0.14 & 14.08$\pm$0.03 & 13.58$\pm$0.05   & 12.35$\pm$0.51 & 8.56$\pm$0.37 & 4.1  & II \\ 
J065929.24$-$043538.0 & 217.937 &$-$0.308& 15.60$\pm$0.06 & 14.49$\pm$0.06 & 13.96$\pm$0.07 & 13.35$\pm$0.04 & 12.94$\pm$0.04   & 10.55          & 8.54$\pm$0.00 & 4.6  & II \\ 
J065940.56$-$044637.1 & 218.121 &$-$0.349& 15.62$\pm$0.07 & 14.24$\pm$0.05 & 13.44$\pm$0.04 & 12.65$\pm$0.03 & 12.04$\pm$0.03   & 10.66          & 7.23$\pm$0.00 & 5.6  & II \\ 
J065948.09$-$044537.8 & 218.121 &$-$0.314& 15.95$\pm$0.08 & 14.60$\pm$0.06 & 13.82$\pm$0.04 & 13.03$\pm$0.05 & 12.40$\pm$0.04   & 10.37          & 7.10$\pm$0.00 & 5.5  & II \\ 
J065937.35$-$044313.5 & 218.065 &$-$0.335& 16.08$\pm$0.09 & 15.08$\pm$0.10 & 14.40$\pm$0.10 & 13.56$\pm$0.13 & 13.24$\pm$0.09   & 8.92           & 6.16$\pm$0.37 & 1.3  & II \\ 
J065927.44$-$035842.5 & 217.386 &$-$0.033& 15.28$\pm$0.05 & 14.14$\pm$0.05 & 13.78$\pm$0.05 & 13.25$\pm$0.03 & 12.75$\pm$0.03   & 10.77          & 5.73$\pm$0.06 & 5.0  & II \\ 
J065929.57$-$043554.9 & 217.942 &$-$0.309& 16.68$\pm$0.16 & 15.52$\pm$0.15 & 14.98$\pm$0.13 & 13.81$\pm$0.05 & 13.23$\pm$0.05   & 11.06$\pm$0.49 & 8.53$\pm$0.00 & 4.9  & II \\ 
J065844.59$-$045610.5 & 218.157 &$-$0.629& 16.23$\pm$0.10 & 15.12$\pm$0.11 & 14.43$\pm$0.08 & 13.58$\pm$0.05 & 13.01$\pm$0.04   & 9.80           & 7.00$\pm$0.00 & 2.7  & II \\ 
J065839.23$-$045541.4 & 218.139 &$-$0.645& 15.61$\pm$0.06 & 14.34$\pm$0.06 & 13.88$\pm$0.05 & 13.28$\pm$0.03 & 12.78$\pm$0.03   & 10.74          & 7.51$\pm$0.37 & 6.3  & II \\ 
J065845.75$-$045624.7 & 218.163 &$-$0.626& 15.99$\pm$0.09 & 14.65$\pm$0.06 & 13.89$\pm$0.06 & 12.69$\pm$0.03 & 12.08$\pm$0.03   & 10.26$\pm$0.53 & 6.95$\pm$0.00 & 5.3  & II \\ 
J065917.52$-$040035.2 & 217.395 &$-$0.084& 16.12$\pm$0.09 & 14.92$\pm$0.07 & 14.29$\pm$0.07 & 13.34$\pm$0.03 & 12.82$\pm$0.03   & 12.31          & 3.04$\pm$0.04 & 5.6  & II \\ 
J065940.58$-$041031.3 & 217.586 &$-$0.074& 14.62$\pm$0.03 & 13.64$\pm$0.03 & 13.02$\pm$0.04 & 12.45$\pm$0.03 & 12.07$\pm$0.03   & 10.45$\pm$0.49 & 6.92$\pm$0.14 & 1.6  & II \\ 
J065918.52$-$040016.9 & 217.393 &$-$0.078& 16.01$\pm$0.08 & 14.28$\pm$0.04 & 13.18$\pm$0.03 & 11.86$\pm$0.02 & 11.10$\pm$0.02   & 12.30          & 3.45$\pm$0.07 & 8.0  & II \\ 
J070139.28$-$031022.2 & 216.920 &0.823   & 16.27$\pm$0.10 & 15.19$\pm$0.07 & 14.77$\pm$0.08 & 13.83$\pm$0.04 & 13.15$\pm$0.03   & 11.13$\pm$0.43 & 8.75$\pm$0.00 & 3.7  & II \\ 
\end{longtable}                                                                                                                                                      
\end{landscape}                                                                                                                                                      
 }

\end{appendix}

\end{document}